\documentclass[%
 reprint,
 amsmath,amssymb,
 aps,
]{revtex4-2}

\usepackage{hyperref}
\hypersetup{
    colorlinks=true,
    linkcolor=blue,
    filecolor=magenta,      
    urlcolor=cyan,
    pdftitle={Overleaf Example},
    pdfpagemode=FullScreen,
    }

\usepackage{booktabs}
\usepackage[table,xcdraw]{xcolor}

\usepackage{dsfont}

\usepackage{xcolor}

\def\ket#1{\left| #1 \right\rangle}

\newcommand{\angstrom}{\textup{\AA}}
\usepackage{graphicx}
\usepackage{dcolumn}
\usepackage{bm}

\begin{document}


\title{Precise control of entanglement in multinuclear spin registers coupled to defects}

\author{Evangelia Takou}
\email{etakou@vt.edu}
\affiliation{Department of 
Physics, Virginia Tech, 24061 Blacksburg, VA, USA}
 \author{Edwin Barnes}%
 \email{efbarnes@vt.edu}
 \affiliation{Department of 
Physics, Virginia Tech, 24061 Blacksburg, VA, USA}
\author{Sophia E. Economou}%
 \email{economou@vt.edu}
\affiliation{Department of 
Physics, Virginia Tech, 24061 Blacksburg, VA, USA}

\begin{abstract}
Quantum networks play an indispensable role in quantum information tasks such as secure communications, enhanced quantum sensing, and distributed computing. Among the most mature and promising platforms for quantum networking are nitrogen-vacancy centers in diamond and other color centers in solids. One of the challenges in using these systems for networking applications is to controllably manipulate entanglement between the electron and the nuclear spin register despite the always-on nature of the hyperfine interactions, which makes this an inherently many-body quantum system. Here, we develop a general formalism to quantify and control the generation of entanglement in an arbitrarily large nuclear spin register coupled to a color center electronic spin. We provide a reliable measure of nuclear spin selectivity, by exactly incorporating into our treatment the dynamics with unwanted nuclei. We also show how to realize direct multipartite gates through the use of dynamical decoupling sequences, drastically reducing the total gate time compared to protocols based on sequential entanglement with individual nuclear spins. We quantify the performance of such gate operations in the presence of unwanted residual entanglement links, capturing the dynamics of the entire nuclear spin register. Finally, using experimental parameters of a well-characterized 27 nuclear spin register device, we show how to prepare with high fidelity entangled states for quantum error correction.
\end{abstract}

\maketitle


\section{Introduction}

Controlling on-demand quantum nodes with high precision and scaling up to build large-scale quantum architectures is the ultimate goal of quantum information processing. Quantum networks are clusters of nodes interconnected via communication channels, which transfer information or distribute entanglement using photons \cite{Wehner2019}. Long-distance connections are established by breaking the transmission distance into smaller segments and creating intermediate entanglement links through quantum repeaters \cite{BriegelPRL1998}. Quantum networks will enable secure communication \cite{MunroPRA2017,BensonNewJPhys2014,LimNatCommun2021,UrsinNature2018} between qubit devices and enhance quantum computing or sensing capabilities \cite{WrachtrupNatCommun2016,CappellaroPRapplied2019,WrachtrupNpj2021} by using entanglement as a resource. Few-node quantum networks in spin-based solid-state platforms have already been realized using NV centers in diamond~\cite{HansonNature2013,PompiliSci21,KalbSci2017}, SiV centers in diamond~\cite{LukinPRL2019,LukinPRB2019}, or quantum dots~\cite{MetePRL17}. Proposals for hybrid architectures complemented by transducers~\cite{NeumanNPJ2021} or modular designs~\cite{MonroePRA2014,MonroeNature2015}  have also been put forward. In defect platforms, the electronic spin serves as the communication qubit, because it features a spin-photon interface, while nearby nuclear spins can serve as long-lived quantum memories.  
 
A challenge with exploiting the long coherence times of the nuclear spins is twofold: (i) the interactions between the nuclear spins and the electronic defect are always on (not switchable) and (ii) the majority of the nuclear spins are located at distant lattice sites, which leads to interactions that are weak compared to the dephasing rate of the defect spin. Fortunately, both these issues can be addressed simultaneously through the use of dynamical decoupling (DD) pulse sequences~\cite{TaminiauPRL2012}. The parameters associated with these DD sequences (specifically, the interpulse spacing) are selected such that, ideally, all nuclear spins except for one are decoupled from the defect. This effectively creates a knob to select a target nuclear spin. By varying the pulse spacing, different nuclear spins can be selected across the register. This approach has led to bold first steps toward distributing entanglement across a network of a few quantum nodes \cite{PompiliSci21,HumphreysNat18}, realizing error-correction schemes~\cite{TaminiauNatNano2014,CramerNatCommun2016,TaminiauNature2022}, performing entanglement distillation~\cite{KalbSci2017}, or implementing quantum repeater protocols \cite{ElkoussPRA2019}.
 
Despite these seminal experimental demonstrations, critical challenges remain in exploiting nuclear spins as quantum memory registers for networks. A key issue is that, due to the many-body nature of this always-coupled system, the electron is never fully decoupled from the remaining nuclear spins, leading to residual electron-nuclear entanglement. This lowers the fidelity of the gates, and can be detrimental in the operation of the network. An additional consideration is that in these DD control protocols, the gates between the defect and each nuclear spin are implemented sequentially, which can lead to impractically long operations in the encoding and decoding steps of quantum error correction. While these issues can be in part addressed by adding controls to the system, e.g., by directly driving the nuclear spins through nuclear magnetic resonance~\cite{BradleyPRX19}, this complicates the experiment significantly, leading to a potentially impractical overhead that could limit scalability.  
 
In this paper, we address these challenges by developing a formalism that allows us to capture the dynamics of the full system. This in turn enables us to both characterize the quality of the electron-nuclear gates and to design DD sequences that can directly create multipartite entangling gates within the defect-nuclear spin register. A key insight in our approach is that the form of the Hamiltonian allows an exact analysis of the whole system in terms of only bipartite dynamics. We use the notion of one-tangles, an entanglement measure that captures quantum correlations between a single spin and a spin ensemble. We present closed-form expressions for the one-tangles of individual nuclear spins in the register and of the defect electronic spin. Remarkably, these one-tangles depend only on two-qubit Makhlin invariants (parameters that quantify and classify the entangling power of two-qubit gates). This critical simplification allows us to systematically determine the DD sequences that maximize or minimize the one-tangles as desired for nuclear spin registers containing up to hundreds of nuclei. We use this approach to find sequences that create entanglement between the electron and a target subset of nuclei while simultaneously decoupling unwanted nuclei. We show that it is possible to perform controlled entangling operations involving three nuclear spins more than four times faster than sequential gate approaches while achieving significantly higher gate fidelities, which capture errors due to the presence of the entire nuclear register. We further reformulate the three-qubit bit-flip code in terms of the multi-spin gates and, using parameters from the well-characterized 27-qubit device by the Delft group, we show that the electron's state can be retrieved with probability $>99\%$. Our approach provides a practical and scalable means for selecting nuclear spins as quantum memory qubits and for designing gates among them that can prepare entangled multipartite states for efficient encoding and decoding steps in quantum error correction protocols.
 
The paper is organized as follows. In Sec.~\ref{Sec:PropertiesSeqs}, we review and generalize existing results on $\pi$-pulse sequences used for controlling single nuclear spins. In Sec.~\ref{Sec:Entang}, we quantify entanglement in the case of a single nuclear spin coupled to the electron, and we present our formalism for the entanglement distribution in the entire nuclear spin register. Finally, in Sec.~\ref{Sec:CR}, we show how to perform multi-spin gates, quantify their gate fidelity in the presence of spectator nuclei, and show how to use these gates for quantum error correction codes.

\section{Controlling a single nuclear spin \label{Sec:PropertiesSeqs}}

The application of periodic trains of pulses on the electron interleaved by free-evolution periods can generate either single-qubit gates on a nuclear spin or entangle it with the electronic spin. This is because dynamical decoupling sequences can modify the effective electron-nuclear hyperfine interaction, allowing one to couple a specific nucleus to the electron while decoupling others. Well-known examples of dynamical decoupling sequences that have been under investigation for many decades include the  Carr-Purcell-Meiboom-Gill (CPMG)~\cite{CarrPurcellPhysRev54,MeiboomGill58,deLangeSci10,TerryMagnRes90} and Uhrig (UDD)~\cite{UhrigNewJPhys08,UhrigPRL07} sequences. In this section, we review and generalize existing results for single nuclear spin control via electronic spin driving. In subsequent sections, we treat the problem of controlling multiple nuclear spins at the same time.

\subsection{Creating electron-nuclear spin entanglement \label{Sec:ResConds}}

We begin with the task of creating electron-nuclear spin entanglement. It was shown in Ref.~\cite{TaminiauPRL2012} that by choosing the pulse spacing to satisfy a certain resonance condition that depends on the hyperfine couplings, it is possible to rotate a target nuclear spin in a way that depends on the electronic spin state. This is done using pulse sequences that are obtained by concatenating a basic ``unit" multiple times. For example, the CPMG sequence can be expressed in terms of $N$ units as $(t/4-\pi-t/2-\pi-t/4)^N$, where $t$ is the duration of the unit, and $\pi$ represents a $\pi$-pulse. The pulses are implemented experimentally via a microwave (MW) drive to directly induce transitions between electronic spin states. The idealized instantaneous $\pi$-pulses, in reality, have finite amplitude and duration; they could be generated using a vector source \cite{BradleyThesis2021}, whose characteristics (e.g. frequency, duration, amplitude) are pre-defined by an arbitrary waveform generator, and their shapes could, for example, be Hermite envelopes~\cite{BradleyPRX19,ChuangRevModPhys2005}.

The Hamiltonian for a single nuclear spin ($I=1/2$) is given by~\cite{Freeman1967}:
\begin{equation}\label{Eq:Hamiltonian}
    \begin{split}
    H&=\frac{\omega_L}{2} \mathds{1}\otimes \sigma_z+\frac{A}{2} Z_e\otimes \sigma_z + \frac{B}{2} Z_e\otimes \sigma_x
    \\&=\sigma_{00}\otimes H_0 +\sigma_{11}\otimes H_1,
    \end{split},
\end{equation}
where $\sigma_j$ are the Pauli matrices, $\omega_L$ is the Larmor frequency of the nuclear spin, and $A$ and $B$ are the parallel and perpendicular components of the hyperfine interaction respectively. The electron spin operator $Z_e$ is defined as $Z_e=s_0|0\rangle\langle 0|+s_1|1\rangle\langle 1|$, where $\ket{0}$ and $\ket{1}$ are the two levels of the electron spin multiplet used to define the qubit, and $s_j$ are the corresponding spin projection quantum numbers. Further, we define $H_j$ as $H_j=1/2[(\omega_L+s_j A)\sigma_z+s_jB\sigma_x]$. From the above Hamiltonian, it follows that the electron-nuclear spin evolution operator after one unit of the pulse sequence is given by
\begin{equation}\label{Eq:U}
U=\sigma_{00}\otimes R_{\textbf{n}_0}(\phi_0)+\sigma_{11}\otimes R_{\textbf{n}_1}(\phi_1),
\end{equation}
where $\sigma_{jj}\equiv |j\rangle\langle j|$ are projectors onto two of the levels in the electron spin multiplet, and $R_{\textbf{n}_j}(\phi_j)=e^{-i\phi_j/2(\boldsymbol{\sigma}\cdot \textbf{n}_j)}$ denotes two different conditional nuclear spin evolution operators specified by rotation axes $\textbf{n}_j$ and angles $\phi_j$. Both $\textbf{n}_j$ and $\phi_j$ in general depend on the electron's spin state and on the pulse sequence. The explicit form of $R_{\textbf{n}_j}(\phi_j)$ in the case of CPMG is found in Appendix~\ref{App:EvolOper}. 

To create entanglement, we need the two rotation operators, $R_{\textbf{n}_j}(\phi_j)$, to differ. It is in fact possible to choose the pulse time $t$ such that the nuclear spin axes are antiparallel, i.e., $\textbf{n}_0\cdot \textbf{n}_1=-1$. At the same time, the coherence function $P_x$, which is the probability for an electron prepared in state $|+\rangle$ to return to this state at time $t$, reaches a minimum. The coherence function can be expressed as $P_x=1/2(1+M)$, where $M=\frac{1}{2}\Re\text{Tr}[R_{\textbf{n}_0}(\phi_0) R_{\textbf{n}_1}^\dagger(\phi_1)]$ [see also Ref.~\cite{TaminiauPRL2012} and Appendix~\ref{App:ResTimes}]. As has been shown in Ref.~\cite{TaminiauPRL2012}, for $\phi_0=\phi_1\equiv \phi$ (which holds for CPMG), $M$ is given by $M=1-\sin^2(\phi/2)(1-\textbf{n}_0\cdot \textbf{n}_1)$. By calculating $M$ analytically using the explicit expressions for the conditional evolution operators $R_{\textbf{n}_j}(\phi_j)$, and by setting $\textbf{n}_0\cdot \textbf{n}_1=-1$, the resonance times can be obtained. For the CPMG, UDD$_3$ and UDD$_4$ sequences, we find that these resonances occur at times
\begin{equation}\label{Eq:ResTimes}
    t_k = \frac{4\pi(2k-1)}{\tilde{\omega}},
\end{equation}
where $\tilde{\omega}=\omega_0+\omega_1$, $\omega_j = \sqrt{(\omega_L+s_jA)^2+(s_j B)^2}$, and $k$ is the order of the resonance. This expression for $t_k$, which is valid for $\omega_L\gg A,B$, combines and generalizes known results. For example, the resonance times of Eq.~(\ref{Eq:ResTimes}) have been shown in Ref.~\cite{TaminiauPRL2012,Dong2020} for $s_0=0$, $s_1=-1$, while in Ref.~\cite{LukinPRL2019} for $s_0=-s_1=1/2$. For the UDD$_4$ sequence we find that there are additional resonances at times $t_k=8 \pi(2k-1)/\tilde{\omega}$, which was also reported in Ref.~\cite{Dong2020}. All resonance times are valid for any electronic spin projection and any type of nuclear spin with $I=1/2$ (e.g. $^{13}$C in diamond/SiC or $^{29}$Si in SiC).

An entangling gate is achieved by iterating the sequence an appropriate number $N$ to accumulate a desired rotation angle on the nuclear spin.  We present the rotation angles for the three pulse sequences in Appendix~\ref{App:RotationAngles}. Sequences with an odd number of pulses in the basic unit need to be repeated twice to ensure the electron returns to its initial state. For CPMG and UDD$_3$, we find that the rotation angles per iteration are equal, i.e., $\phi_0=\phi_1$. One way to generate an entangling gate is to set the unit time equal to a resonance time and repeat the sequence such that it leads to a total angle of $\pi/2$, and hence, implements a CR$_x(\pi/2)$ gate \cite{TaminiauPRL2012,TaminiauNatNano2014}. This is possible since the evolution operator after $N$ repetitions of the basic unit retains the form of Eq.~(\ref{Eq:U}) with $\phi_j$ replaced by the total rotation angle, $\phi_j(N)$, whereas the dot product $\textbf{n}_0\cdot \textbf{n}_1$ is independent of $N$ at resonance. However, this latter feature does not hold for any sequence. In principle, one can realize entangling operations beyond CR$_x(\pi/2)$, which we will explore later on in Sec.~\ref{Sec:CR}. 

\begin{figure}[!htbp]
    \centering
    \includegraphics[scale=0.50]{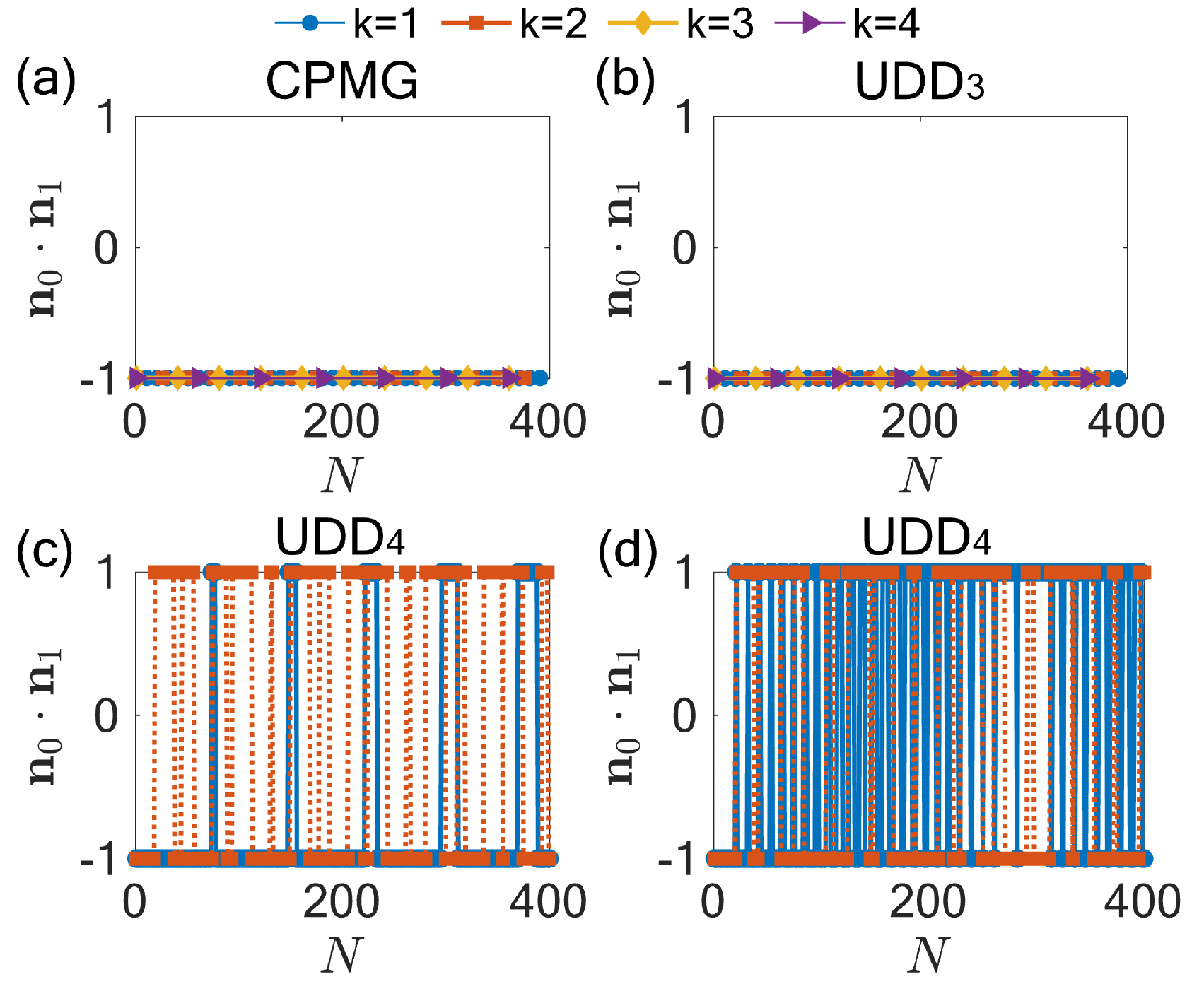}
    \caption{Dot product of nuclear spin rotation axes for (a) CPMG, (b) UDD$_3$, (c,d) UDD$_4$ versus the number of iterations $N$ of the basic pulse sequence unit at the first four ($k\in[1,4]$) resonances of a target spin [$(A,B,\omega_L)=2\pi\cdot(80,25,314)$~kHz] for an electronic spin with $S=1/2$. For CPMG and UDD$_3$, $\textbf{n}_0\cdot \textbf{n}_1$ is independent of $N$ since the rotation angles $\phi_0$ and $\phi_1$ per iteration are equal. For UDD$_4$, the dot product jumps between $-1$ and $+1$ due to the different rotation angles $\phi_j$. In the ranges where $\textbf{n}_0\cdot \textbf{n}_1=1$, it holds that $\phi_0\approx \phi_1$, and the rotation of the nuclear spin is unconditional on the electron. In (c) we consider the resonance time $4\pi(2k-1)/\tilde{\omega}$ and in (d) the time $8\pi(2k-1)/\tilde{\omega}$. The times for (a) are $t_k=(3.1822,~9.5465,~15.9108,~22.2751)~\mu$s (b) $t_k=(3.1850,~9.5537,~15.9124,~22.2805)~\mu$s, (c) $t_k=(3.1857,~9.5481,~15.9169,~22.2737)~\mu$s, (d) $t_k=(6.3661,~19.0883,~31.8190,~44.5509)~\mu$s. For the UDD sequences, we optimized the time around the resonance.  }
    \label{fig:DotProd}
\end{figure}

The UDD$_4$ sequence yields a more complicated evolution of the nuclear spin since it rotates by a different amount, depending on the electron's state (i.e., $\phi_0\neq\phi_1$). This condition leads to a non-trivial feature based on which the dot product of its rotation axes depends on $N$. Thus, even if one fixes a resonance time for the basic UDD$_4$ unit, the nuclear rotation axes can switch from antiparallel to parallel for some $N$. This feature is shown in Figs.~\ref{fig:DotProd}(c), (d) near the resonance time $t_k=4\pi(2k-1)/\tilde{\omega}$ and $t_k=8\pi(2k-1)/\tilde{\omega}$ respectively, for the first four UDD$_4$ resonances. In Appendix~\ref{App:UDD4jumps}, we show that $\textbf{n}_0\cdot \textbf{n}_1=1$ at values of $N$ where the rotation angles $\phi_0$ and $\phi_1$ become equal; since the axes are parallel in these ranges, the nuclear spin undergoes an unconditional rotation, and no entanglement is generated. 

The jumps in $\textbf{n}_0\cdot \textbf{n}_1$ in the case of UDD$_4$ appear because we restrict the value of the rotation angles in $[0,\pi]$; if the angles are in $[-\pi,0]$, we make them positive, and reverse the corresponding signs of the rotation axes $\textbf{n}_j$ for consistency. Alternatively, if the rotation angles are not restricted in this way, the dot product remains fixed at $\textbf{n}_0\cdot \textbf{n}_1=-1$ for all $N$. However, for some $N$, it could happen that $\phi_0=-\phi_1$ (modulo $2\pi$), which means that such $N$ cannot produce an entangling gate. It would then be misleading to claim there is a resonance whenever $\textbf{n}_0\cdot \textbf{n}_1=-1$ for UDD$_4$. Thus, we fix the convention $\phi_j \in[0,\pi]$ to ensure that we find the right $N$ to produce conditional rotations on the nuclear spins. This convention is not necessary for CPMG and UDD$_3$, as it always holds that $\phi_0=\phi_1$, and we can reliably identify $N$ to create entangling gates. No matter which convention is used for the rotation angles of CPMG or UDD$_3$, the dot product shows no dependence on $N$ [Figs.~\ref{fig:DotProd}(a), (b)].

It is important to note that, in addition to implementing gates, $\pi$-pulse sequences can also average out the interactions of the electron with unwanted spins, ensuring some degree of selectivity with a target spin. Higher-order resonances were proven to be more effective in targeting a desired nuclear spin~\cite{Dong2020,BourassaNatMater2020}. In turn, this implies that long sequences are required to achieve enhanced selectivity. In some cases, the sequences average out even the interaction with a target nucleus, rendering such spins uncontrollable, or introducing the need for more sophisticated approaches, such as decoherence protected subspaces~\cite{vanderSarNat2012} (which also require direct driving of nuclear registers). These issues will also be discussed further later on when we talk about simultaneous control of multiple nuclei.

\subsection{Implementing single-qubit gates on a nuclear spin \label{SubSec:TrivialEvol}}

We can use similar ideas to determine how to implement single-qubit gates on a nuclear spin without entangling it with the electron. Let us illustrate this in the case of CPMG. The CPMG sequence yields a rather simple equation for the rotation axes dot product of a single nuclear spin, which reads:
\begin{equation}\label{Eq:trivialEvol}
    1-\textbf{n}_0\cdot \textbf{n}_1=\frac{4\sin^2(\theta_0-\theta_1)\sin^2(\omega_0 t/8)\sin^2(\omega_1 t/8)}{\sin^2(\phi/2)},
\end{equation}
where $\cos\theta_j=(\omega_L+s_j A)/\omega_j$. This expression is exact for $s_jB<<\omega_j$, or fairly in the limit $\cos\theta_j \rightarrow 1$. Eq.~(\ref{Eq:trivialEvol}) is a generalization of the inner product of Ref.~\cite{TaminiauPRL2012}, with the difference that it was presented there for an electron spin $S=1$ (with the choice $s_0=0$ and $s_1=-1$). The nuclear spin  evolves independently of the electron when $\textbf{n}_0\cdot \textbf{n}_1=1$ and $\phi_0=\phi_1$. For the CPMG sequence, it always holds that $\phi_0=\phi_1$. Thus, using Eq.~(\ref{Eq:trivialEvol}), and by requiring that $\textbf{n}_0\cdot \textbf{n}_1=1$, we find two conditions for the decoupled evolution:
\begin{equation}\label{Eq:circle}
    \left(A+\frac{\omega_L}{s_j}\right)^2+B^2=\left(\frac{8\kappa \pi}{s_j t}\right)^2,
\end{equation}
which are the equations of a circle with center $C=(-\omega_L/s_j,0)$ and radius $R=8\kappa\pi/s_jt$ [with $\kappa\in\mathbb{Z}$ and $t$ being the time of one CPMG unit]. Note that for a $S=1$ defect electron spin, and if $s_j=0$, the decoupled evolution happens at times $t=8\kappa \pi/\omega_L$ for all nuclei. 
Using Eq.~(\ref{Eq:circle}), one can identify nuclei that do not affect the gate fidelity of target nuclear spins, as the former show no correlations with the electron. Notice that these conditions are independent of the number of repetitions of the sequence, as the dot product itself does not depend on $N$. In addition, since the evolution operator of the system is defined by the rotation each spin undergoes, this feature continues to hold in the total system. We will use the condition for decoupled evolution in Sec.~\ref{SubSec:CR2} to show that such spins have no effect on the gate operations with target nuclei.
    \begin{figure}[!htbp]
        \centering
        \includegraphics[scale=0.46]{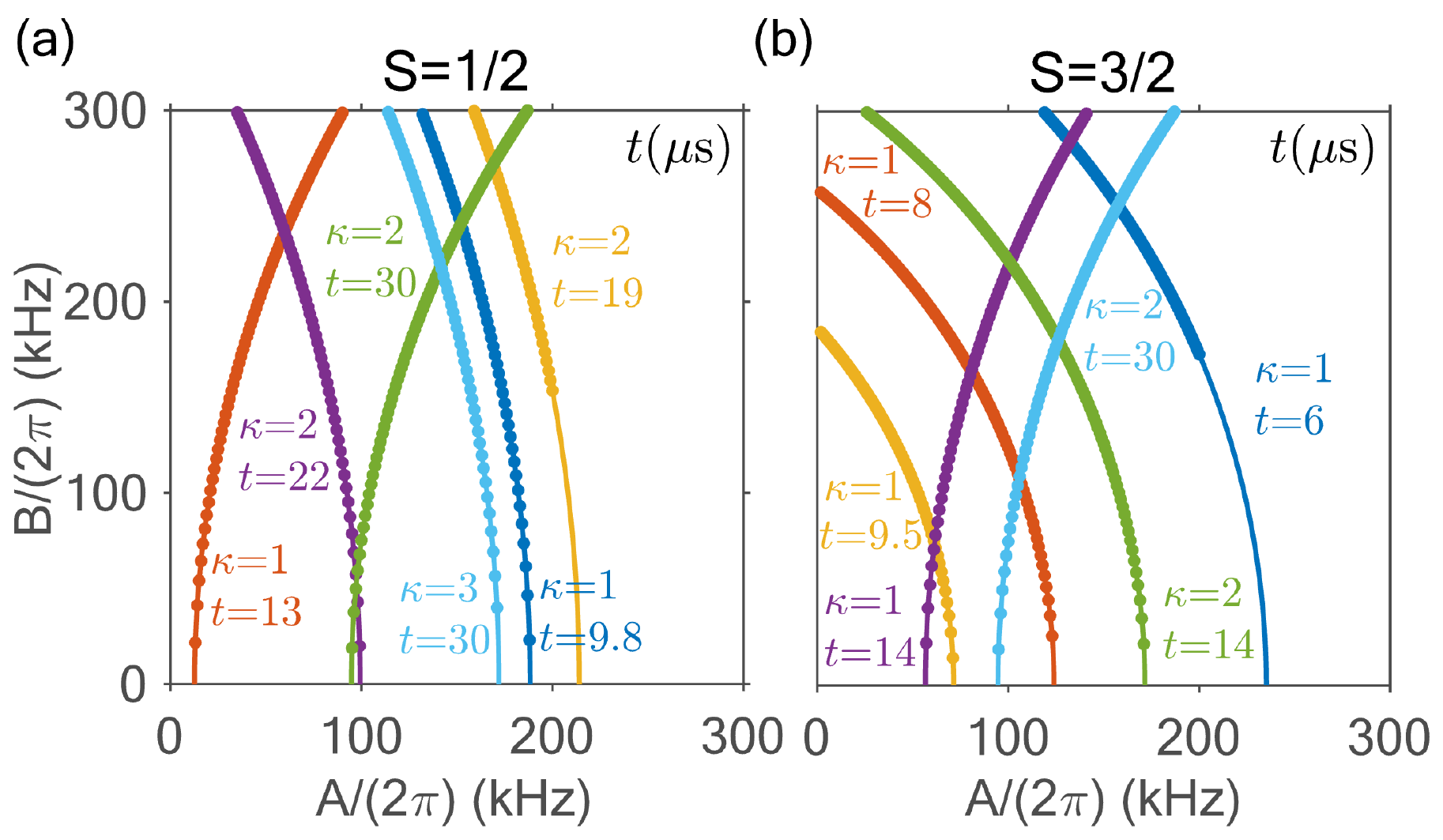}
        \caption{Hyperfine parameters of nuclear spins that undergo a trivial evolution under the CPMG sequence. Each circle corresponds to a constant time of one CPMG unit and different value of $\kappa$ [see main text]. In (a) we have selected the electron's spin projections $s_0=-s_1=1/2$ and in (b) $s_0=3/2$, $s_1=-1/2$. The Larmor frequency was considered to be $\omega_L=2\pi \cdot 314$~kHz. For illustration purposes, we have shown mainly times $t\in\mathbb{Z}^+$, but $t$ could also take any positive non-integer values. }
        \label{fig:TrivialEvol}
    \end{figure}
For now we stress that Eq.~(\ref{Eq:circle}) is valid for $(8\kappa \pi/s_j t)^2>(A+\omega_L/s_j)^2$, while we also constrain the $\kappa/t$ range such that $A,B \leq 2\pi \cdot 300$~kHz i.e., such that the nuclei are weakly coupled with the electron. Some examples for an electron-spin $S=1/2$ ($s_0=-s_1=1/2$) and $S=3/2$ ($s_0=3/2, s_1=-1/2$) are shown in Fig.~\ref{fig:TrivialEvol}(a) and Fig.~\ref{fig:TrivialEvol}(b) respectively. One notices that the times $t$ of the basic sequence exceed a few $\mu$s. In turn, this implies that the condition of the trivial evolution is strictly satisfied for $k\geq 2$ CPMG resonances of the spins with HF parameters shown in Fig.~\ref{fig:TrivialEvol}. In Appendix~\ref{App:Minimization_Tangle} we further show that trivial evolution can occur for shorter times of the basic unit, although the triviality is only approximate in this case.

\section{Quantifying entanglement in the electron-nuclear spin system \label{Sec:Entang}} 

Controlling multiple nuclear spins is usually done by applying additional radio-frequency pulses that drive the nuclear spins directly to facilitate entangling gates, either in terms of speed or precision, or even to reduce cross-talk~\cite{BradleyPRX19}. It is also possible to control multiple nuclear spins by driving only the defect electronic spin. The most straightforward way to do this is by implementing entangling gates sequentially using the techniques for addressing individual nuclear spins described in the previous section. However, the slowness of this approach can result in low entanglement and gate fidelities due to the electron's dephasing, as errors on the electron spread to the nuclei. This issue can in principle be addressed by applying dynamical decoupling on the electron or nuclei while new entanglement links are generated~\cite{PompiliSci21}; 
reaching long coherence times, however, requires a large number of pulses (e.g. for coherence $>1$s for an NV electronic spin, $10240$ pulses are required \cite{AbobeihNatCommun2018}). Hence, as the number of target nuclear spins grows, the experimental overhead increases significantly. 

In what follows, we show that these challenges can be largely sidestepped by creating multi-nuclear entanglement simultaneously rather than sequentially. To see how this works, we first discuss how to quantify multi-spin entanglement in these types of defect spin systems. We first consider measures of entangling power for a single nuclear spin coupled to the electron and then generalize this to multiple spins using the concept of one-tangles. In subsequent sections, we then show how to employ these measures to guide the design of multi-nuclear spin entangling gates.

\subsection{Disjoined picture}

The joint evolution of the electron and a single nuclear spin can be described via the Makhlin (or local) invariants \cite{Makhlin2002}, typically denoted as $G_1$ and $G_2$. These invariants classify all two-qubit operations into distinct entangling classes, such that gates sharing the same local invariants belong to the same entangling class. This property stems from the fact that local operations do not change the amount of entanglement between two parties. Entangling gates that give rise to maximum correlations are known as perfect entanglers; examples include the CNOT and CZ gates, which are locally equivalent. Makhlin invariants are suitable for classifying two-qubit gates; a more general metric that omits details of the gate structure and focuses instead on the entanglement it can generate is the entangling power~\cite{Zanardi2000}. 

For any arbitrary $\pi$-pulse sequence, the electron-nuclear evolution operator after $N$ repetitions of the sequence retains the form of Eq.~(\ref{Eq:U}), with $\phi_j$ replaced by the total rotation angle, $\phi_j(N)$. This special form of the evolution operator allows us to find the analytical forms of $G_1$ and $G_2$ as a function of $N$:
\begin{equation}\label{Eq:G1}
\resizebox{0.97\hsize}{!}{
   $G_1=\Big(\cos\frac{\phi_0(N)}{2}\cos\frac{\phi_1(N)}{2}+n_{01}\sin\frac{\phi_0(N)}{2}\sin\frac{\phi_1(N)}{2}\Big)^2$},
\end{equation}
\begin{equation}
\begin{split}\label{Eq:G2}
    &G_2 =1+n_{01}\sin\phi_0(N)\sin\phi_1(N)+
    \\&
    2\left(\cos^2\frac{\phi_0(N)}{2}\cos^2\frac{\phi_1(N)}{2}+
    n_{01}^2\sin^2\frac{\phi_0(N)}{2}\sin^2\frac{\phi_1(N)}{2}\right),
    \end{split}
\end{equation}
where $n_{01}\equiv \textbf{n}_0\cdot \textbf{n}_1$, and with $G_1\in[0,1]$ and $G_2\in[1,3]$. Based on these ranges, one notices that $\pi$-pulse sequences can only generate perfect entangling gates in the CNOT-equivalent class, for which it holds that $(G_1,G_2)=(0,1)$. Under the resonance condition ($n_{01}=-1$), the first Makhlin invariant simplifies to $G_1=\cos^2\frac{\phi_0(N)+\phi_1(N)}{2}$, and requiring $G_1=0$ gives the number of sequence iterations needed to obtain a controlled gate. To estimate the number of repetitions $N$, we only need to know the rotation angles in one iteration. The minima of $G_1$ are located at $N=(2\kappa+1)\pi/(\phi_0+\phi_1)$. In general, $G_1$ can be zero for other $N$ as well, as long as $\textbf{n}_0\cdot \textbf{n}_1\leq 0$. We provide the analytical expressions for $N$ for this general case in Appendix~\ref{App:Nestimate} and use these conditions to identify nuclear spin candidates to realize simultaneous controlled gates in Sec.~\ref{Sec:CR}.

\begin{figure}[!htbp]
    \centering
    \includegraphics[scale=0.5]{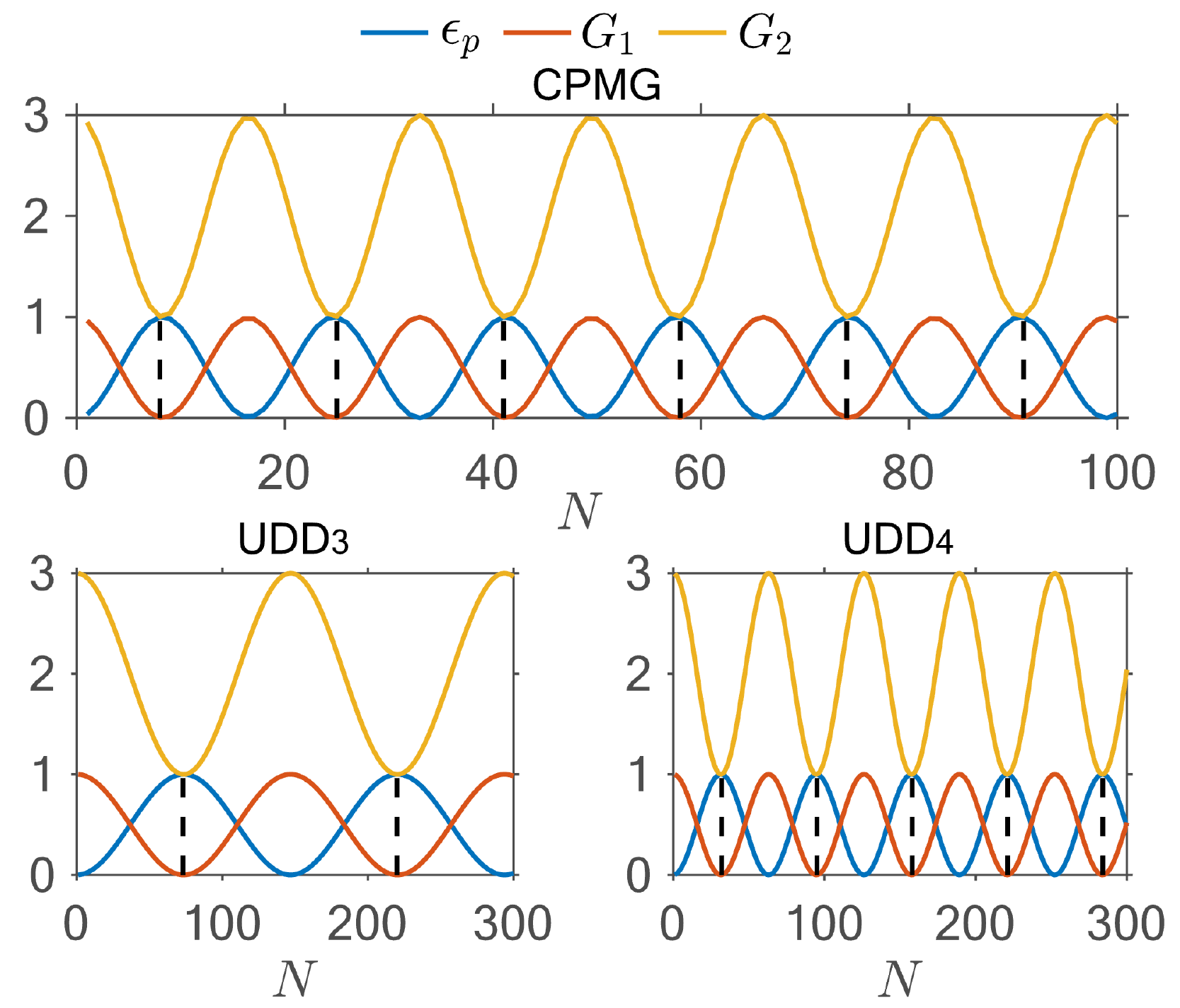}
    \caption{Entangling power (blue) and Makhlin invariants ($G_1$: red, $G_2$: yellow) as a function of the number of repetitions of the CPMG (top), and UDD$_3$ or UDD$_4$ (bottom) sequences for a single nuclear spin. The dotted lines correspond to the analytically expected minima of $G_1$ (see text). We considered the $k=1$ resonance for each sequence. The times for the UDD$_n$ sequences were optimized around the analytical resonance time [$(t_{\text{CPMG}},t_{\text{UDD}_3},t_{\text{UDD}_4})=(3.1811,3.1852,3.1862)~\mu$s]. For the nuclear spin, we set  $(A,B,\omega_L)=2\pi\cdot(60,30,314)$~kHz, and for the electron's spin projections $s_0=-s_1=1/2$. }
    \label{fig:EPG1G2}
\end{figure}
It has been shown that the entangling power of a two-qubit operator can be expressed in terms of $G_1$ as~\cite{Balakrishnan2010}:
\begin{equation}\label{Eq:ep}
    e_p = \frac{2}{9}(1-|G_1|).
\end{equation}
It is clear that for $G_1=0$ the entangling power is maximized and saturates to $2/9$ for the two-qubit case. In Fig.~\ref{fig:EPG1G2}, we show the entangling power (scaled by $2/9$) and Makhlin invariants for the CPMG, UDD$_3$, and UDD$_4$ sequences. The vertical lines correspond to the minima of $G_1$. We notice that the period of oscillations is larger for CPMG since the angle per iteration is greater compared to the UDD$_n$ sequences (see Appendix~\ref{App:CPMGvsUDDRotAngle} and Ref.~\cite{Dong2020}).

\subsection{Assessing multi-spin entanglement via one-tangles\label{SubSec:One-tangles}}

To understand the entanglement distribution in the total system (consisting of the electron and multiple nuclei), we need to extend the notion of the two-qubit entangling power. To this end, we employ the one-tangles~\cite{Coffman2000,Bengtsoon_2006}, which measure the total amount of entanglement in a state by considering all possible bipartitions of the system. That is, by fictitiously dividing the total system into subsystems, one can quantify the degree of correlations between the subsystems (also known as the bipartition entanglement). We choose to use the one-tangle as the entanglement metric, which means for each bipartition, we separate only one qubit (electron or nuclear spin) from the rest of the system. 

One-tangles carry only the information of the entanglement capacity in the system and cannot distinguish states that belong to different families (e.g., W states versus GHZ states for the tripartite case)~\cite{Linowski_2020}. Such a metric is convenient since we are interested in the general evolution of the system rather than generating particular entangled states. 

Similar to the two-qubit entangling power, the one-tangles are defined through the linear entropy. For a pure state $|\psi\rangle$, the one-tangle reads:
\begin{equation}\label{Eq:one-tangle}
    \tau_{g|g'}(|\psi\rangle):=1-\text{tr}[\rho^2_{g'}],~ \rho_{g'}=\text{tr}_g[|\psi\rangle\langle \psi|],
\end{equation}
where $g|g'$ denotes a bipartition of the system. Some authors include an overall multiplicative factor of $2$ for the linear entropy; we choose not to follow this convention as it simply redefines the bounds of the linear entropy and does not affect our following analysis.  

Eq.~(\ref{Eq:one-tangle}) in its current form is not particularly useful for quantifying the entanglement of multi-nuclear operations since it depends on the initial state. We must therefore average over initial states. In particular, we will use the bipartition entangling power, which is defined as the average of the one-tangle over all initial product states. This average can be computed by averaging over single-qubit unitaries applied to an arbitrary initial product state, i.e., $\epsilon_{g|g'}(U):=\langle\tau_{g|g'}[U( U_i^{\otimes_i} |\psi_0\rangle)]\rangle_{U_i}$.  In Ref.~\cite{Linowski_2020}, it was shown that the entangling power (with one-tangles as the measure) for a bipartition $p|q$ of the system is given by
\begin{equation} \label{Eq:one-tangleGen}
    \epsilon_{p|q}(U)=1-\left(\prod_{i=1}^n\frac{d_i}{d_i+1}\right)\sum_{x'|y'}\text{tr}\Big[(\text{tr}_{px'}[|U\rangle\langle U|])^2\Big],
\end{equation}
where $d_i=2$ is the dimension of each qubit subsystem. The state $\ket{U}$ is defined in the context of the Choi-Jamiolkowski isomorphism~\cite{JAMIOLKOWSKI1972275,CHOI1975285}, which maps any projector living in a $d$-dimensional Hilbert space ($\mathcal{H}_d$) into a state vector in an extended space ($\mathcal{H}_{d^2}\equiv \mathcal{H}\otimes \mathcal{H}'$), i.e. $|i\rangle\langle j'|\mapsto |ij'\rangle$. In our case, $d=2^n$ where $n$ is the number of qubits, including the electron and the nuclei. $x'|y'$ denotes a bipartition in the secondary system of the total extended space. The summation is performed over all $2^n$ bipartitions in $\mathcal{H}'$. For example, for the tripartite case we have $x'|y'\in\{1'2'3'|\cdot,1'2'|3',1'3'|2',2'3'|1',1'|2'3',2'|1'3',3'|1'2',\cdot|1'2'3'\}$, where `$\cdot$' is the empty bipartition. Eq.~(\ref{Eq:one-tangleGen}) is applicable for multipartite unitary gates, with $q$ referring to a single qubit partitioned from the $d$-dimensional Hilbert space, $\mathcal{H}_d$, and $p$ referring to the remaining $(d-1)$-dimensional subsystem. As an example, for 4 qubits in total, $p|q$ can take the values $p|q\in\{123|4,124|3,134|2,234|1\}$. 

In the case of $\pi$-pulse sequences, the evolution operator has a special form given by:
\begin{equation}
    U=\sum_{j\in\{0,1\}}\sigma_{jj} \otimes_{l=1}^L R_{\textbf{n}_j}^{(l)}(\phi_j^{(l)}),
\end{equation}
where $L$ is the total number of nuclear spins, and for conciseness we refer to $\phi_j^{(l)}(N)$ as simply $\phi_j^{(l)}$. The evolution operator is therefore defined by the evolution of each nuclear spin in the disjoined picture (see Appendix~\ref{App:EvolOper} for a proof). This feature allows us to obtain analytical expressions for the average of the one-tangles for any number of nuclear spins. However, we need to distinguish the case when either a single nuclear spin or the electron is partitioned from the rest of the system. For brevity, we will refer to these types of average one-tangles as the one-tangle of a nuclear spin and the one-tangle of the electron, respectively. 

Starting with the one-tangle of a single nuclear spin, we find that it is given by (see Appendix~\ref{App:DerOneTangle})
\begin{equation}\label{Eq:NuclearTangle}
\epsilon_{p|q}^{\mathrm{nuclear}}=\frac{2}{9}(1-G_1),
\end{equation}
which holds for $n\geq 3$ qubits. For $n=2$, the average of the one-tangle is the two-qubit entangling power of Eq.~(\ref{Eq:ep}). $G_1$ is given by Eq.~(\ref{Eq:G1}). Note that as is expected, the one-tangle of a nuclear spin does not depend on other quantities besides those that determine its evolution (due to the tensor product form of the total evolution operator $U$). In the case when the electron is partitioned from the system the one-tangle reads (see Appendix~\ref{App:DerOneTangle})
\begin{equation}\label{Eq:ElectronTangle}
    \epsilon_{p|q}^{\mathrm{electron}}=\frac{1}{3}-\frac{1}{3^n}\prod_{i=1}^{n-1}(1+2G_1^{(i)}),
\end{equation}
where $G_1^{(j)}\equiv G_1(\phi_0^{(j)},\phi_1^{(j)},n_{01}^{(j)})$ contains the information of the evolution of the $j$-th nuclear spin. The one-tangle of the electron now includes contributions from the evolutions of each nuclear spin; due to the always-on nature of the HF interaction, the electron can be correlated with all nuclei. On the other hand, we see from Eq.~\eqref{Eq:NuclearTangle} that a single nuclear spin can only have explicit correlations with the electron and evolves independently of all other nuclei (assuming no inter-nuclear spin interactions). 
 
Remarkably, the expressions for the one-tangles, Eq.~(\ref{Eq:NuclearTangle}) and Eq.~(\ref{Eq:ElectronTangle}), allow us to study the entanglement distribution in an arbitrarily large nuclear spin register. Together with the knowledge of the evolution of each nuclear spin in the disjoined picture, we can simulate efficiently a large number of nuclei and obtain complete information about the dynamics of the system. The simplicity of Eqs.~\eqref{Eq:NuclearTangle} and \eqref{Eq:ElectronTangle} is what allows us to obtain a detailed understanding of how entanglement gets distributed throughout the system for various pulse sequences, as we discuss in the remainder of the paper.

\begin{figure}[!htbp]
    \centering
    \includegraphics[scale=0.52]{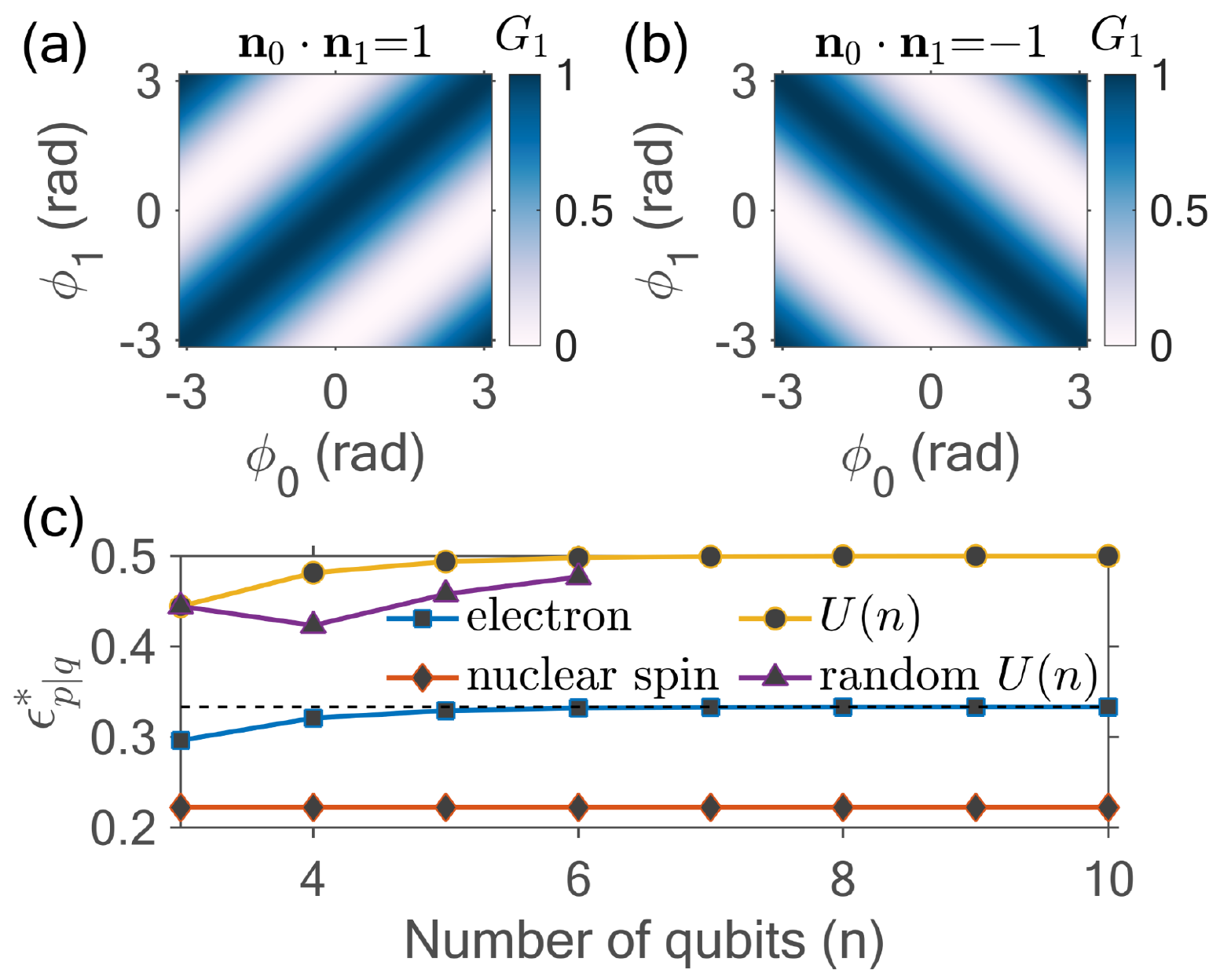}
    \caption{Function $G_1$ versus the rotation angles $\phi_j$ for the case of $\textbf{n}_0\cdot\textbf{n}_1=1$ (a) and $\textbf{n}_0\cdot\textbf{n}_1=-1$ (b). (c) Maximum one-tangles as a function of the number of qubits for the case when the electron (blue) or a single nuclear spin (red) is partitioned from the rest of the system. The yellow line is the theoretical maximum bound for a perfect $U(n)$ entangler, while the purple line is the numerical bound we found for randomly generated $U(n)$, obtained by retaining the maximal value over 100 random unitaries for $n=4,5$ and 5 random unitaries for $n=6$. For $n=3$, we construct a $U(n)$ from an absolutely maximally entangling (AME) state. Such $U(n)$ saturates the bound, if the AME$(2n,d)$ state exists (in this case $d=2$).}
    \label{fig:Bounds}
\end{figure}

One thing we can immediately see from Eq.~\eqref{Eq:NuclearTangle} is that the one-tangle of a nuclear spin is minimized when the function $G_1$ is maximized. This can happen when the nuclear spin undergoes a trivial evolution, namely when $\textbf{n}_0\cdot \textbf{n}_1=1$ ($\textbf{n}_0\cdot \textbf{n}_1=-1$) and $\phi_0=\phi_1$ ($\phi_0=-\phi_1$). The range of the function $G_1$ is shown in Fig.~\ref{fig:Bounds}(a) and Fig.~\ref{fig:Bounds}(b) for the cases of $\textbf{n}_0\cdot \textbf{n}_1=\pm 1$ respectively. Whenever $G_1=0$, the one-tangle of a nuclear spin is maximal, whereas when $G_1=1$, the nuclear spin decouples from the system. In Fig.~\ref{fig:Bounds}(c), we show the maximum one-tangle when a single nuclear spin (red) or the electron (blue) is separated from the rest of the spins. As expected, the maximum nuclear one-tangle is independent of the number of total qubits in the system and saturates to the value $2/9$, which also holds for two-qubit operations. On the other hand, the electron's one-tangle shows an increase with the number of qubits until it becomes independent of $n$ and saturates close to $1/3$. 

In light of these results, it is interesting to ask whether it is possible to achieve maximal entangling power by applying $\pi$-pulses to this central spin system.
In Fig.~\ref{fig:Bounds}(c), we also show the bound of the bipartition entanglement for an arbitary $n$-qubit gate $U(n)$ (yellow), which is calculated according to~\cite{Linowski_2020}:
\begin{equation}\label{Eq:OneTangleBound}
    \epsilon_{p|q}^{\text{max}}=1-\prod_i \frac{d_i}{d_i+1}\sum_{x'|y'}\frac{1}{\text{min}[d_{px'},d_{qy'}]},
\end{equation}
where $d_{px'}$ and $d_{qy'}$ are the dimensions of the subsystems $px'$ and $qy'$. Interestingly, this bound is never reached by $\pi$-pulse sequences. However, this upper bound is not always tight. A necessary requirement for the bound to be tight is that the CP-maps associated with $U(n)$ are unital \cite{Zanardi2000}, which means that they map maximally mixed states onto maximally mixed states. This condition alone is not sufficient, 
since as was shown in Ref.~\cite{Zanardi2000}, for the two-qubit case, the bound given by the linear entropy (which is 1/3) is never saturated, and the well-known perfect entanglers, such as CNOT, can only reach the value of 2/9. The saturation of the bound occurs when the matrix elements of $U(n)$ can be obtained from so-called absolutely maximally entangling states, known as AME($2n,d$), if these exist~\cite{Linowski_2020}. For $d=2$ (i.e., qubit subsystems), AME($n,d$) states exist only for $n=3,5,6$~\cite{HuberPRL2017}. In Fig.~\ref{fig:Bounds}(c), we show that the bound is indeed saturated for $n=3$ [for which AME($2n,d$) exists], if we construct such $U(n)$ based on Ref.~\cite{Linowski_2020}, for an AME$(2n,d)$ state found in Ref.~\cite{Raissi2018}. For $n=4,5,6$, we generated random $n$-qubit unitaries $U(n)$ and calculated the maximum value of one-tangles; the results are depicted with a purple line.  Although we have not sampled a large number of $U(n)$, we see that the maximum bipartition entanglement of random unitaries exceeds the bound of the one-tangles corresponding to $\pi$-pulse sequences. Therefore, the multipartite controlled gates generated by $\pi$-pulse sequences applied to this central spin system do not saturate the one-tangle bound for $n\geq 3$, and hence the amount of entanglement they can create is limited.

We now illustrate the utility of Eqs.~\eqref{Eq:NuclearTangle} and \eqref{Eq:ElectronTangle} by using them to design electron-nuclear entangling gates that avoid unwanted nuclei.
We first consider the simplest example of two nuclei under the CPMG sequence, for an electron spin $S=1/2$. We fix the HF parameters of the target spin to be $(A,B)=2\pi\cdot(60,30)$~kHz, and allow the HF parameters of the second spin to vary in the range $2\pi\cdot[10,200]$~kHz. The
nuclear spin Larmor frequency is set to be $\omega_L=2\pi\cdot314$~kHz; for $^{13}$C atoms this corresponds to a magnetic field of $B\approx 293.46$~G. Depending on the defect electronic spin, the $B$-field should be chosen such that it ensures the MW qubit transitions are far from anti-crossings and hence, leakage outside of the electronic qubit subspace is suppressed~\cite{BradleyThesis2021}. 

\begin{figure}[!htbp]
    \centering
    \includegraphics[scale=0.51]{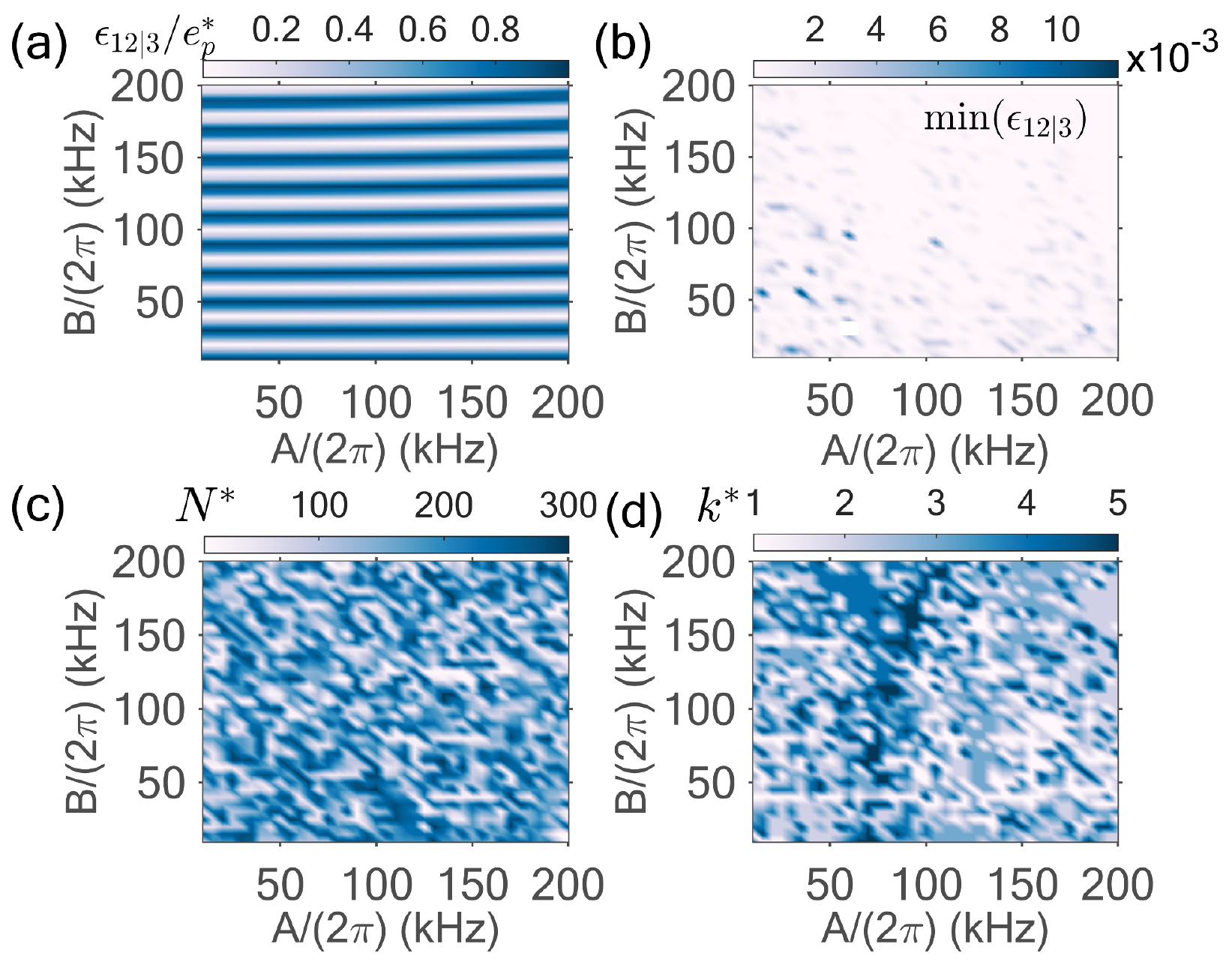}
    \caption{Controlling a target nuclear spin with parameters $(A,B,\omega_L)=2\pi\cdot(60,30,314)$~kHz, in the presence of an unwanted spin with HF parameters $\in 2\pi\cdot[10,200]$~kHz. (a) One-tangle of the unwanted spin scaled by the maximum bound of $2/9$. The time of one sequence unit is the first resonance of the target spin, and the number of iterations is $N=25$, which maximize its one-tangle.  (b) Minimization of the unwanted spin's one-tangle using the first five $(k=1,\dots,5)$ resonances of the target spin, and up to 300 pulses on the electron. Optimal sequence iterations (c) and optimal resonance (d) to minimize the unwanted spin's one-tangle, while keeping the one-tangle for the target spin maximal. In all plots, we considered an electron spin $S=1/2$, and the CPMG sequence. }
    \label{fig:OneTangle2Spins}
\end{figure}

In  Fig.~\ref{fig:OneTangle2Spins}(a), we select the first resonance of the target spin and $N=25$ sequence iterations, which maximize its one-tangle, and show the one-tangle of the unwanted spin (scaled by the maximum value of $2/9$). In the ranges where the one-tangle of the unwanted spin is minimal, we successfully decouple it from the rest of the spins. We have verified that these ranges correspond to nuclear spins whose HF parameters approximately satisfy the condition for trivial evolution; we further depict this behavior for an $S=1$ electron system in Appendix~\ref{App:Minimization_Tangle}.

Based on Fig.~\ref{fig:OneTangle2Spins}(a), we would conclude that certain unwanted nuclei cannot be decoupled, as they show non-zero entanglement with the rest of the system. If one wishes to target a specific spin with high selectivity then, different resonance times and sequence iterations need to be considered. Note that this effect would be completely missed in prior formulations of this problem, and the issue of insufficient decoupling would only appear in numerics, where the simulations would have to be repeated for all the different physically relevant hyperfine couplings. In Fig.~\ref{fig:OneTangle2Spins}(b), we show the minimal value of the unwanted spin's one-tangle (excluding the case of same HF parameters for the unwanted and target nuclei), which is optimized over the first five resonances of the target spin and up to 300 repetitions of the sequence. We search only over iterations that generate maximal entanglement between the target nucleus and electron, which we obtain from the minima of $G_1$. The optimal iterations and resonances are shown in Fig.~\ref{fig:OneTangle2Spins}(c) and Fig.~\ref{fig:OneTangle2Spins}(d), respectively. The optimization yields minimum one-tangles on the order of $\sim 10^{-3}$ for the unwanted spin, providing isolation for the electron-target nuclear spin system. We conclude that using the analytical expressions of the one-tangles to minimize unwanted one-tangles via optimization of the parameters of the $\pi$-pulse sequence provides a faithful metric of selectivity with a single target spin.  

Lastly, it is interesting to note that Fig.~\ref{fig:OneTangle2Spins}(a) reveals that the unwanted spin's one-tangle can be maximal (depending on its  HF parameters) at the same time, $t$, and repetitions $N$ we chose for the target spin. This feature is further studied in Sec.~\ref{Sec:CR} and paves the path to identifying nuclei that synchronously undergo controlled gates.

\section{Synchronous controlled gates on multiple nuclei \label{Sec:CR}}

\subsection{Maximization of multiple one-tangles \label{SubSec:CR1}}

As we saw in Sec.~\ref{SubSec:One-tangles}, one-tangles corresponding to different nuclei can be maximized/minimized simultaneously and for the same number of repetitions of the sequence unit. This suggests that instead of generating entanglement with single spins sequentially, one can simultaneously entangle multiple nuclei with the electron. In this section, we confirm that this is indeed the case. 

To see how such direct generation of multi-spin entanglement is possible, we devise a simple strategy of
identifying nuclei whose one-tangles become simultaneously maximal. 
To demonstrate our method, we select nuclei randomly from the HF range $ 2\pi \cdot[10,200]$~kHz. There are two relevant parameters we need to decide how to fix; the time, $t$, of one unit of the sequence, and the repetitions, $N$. We fix $t$ by setting it equal to a chosen resonance of the first randomly selected nucleus. For this nucleus, we find the iterations that maximize its one-tangle, based on the minima of $G_1$, and store these into the set $\tilde{N}^{(1)}$. Since the time we choose does not in principle coincide with a resonance of other randomly selected nuclei (as the HF parameters differ), it will in general hold that $\textbf{n}_0\cdot \textbf{n}_1\neq -1$ for these nuclei, meaning that we need a reliable way of estimating iterations that maximize their one-tangles. As long as $\textbf{n}_0\cdot \textbf{n}_1\leq0$ for a single nuclear spin, the one-tangle can be maximal for some $N$. We explain how we find the maxima [analytically for CPMG and UDD$_3$; numerically for UDD$_4$] in Appendix~\ref{App:Nestimate}. Based on the maxima, we assign to each nucleus a set $\tilde{N}^{(j)}$, similar to what we did for the first nucleus. Then, we search for a common intersection i.e., one number of iterations of the sequence that belongs to multiple sets [$\bigcap\limits_{j=1}^{n-1}\tilde{N}^{(j)}$]. The first set we fix is that of the first randomly chosen spin, and then we test its intersection with the remaining sets. Nuclear spins whose sets have zero intersection with this initial fixed set are removed. In the end, we obtain a particular value of iterations ($N^*$) and nuclear spin candidates that can participate in a multipartite gate.
\begin{table}[!htbp]
    \centering
    \scalebox{0.96}{
    \begin{tabular}{|c|c|c|}\hline
         HF range $(\frac{A}{2\pi},\frac{B}{2\pi})$ & Distance from  & Atoms  \\                (MHz)       & vacancy site ($\angstrom$) & \\
         \hline
         100-200 ~\cite{GaliPRB2008} & 1.61 &  $^{13}$C   \\
         & (1st neighbor) & (NV diamond)
         \\
         \hline 
          $\sim$ 10-20~\cite{GaliPRB2008}  & 3.86  &  $^{13}$C \\
          $(19.4,13.9)$~\cite{FeltonPRB2009}& (3rd neighbor) & (NV diamond, $\text{C}_g$ ~\cite{FeltonPRB2009}) 
          \\ \hline
          $\sim$4 \cite{BossPRL2016} & 3 & $^{13}$C  \\
          &(sites G,H)& (NV diamond)
          \\
          \hline
          $\sim$ 2 ~\cite{GaliNewJ2011} & 5 & $^{13}$C   \\
          & & (NV diamond) \\
          \hline
          \hline
          HF range $(\frac{A}{2\pi},\frac{B}{2\pi})$ & Distance from & Atoms \\
          (kHz) & vacancy site ($\angstrom$) &  \\
          \hline
          60-120  ~\cite{ZopesNatCommun2018} & 6.8 & $^{13}$C \\
          & &(NV diamond) \\
          \hline
          20-50 ~\cite{ZopesNatCommun2018} & 8-9 & $^{13}$C  \\
          & & (NV diamond)\\ 
          \hline
          2-20  ~\cite{ZopesNatCommun2018} & 11.5 & $^{13}$C  \\
          & & (NV diamond)\\
          \hline
          $(10,29)$~\cite{NagyNatCommun2019} & 11.6 & $^{29}$Si \\
          & & (SiC) \\
          \hline
          $(0.65,11.45)$~\cite{BourassaNatMater2020} & 12.4 & $^{29}$Si\\
          & & (SiC) \\
          \hline
    \end{tabular}}
    \caption{Range of hyperfine parameters and corresponding distances from the vacancy site for $^{13}$C atoms and $^{29}$Si atoms in diamond or SiC. Explicit values $(A,B)$ are shown in parentheses, otherwise we provide approximate ranges.}
    \label{tab:HFparams}
\end{table}

\begin{figure*}[!htbp]
    \centering
    \includegraphics[scale=0.65]{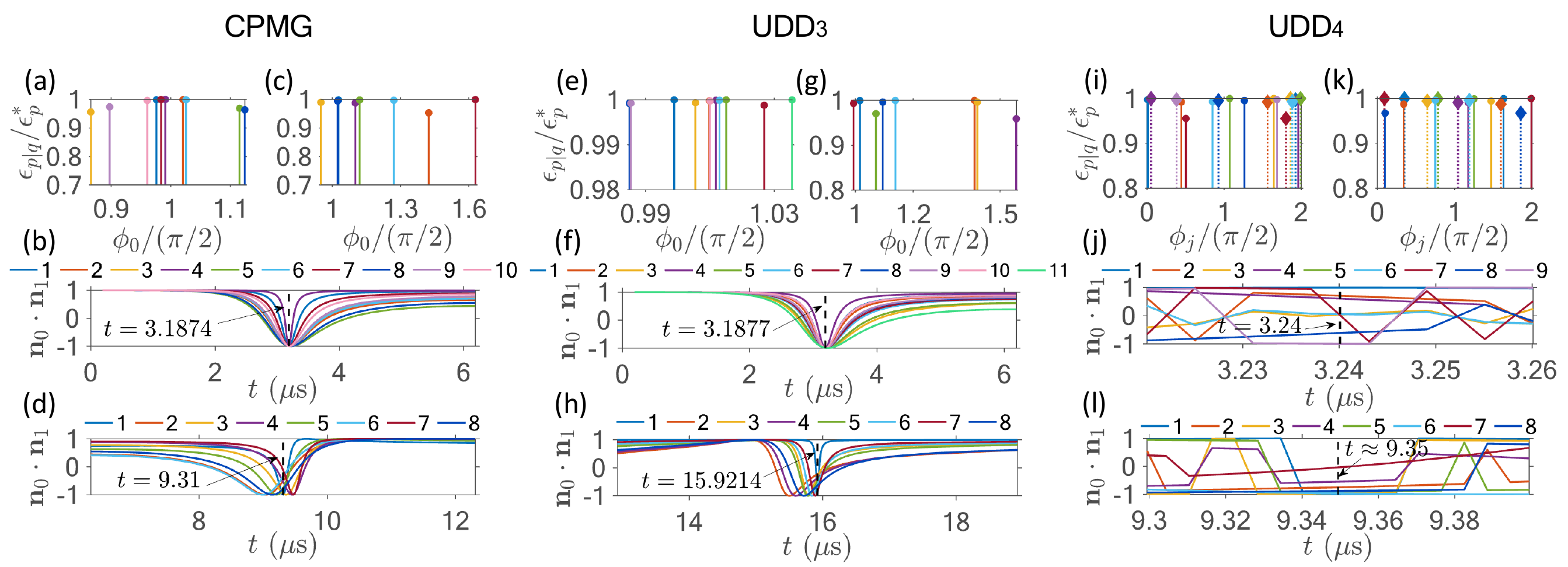}
    \caption{Nuclear one-tangles versus accumulated rotation angle for: CPMG (a) [$(N^*,k)=(56,1)$] and (c) [$(N^*,k)=(8,2)$], UDD$_3$ (e) [$(N^*,k)=(487,1)$] and (g) [$(N^*,k)=(93,3)$], and UDD$_4$ (i) [$(N^*,k)=(252,1)$] and (k) [$(N^*,k)=(41,2)$]. Dot product of nuclear rotation axes close to: $k=1$ CPMG resonance (b), $k=2$ CPMG resonance (d), $k=1$ UDD$_3$ resonance (f), $k=3$ UDD$_3$ resonance (h), $k=1$ UDD$_4$ resonance (j) and $k=2$ UDD$_4$ resonance (l). The vertical lines in the dot product panels denote the common time of the basic unit used to evaluate the corresponding one-tangles, and is the resonance of nuclear spin ``1''. In (i) and (k) the lines with circles correspond to $\phi_0$ and the lines with diamonds to $\phi_1$.  The nuclear spins for each set $(N^*,k)$ of each sequence are different and are provided in Tables~\ref{tab:2} (CPMG),~\ref{tab:3} (UDD$_3$),~\ref{tab:4} (UDD$_4$), of Appendix~\ref{App:HFparams}.  }
    \label{fig:MaximizationTangles}
\end{figure*}
In the simulations that follow, we assume an electron spin $S=1/2$, that could correspond to SiV$^-$ or SnV$^-$ defect in diamond~\cite{HeppPRL2014,RugarACSPhot2020,RugarNanoLet2020,RugarPRX2021}. We further set the nuclear Larmor frequencies to be $\omega_L=2\pi\cdot 314$~kHz. The HF range $2\pi \cdot[10,200]$~kHz we choose for the nuclei would for instance correspond to the median of the HF distribution for an isotopic concentration of $\sim 10^{-3}$ in SiC~\cite{BourassaNatMater2020}. Such nuclei are weakly coupled since the HF parameters are smaller than $1/T_2^*$, which is typically a few MHz \cite{MazeNewJ2012,Liu2012} for NV centers, or in general for $A,B\ll 1$~MHz \cite{GalliNpj21} ($\sim$1~MHz is also the electron linewidth for the neutral divacancy in SiC~\cite{BourassaNatMater2020}). For HF strengths $>2\pi\cdot 6$~kHz, the nuclei are within a distance of $R<15~\angstrom$ from the vacancy site, while for strengths on the order of $2\pi\cdot 1$~kHz, they are within $R\sim 25~\angstrom$~\cite{WangSciRep15}. More precise ranges of HF values and distances from the vacancy are shown in Table~\ref{tab:HFparams}. The HF values for our following simulations, and estimations of the nuclear positions relative to the vacancy, are listed in Appendix~\ref{App:HFparamsRandom}. To ensure that the spins selected via random generation are distinct, we give a bound on how different the HF values should be, e.g. for CPMG, we require that at least one of the HF values differs by at least $2\pi\cdot 25$~kHz from the rest. This bound is set to a reasonable value so that we generate enough nuclei within the HF range, but with distinct enough HF values. In the following, we study two different resonances for CPMG, UDD$_3$, or UDD$_4$, and for each resonance, we perform a distinct random generation of nuclei.

Considering the first resonance ($k=1$ of one of the target spins) and using the CPMG sequence, we show ten nuclear spin one-tangles [Fig.~\ref{fig:MaximizationTangles}(a)] that are maximized for a unit time $t=3.1874~\mu$s. In Fig.~\ref{fig:MaximizationTangles}(b) we show the dot product of the rotation axes of each of the ten nuclei. It is apparent that the axes of each spin are nearly antiparallel since, for $k=1$, the individual resonance times have only a small deviation from $t=3.1874~\mu$s. Consequently, the only way for the one-tangles to be maximized is that the nuclei rotate with $\phi_0(N^*)=\phi_1(N^*)\approx \pi/2$ [see Eq.~(\ref{Eq:G1})] and hence, the realized gate is close to a multipartite CR$_x(\pi/2)$. It is interesting to notice that, based on Table~\ref{tab:2}, nuclear spins 6 and 7, 2 and 5, 4 and 8, as well as spins 1 and 9, have similar $A$ values. In Ref.~\cite{TaminiauNatNano2014} it was reported that two weakly coupled nuclear spins (one of them was a spectator unwanted nucleus) showed similar $A$ values, and thus the controlled gate on one of them also rotated the other one (potentially leading to unwanted residual entanglement), but this effect was not quantified in their quantum error-correction scheme. 

In Figs.~\ref{fig:MaximizationTangles}(c),(d) we again show nuclear spin one-tangles and rotation axis dot products, but now for the $k=2$ resonance. As the order of the resonance increases, the individual resonance times show a larger dispersion, leading to nuclear rotation axes that deviate from being antiparallel. For multiple nuclei to be (close to) maximally entangled with the electron, they would then have to compensate for this feature by rotating by an angle $\phi_0$ that differs from $\pi/2$ [Fig.~\ref{fig:MaximizationTangles}(c)].

We can perform a similar analysis for the UDD$_3$ sequence for which again, the rotation angle of each nucleus is independent of the electron's state, i.e., $\phi_0=\phi_1$. The basic UDD$_3$ unit now contains an odd number of pulses and thus, needs to be repeated twice. For this reason, the UDD$_3$ angle per iteration is smaller than those of CPMG or UDD$_4$ (see Appendix~\ref{App:CPMGvsUDDRotAngle} and Ref.~\cite{Dong2020}), implying higher precision on the accumulated angle, but slower multipartite gates. This behavior is verified in Fig.~\ref{fig:MaximizationTangles}(e), where we plot the one-tangles of eleven nuclear spins versus their accumulated rotation angle, which is very close to $\pi/2$.  As the first resonance is very sharp [see Fig.~\ref{fig:MaximizationTangles}(f)], the nuclear rotation axes are very close to antiparallel. This gives rise to very high entanglement but a long sequence with $N^*=487$ repetitions. However, one can impose restrictions on the total time and still find very high one-tangles for the $k=1$ UDD$_3$ resonance.

On the other hand, for $k=3$ [Fig.~\ref{fig:MaximizationTangles}(h)], the resonance is broader, and hence, the rotation angles of the target nuclei deviate in general from $\pi/2$ [Fig.~\ref{fig:MaximizationTangles}(g)], similar to what we observed for CPMG. The $k=1$ UDD$_3$ resonance leads to higher entanglement since the unit time is smaller than for $k=3$, implying greater precision in the accumulated rotation angle per iteration. Of course, one reason for the difference between the two resonances is the random selection of HF values, which is distinct in the two cases. In addition, the chosen number of sequence repetitions might not be optimal for $k=3$. It is not surprising that particular resonances and iterations can lead to better nuclear spin control, as the rotation angle depends both on the sequence time and $N$. Since $N$ takes discrete values, this implies that features of over- or under-rotation result in imperfect entanglement.

Lastly, we consider the UDD$_4$ sequence. In this case, the rotation angle of each spin depends on the electron's state, and we cannot estimate analytically the maxima of one-tangles; instead, we identify them via numerical search. In Fig.~\ref{fig:MaximizationTangles}(i) we show the one-tangles versus the rotation angles ($\phi_j$) for nine nuclei selected from the randomly distributed ensemble, for $k=1$ [lines with circles (diamonds) show $\phi_0$ ($\phi_1$)]. The dot product of the nuclear axes is shown in Fig.~\ref{fig:MaximizationTangles}(j). Even though the dot product shows nontrivial jumps (due to $\phi_0\neq\phi_1$), one can still obtain appreciable entanglement with multiple nuclei. The one-tangles in Fig.~\ref{fig:MaximizationTangles}(k) and the dot products in Fig.~\ref{fig:MaximizationTangles}(l) correspond to the $k=2$ resonance. The entangling operations for UDD$_4$ are in general faster than for UDD$_3$, since the former induces a larger nuclear spin rotation. An interesting feature that emerges from $\phi_0\neq\phi_1$ is that the nuclei undergo a more complicated evolution, and entanglement generation can occur for multiple sets of rotation angles and axes. For example, we see that both in Fig.~\ref{fig:MaximizationTangles}(i) and Fig.~\ref{fig:MaximizationTangles}(k) it can happen that $(\phi_0(N^*),\phi_1(N^*))\approx(0,\pi)$ (or vice-versa) realizing a CR$(\pi$) operation with that particular nuclear spin [see Table~\ref{tab:4} in Appendix~\ref{App:HFparamsRandom}]. This is not surprising, since based on Eq.~(\ref{Eq:G1}) for $\textbf{n}_0\cdot \textbf{n}_1=0$ [see spin ``7'' in Fig.~\ref{fig:MaximizationTangles}(l)], $G_1=0$ if either $\phi_0(N)$ or $\phi_1(N)$ is $(2\kappa+1)\pi$.

\subsection{Effect of unwanted spins on gate-fidelity \label{SubSec:CR2}}

Using the language of one-tangles, we showed that it is possible to realize direct multipartite gates, providing a speed-up compared to sequential entanglement-generation schemes. However, the gate fidelity could still be affected by unwanted nuclei, especially if these become entangled with the electron. We now examine this issue.

To keep the discussion general, let us consider $L$ nuclear spins in total, with $K$ of them corresponding to the target nuclei that show maximal one-tangles. The $L-K$ unwanted nuclei affect the target gate since in general, they have a non-zero degree of entanglement with the electron. This means that projecting the evolution operator onto the target subspace would result in a non-unitary gate. In Appendix~\ref{App:Kraus} we show how this can be avoided by using the Kraus operator representation of the partial trace channel, based on which we can work directly with the total evolution operator and do not need to specify an initial state for the system. The operator-sum representation~\cite{nielsen_chuang_2010} allows us to derive an analytical expression for the gate fidelity of the target subspace. As a target gate $U_0$ we consider the evolution operator of the $K$ target spins in the absence of the unwanted spins, i.e.,
\begin{equation}\label{Eq:TargetGate}
    U_0=\sum_{j\in\{0,1\}}\sigma_{jj}\otimes_{k=1}^{K}R_{\textbf{n}_j^{(k)}}(\phi_j^{(k)}).
\end{equation}
Using the analytical expressions for the Kraus operators, we find that the target subspace gate fidelity reads:
\begin{equation}\label{Eq:Fid}
    F=\frac{1}{2^{K+1}+1}\Big(1+2^{K-1}\sum_{k=0}^{2^{L-K}-1}\Big|\sum_{j\in\{0,1\}} c_{j}^{(k)}p_{j}^{(k)}\Big|^2\Big),
\end{equation}
where $c_{j}^{(i)}$ and $p_{j}^{(i)}$ are given in Appendix~\ref{App:Kraus}. The summation is performed over the $2^{L-K}$ Kraus operators of the unwanted subspace. The expression of the gate fidelity depends solely on the parameters describing the unwanted spins' evolution since we assumed that $U_0$ is the evolution that would occur in the absence of any unwanted spins. The gate fidelity is clearly maximized when $\sum_{k=0}^{2^{L-K}-1}\Big|\sum_{j\in\{0,1\}} c_{j}^{(k)}p_{j}^{(k)}\Big|^2=2^2$. This happens when the unwanted spins evolve trivially (i.e., independently of the electron's spin state), which is an immediate consequence of the minimization of unwanted nuclear spin one-tangles.

\begin{figure*}[!htbp]
    \centering
    \includegraphics[scale=0.75]{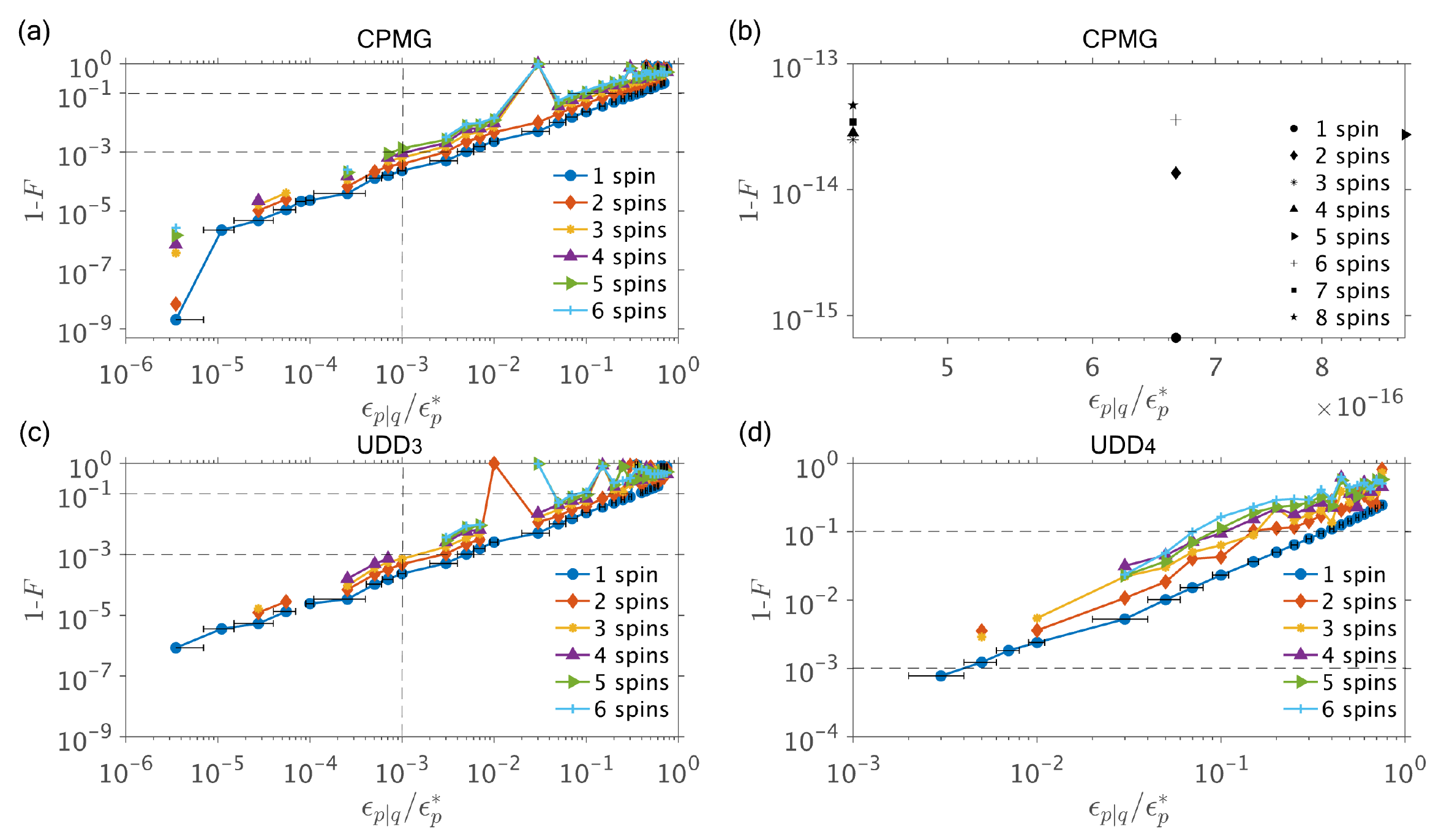}
    \caption{Gate error $1-F$ as a function of one-tangles of unwanted nuclear spins for (a),(b) the CPMG, (c) the UDD$_3$, and (d) the UDD$_4$ sequences. The labels in all graphs show up to how many spins were
    ``traced-out'' from the total system. The unwanted spins have one-tangles in the range [0,0.76]. The error-bars of the blue points show the intervals of one-tangles to which we assign unwanted spins and are the same for all differently colored lines. In the ranges where we cannot find up to 6 unwanted spins, we depict a smaller number of them. In (b) we have used the condition for the trivial evolution to identify unwanted nuclei which do not introduce any gate error. The dashed lines in the plots serve as a guide to the eye.}
    \label{Fig:Infid}
\end{figure*}

To understand the impact of an unwanted spin bath on the target evolution we consider as our target nuclear spins three different groups from Sec.~\ref{SubSec:CR1}: i) those we identified at the $k=2$ CPMG resonance, ii) those at the $k=3$ UDD$_3$ resonance and iii) those at the $k=2$ UDD$_4$ resonance. For each case, we construct an ensemble of unwanted nuclear spins with randomly distributed HF parameters and identify those with one-tangles in the range $[0,0.76]$. As the target gate operation, we consider the evolution of the target spins of Eq.~(\ref{Eq:TargetGate}), isolated from unwanted nuclei. The gate error arises once we introduce unwanted nuclei, let them interact with the electron, and then trace them out to obtain the effective evolution in the target subspace. In reality, we never assume an initial state or trace out nuclei since we can use Eq.~(\ref{Eq:Fid}) to find the gate error by using only the information of the unwanted spins' evolution.

As an example, we gradually build up a bath of six unwanted, spectator nuclei by adding one of them at a time, in each case examining the impact on the gate error. To do this, we start with an ensemble of $3\times10^5$ nuclear spins with randomly distributed HF parameters (with a tolerance of at least $3$~kHz difference for at least one of the HF components to ensure we have sufficiently distinct nuclei) such that their one-tangles span the range $[0,0.76]$. Since each unwanted spin has a different one-tangle, we divide the range $[0,0.76]$ into smaller intervals and assign the nuclear spin one-tangles into these intervals. In Fig.~\ref{Fig:Infid}(a), we depict the infidelity $1-F$ corresponding to the CPMG sequence as a function of the one-tangle interval. For each interval, we gradually increase the number of unwanted nuclei that contribute to the infidelity, starting from 1 and increasing up to 6. Due to the random distribution of HF parameters, it might be the case that there are fewer than six spins in some of these intervals (especially for low values of the one-tangle), in which case we show the gate error as we ``trace out'' a smaller number of spins. As expected, the gate error grows as we increase the size of the nuclear spin environment or as its entanglement with the target subsystem becomes substantial (as indicated by the magnitude of the one-tangle). However, some nuclei can evolve trivially under the CPMG sequence, in particular those whose HF parameters obey the conditions for trivial evolution shown in Sec.~\ref{SubSec:TrivialEvol}. In Fig.~\ref{Fig:Infid}(b), we show the gate error versus the one-tangles of unwanted spins that satisfy the condition for trivial evolution. All one-tangles are trivially zero, leading to a vanishing gate error.

In Fig.~\ref{Fig:Infid}(c) and in Fig.~\ref{Fig:Infid}(d) we show the infidelity of the multipartite gate under the UDD$_3$ or UDD$_4$ evolution. We notice that for UDD$_4$, the one-tangles are distributed at higher values. This is a direct consequence of the more complicated dynamics that the nuclei undergo for this sequence. Recall that multiple conditions allow nuclei to entangle with the electron due to the fact that their individual rotation angles $\phi_0$ and $\phi_1$ are different. 

It is interesting to note that for some values of one-tangles, the gate error shows jumps and becomes very large. It is not surprising that this is possible even at relatively small values of one-tangles ($\sim 10^{-2}$) [see Fig.~\ref{Fig:Infid}(c)]. The reason for this behavior is that the unwanted spins could cause the evolution to deviate from the ideal isolated evolution of Eq.~(\ref{Eq:TargetGate}). However, the resulting gate may have a larger overlap with other target gates. Here we choose not to optimize over the resulting gate, as we want to show the overall tendency of the target subspace gate error as the entanglement of unwanted spins with the remaining system increases. In Appendix~\ref{App:Kraus}, we provide a modified gate fidelity formula if one wishes to optimize over single-qubit gates acting on the target nuclei. 

Although we have not optimized over the sequence parameters and target spin HF parameters, we see that a CPMG sequence with only $N^*=8$ repetitions and a total time of $T\approx 74.5~\mu$s is still capable of entangling eight different nuclear spins with the electron and preserving the multipartite gate operation in general on par with UDD$_4$. However, both UDD sequences are longer in this scenario and require a larger number of sequence iterations than CPMG ($T=1.48$~ms and $N^*=93$ for UDD$_3$, while $T \approx 0.38$~ms and $N^*=41$ for UDD$_4$). Even though we do not compare directly the sequences (as their parameters differ), we see that resorting to long sequences does not necessarily imply enhanced protection of the target evolution. Moreover, in an experimental setup, it is preferable to use a smaller number of sequence iterations to limit potential pulse errors. Experimentally and numerically, it has been shown that CPMG outperforms UDD$_6$~\cite{deLangeSci10} in decoupling capabilities, which is in agreement with a soft cut-off Lorentzian noise spectrum. Further comparison of the gate performance for CPMG, UDD$_3$, and UDD$_4$ can be found in Appendix~\ref{App:GateErrorComp}, where we average over eight different ensembles of randomly generated unwanted nuclei for each sequence.

\subsection{Multipartite gates in a 27 nuclear spin register \label{SubSec:27Spins}}

\begin{figure*}[!htbp]
    \centering
    \includegraphics[scale=0.76]{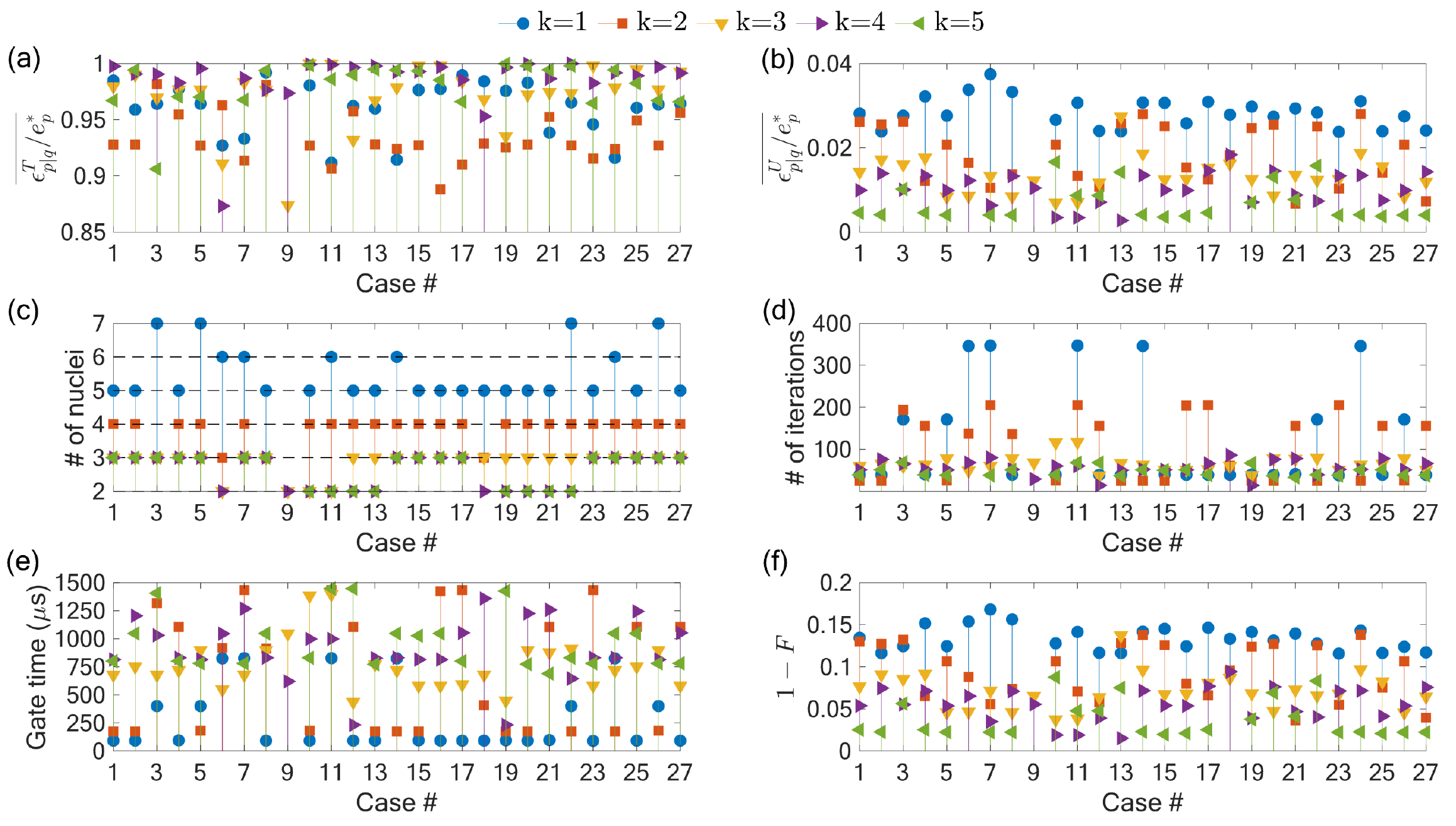}
    \caption{Multipartite gates in a 27 nuclear spin register using the CPMG sequence. Each case $\#$ corresponds to a different set of CPMG unit time and number of iterations. (a) Mean value of target nuclei one-tangles for each of the 27 cases. (b) Mean value of unwanted nuclei one-tangles. (c) Number of target spins to realize the multipartite gates. (d) Number of iterations and (e) gate time of the multipartite gate. (f) Gate error due to residual entanglement with unwanted nuclear spins. For each of the 27 different realizations, the CPMG unit time is optimized close to the resonance time of each of the 27 nuclei. $k$ indicates the number of resonances.     }
    \label{fig:27Spins}
\end{figure*}

Up to this point, we have studied the qualitative behavior of multipartite gates for randomly distributed nuclear spins. In this section we consider an ensemble of 27 $^{13}$C atoms in an NV center ($S=1$) in diamond, using HF parameters experimentally determined via 3D spectroscopy by the Delft group~\cite{TaminiauNat2019,BradleyThesis2021}. To showcase the performance of multipartite gates, we will consider the CPMG sequence. We set the magnetic field to $B=403$~G~\cite{BradleyThesis2021}, which translates into a Larmor frequency of $\omega_L\approx 2\pi \cdot 432$~kHz for the $^{13}$C nuclei. We further select the electron's spin projections to be $s_0=0$ and $s_1=-1$.

To identify target nuclear spins, we could use our analytical expressions to find the number of iterations that maximize multiple one-tangles. Instead, to perform a more rigorous search, we optimize both the time of the CPMG unit and the number of iterations. We explore 135 different realizations (27 cases for each $k\in[1,5]$); in each case, we choose a resonance time of one of the 27 nuclei and vary it within $\pm 0.25~\mu$s. We further perform a search on the number of iterations by constraining the total time of the gate to be $\leq 1.5$~ms. In this way, we restrict the gate time within $T_2^*$ of the nuclei, which ranges from 3 to 17 ms~\cite{BradleyThesis2021}. For each realization, we select the time and number of iterations that ensure: i) one-tangles of target nuclei $>0.8$, ii) one-tangles of unwanted nuclei $<0.14$, iii) mean value of unwanted one-tangles $<0.1$. After we find the potential sets of $(t,N^*)$ which fulfill all the above requirements, we choose a set that can simultaneously entangle two or more nuclear spins with the electron. If no such set exists, we ignore that case. In the end, we calculate the gate fidelity of the target subspace for each of the groups of $(t,N^*)$ in the presence of the remaining unwanted spectator nuclei.

The computation of nuclear spin one-tangles requires only the information of the independent evolution of each nucleus. Hence, this allows us to simulate many nuclear spins without computational difficulty. The gate fidelity, on the other hand, involves $2^{L-K}$ Kraus operators ($L=27$ and $K$ is the number of target nuclei), which translates into $2\times 2^{L-K}$ additions [see Eq.~(\ref{Eq:Fid})]. As an example, a single run for $K=7$ target spins and thus, 20 unwanted spins ($\sim 2\times 10^6$ additions) calculates the gate fidelity within $\sim 8$ seconds, but for $K=2$ ($\sim 67 \times 10^6$ additions) it takes $\sim 4.5$~mins (computational times are w/o parallel computing). However, it is still advantageous that we can do such computations without explicitly defining the Kraus operators.

We display our results in Fig.~\ref{fig:27Spins}. In Fig.~\ref{fig:27Spins}(a) we show the mean of target one-tangles, while in Fig.~\ref{fig:27Spins}(b) we show the mean of the unwanted one-tangles for 27 different realizations, and resonances $k\in[1,5]$. As expected, higher-order resonances in principle give rise to lower residual entanglement with unwanted spins \cite{Dong2020}. In Fig.~\ref{fig:27Spins}(c) we show the number of target nuclei, whose one-tangle mean is the one in Fig.~\ref{fig:27Spins}(a). In general, as the order of the resonance $k$ increases, nuclei tend to decouple more efficiently since the resonant times show larger dispersion, and hence, the number of target nuclei decreases. In Fig.~\ref{fig:27Spins}(d) and Fig.~\ref{fig:27Spins}(e), we show the number of iterations and total gate time. Higher-order resonances require fewer sequence repetitions since the accumulated nuclear rotation angle per iteration is larger. Finally, in Fig.~\ref{fig:27Spins}(f), we show the gate error of the entangling operation. The first resonance yields the highest error since the spectator nuclei have larger residual entanglement with the target spins. The optimization tries to balance the trade-off between maximum achievable entanglement (i.e., target one-tangles $>0.8$) and minimum gate error. Requiring lower values of individual unwanted one-tangles could reduce the gate error more. 

We should further comment that the HF parameters of the 27 nuclear spins are smaller than the randomly generated ones in Sec.~\ref{SubSec:CR1} (see Appendix~\ref{App:HFparamsRandom} and Appendix~\ref{App:27Spins}). It is then a natural consequence that the gate times for the multipartite gates presented in this section are longer. Experimentally, one could identify better candidates for target nuclei to maximize the entanglement in the nuclear spin register while satisfying time constraints. Using target nuclear spins with a bit larger HF parameters could reduce the total gate time. In addition, over- or under- rotation errors that cause the one-tangles of the target nuclei to deviate from their maximum values could potentially be remedied by direct driving of a few nuclear spins or by using hybrid sequence protocols as in Ref.~\cite{Dong2020}. However, our results indicate that multipartite entangling operations can be reliably implemented with gate fidelities above $0.95$ for $k>1$ even without such measures.

\subsection{Speed-up of controlled-gates for QEC  \label{SubSec:SeqVsMulti}}

Practical applications, such as quantum error correction (QEC), require gate durations to be much smaller than $T_2^*$ of the spins which participate in the protocol to ensure reliable performance. Many QEC schemes require repeating a sequence of operations and/or measurements multiple times, and thus it is crucial to perform the gate operations fast; for example, one QEC cycle of Ref.~\cite{CramerNatCommun2016} lasted for $\sim 2.99$~ms. More specifically, for the three nuclei that participated in this QEC scheme~\cite{CramerNatCommun2016}, the durations of each sequential electron-nuclear entangling gate were $980~\mu$s, $400~\mu$s, and $1086~\mu$s, respectively. The accumulation of errors due to decoherence during long gates could be partially alleviated by applying refocusing pulses to extend coherence times~\cite{BradleyPRX19}. However, such techniques add to the experimental overhead, making it desirable to use them only sparingly or not at all if possible; such methods can be avoided if we can accelerate the entangling gates by involving multiple nuclei in the operation simultaneously.
\begin{figure*}[!htbp]
    \centering
    \includegraphics[scale=0.7]{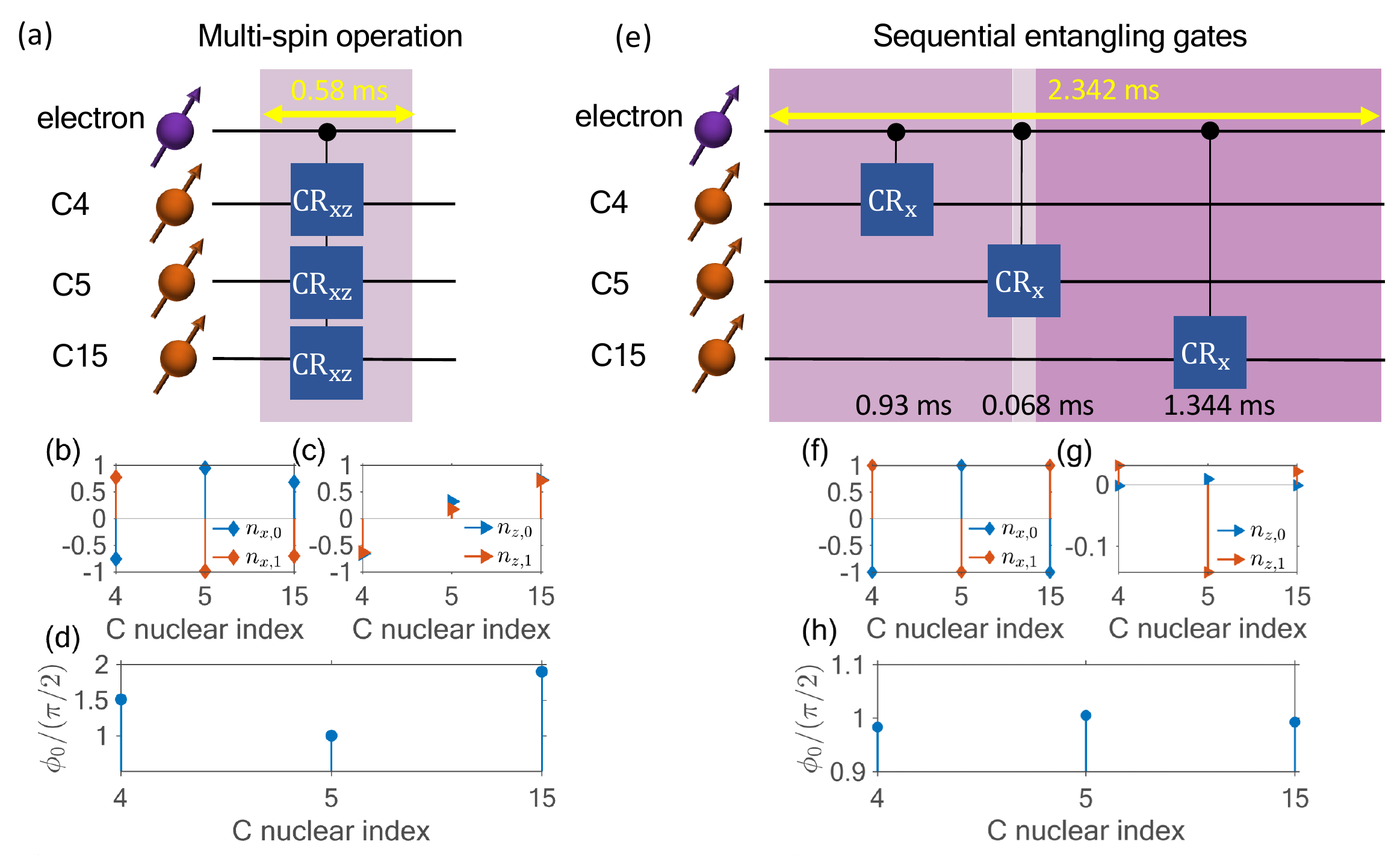}
    \caption{Comparison of synchronous multi-spin gate scheme with the sequential entanglement protocol. Circuit diagram for (a) multi-spin entangling gate operation and for (e) sequential entangling protocol. The $x$-axis components (b), (f) and the $z$-axis components (c), (g) of the $^{13}$C nuclear spin rotations are shown. The subscripts ``0'' and ``1'' on the axis components refer to the nuclear rotations $R_{\textbf{n}_j}$. (d), (h) The rotation angle of each nucleus. (b), (c) and (d) correspond to the multi-spin operation, while (f), (g), and (h) correspond to the sequential entangling protocol. The exact parameters of the rotation axes and rotation angles are given in Table~\ref{tab:11} of Appendix~\ref{App:SeqVsMulti}. Lighter shading in (a) and (e) indicates shorter gate durations. }
    \label{fig:SeqVsMulti}
\end{figure*}

To demonstrate the advantages offered by the synchronous controlled gates, we select as an example case $\#$ 23 for $k=3$ of the previous section [see Table~\ref{tab:8} of Appendix~\ref{App:27Spins}]. For this realization, we entangle simultaneously nuclei $\{\text{C4, C5, C15}\}$ with the electron, with total gate time $T=582.22~\mu$s, individual one-tangles $\epsilon_{p|q}^{\text{nuclear}}=\{0.99994,0.99662,0.99756\}$ (scaled by $2/9$), and a gate error due to residual entanglement with the remaining 24 unwanted nuclei of $1-F=0.067977$. To compare the performance of this direct multi-spin operation against sequential entanglement protocols, we perform another simulation where we entangle each C$j$ nucleus [$j\in\{4,5,15\}$] one at a time with the electron starting with C4. The constraints we impose on the sequential entangling gates are similar to those in the multi-spin case, such that the comparison of the two methods is fair. More details about the constraints and the optimal sequential gates can be found in Appendix~\ref{App:SeqVsMulti}. For now, we stress that we restrict the duration of each entangling gate to be within 1.5~ms (the total gate time of all three gates can exceed 1.5~ms), to allow for potentially enhanced selectivity for each nucleus and a more direct gate fidelity comparison with the multi-spin entanglement protocol.

For each C$j$ nucleus, we search over the first ten resonances ($k\in[1,10]$) and number of CPMG iterations that satisfy our constraints and choose the optimal CR$_x(\pi/2)$ gates [see Table~\ref{tab:10} of  Appendix~\ref{App:SeqVsMulti}]. For C4, we find that the optimal gate time is $T\approx 0.93$~ms with an error due to residual entanglement of $1-F=0.1133$. By performing only this single entangling gate, we already exceed the gate time of $\sim 0.58$~ms of the multipartite operation. For C5, we find that a CR$_x(\pi/2)$ gate can be performed at the shortest gate time of $\sim 68~\mu$s, which leads to a gate error of $1-F=0.1045$. The results for C15 are rather surprising; although we search over ten difference resonances, the best CR$_x(\pi/2)$ gate we can achieve is long ($\sim 1.344$~ms), and the error ($1-F=0.1421$) is larger than the other two entangling gates. 

Overall, we see that the sequential gates for the $\{\text{C4, C5, C15}\}$ set lead to significant gate error since these fail to decouple each nucleus from the remaining spin bath effectively. The total gate time of the sequential entangling operations is $\sim 2.342$~ms, already four times larger than the gate time of the multipartite gate on $\{\text{C4, C5, C15}\}$. Further, the sets we identified as target spins for the multi-spin gates in Sec.~\ref{SubSec:27Spins} contain nuclei, which when attempted to be addressed individually, lead to electron-nuclear entangling gates that suffer from cross-talk arising from the other nuclear spins of the set. Indeed, this is verified by the gate error sources we identified [see Table~\ref{tab:8} of Appendix~\ref{App:27Spins}]; for example, the infidelity of the C4 entangling gate is due to nonzero residual entanglement of the electron with the C15 nucleus. Similar observations hold for the errors of the other two sequential gates. Thus, our formalism not only provides a faithful metric of nuclear spin selectivity but identifies cross-talk issues and optimal nuclear spin candidates for performing entangling gates within time constraints.

In Fig.~\ref{fig:SeqVsMulti} we compare the multi-spin protocol with the sequential entanglement generation scheme. In the latter case, the gates are very close to CR$_x(\pm\pi/2)$ [see Figs.~\ref{fig:SeqVsMulti}(f), (g), (h) and Table~\ref{tab:11} of Appendix~\ref{App:SeqVsMulti}]. The gates acting on the nuclei in the multipartite case, in principle, have both nonzero $x$- and $z$-axis components [see Figs.~\ref{fig:SeqVsMulti}(b), (c), (d) and Table~\ref{tab:11} of Appendix~\ref{App:SeqVsMulti}]. Although the gates of the two approaches are different, they are equivalent up to local rotations.  

\subsection{Three-qubit bit-flip code}

Let us now consider a three-qubit measurement-free QEC protocol that does not require stabilizer measurements or ancillary qubits, and can correct a single bit- or phase-flip error~\cite{CramerThesis2016}. Our goal is to protect the initial state of the electron. Using two nuclei which we assume have been initialized into the $|1\rangle$ state, we will show how to use the CR$_{xz}$ multi-spin operations to recover the electron's state from a single bit-flip error. We will also compare the performance of this approach with the sequential entangling gate protocol.
\begin{figure*}[!htbp]
    \centering
    \includegraphics[scale=0.75]{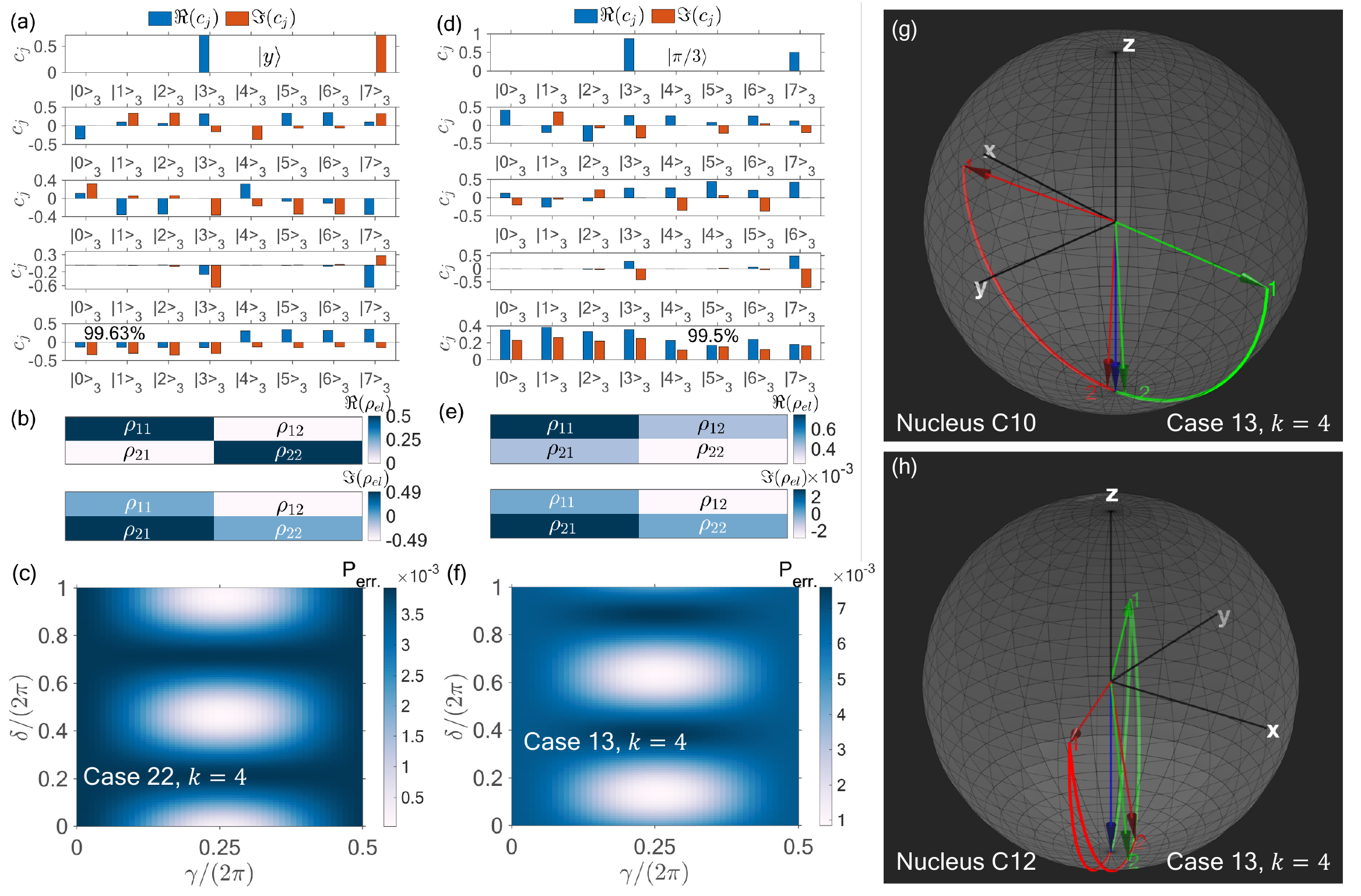}
    \caption{Three-qubit bit-flip code using the CR$_{xz}$ multi-spin operations. The electron's state is encoded into three physical qubits, two of which correspond to the $^{13}$C atoms C10 and C12. 
    (a) Recovery of the electron state $|y\rangle$  [case $\# 22$, $k=4$ of Fig.~\ref{fig:27Spins}]. (d) Recovery of the electron state $|\pi/3\rangle=\cos(\pi/6)|0\rangle+\sin(\pi/6)|1\rangle$ [case $\# 13$, $k=4$ of Fig.~\ref{fig:27Spins}]. From top to bottom the panels in (a), (d) show the coefficients of: initial, encoded, error, decoded, and corrected state. Blue (red) bars are the real (imaginary) parts of the coefficients. The probability to measure the electron in $|y\rangle$ in (a) is 99.63$\%$ while to measure it in $|\pi/3\rangle$ in (d) is $99.5\%$. Real and imaginary components (b), (e) of the final reduced density matrix of the electron verify the recovery of the initial state. Error probability $1-|\langle \psi_{\text{el},0}|\psi_{\text{final}}\rangle|^2$ of measuring incorrect state of the electron at the end of the QEC code for case $\#$22 and $k=4$ (c) and for case $\#$13 and $k=4$ (f). $|\psi_{\text{el},0}\rangle$ is defined as $\cos(\gamma/2)|0\rangle+e^{i\delta}\sin(\gamma/2)|1\rangle$.
    Evolution of $^{13}$C atoms C10 (g) and C12 (h) up to the decoding step, when the electron undergoes a bit-flip. The initial state is $|1\rangle$ for each nucleus (blue arrow). The nuclei follow the green curve evolution if the electron is initially in $|0\rangle$ or the red curve evolution if it is initially in $|1\rangle$. After an error happens on the electron and we perform the decoding, the nuclei approximately return to $|11\rangle$ such that the subsequent Toffoli gate corrects the bit-flip error.}
    \label{fig:3QubitQEC}
\end{figure*}

The QEC protocol consists of three parts: i) the encoding of the electron's physical state into a logical state, ii) the decoding, and iii) the correction. The latter is performed by decomposing the three-qubit Toffoli gate (controlled on the nuclei) using single- and two-qubit gates~\cite{CramerThesis2016}. 
The entire QEC circuit of the sequential protocol can be found in Appendix~\ref{App:Toffoli} and Ref.~\cite{CramerThesis2016}. Such a measurement-free QEC protocol has been realized experimentally in Ref.~\cite{TaminiauNatNano2014}, where very high theoretical fidelities (in excess of 99$\%$) of electron-nuclear entangling gates were reported. However, in Ref.~\cite{TaminiauNatNano2014} it was mentioned that these estimates did not account for the presence of unwanted nuclei, which leads to extra loss of electron coherence. Here we show explicitly that the presence of the unwanted spin bath can have a significant impact on the implementation of target operations, especially when it undergoes substantial entanglement with the electron. 

In the following analysis, we consider that only the electron and the two nuclei that are part of the protocol are present since we cannot simulate the full density matrix of 28 qubits. Although we ignore the presence of the remaining nuclei, our analysis is complete as will provide the gate errors that capture residual entanglement links with nuclei from the entire register. 

To explain the principles of the multi-spin three-qubit QEC protocol, suppose that we wish to recover an arbitrary state of the electron from an X-error that happens after the encoding. We implement the encoding and decoding using the CR$_{xz}$ gate. In the absence of errors, the encoding and decoding gates need to combine to flip the initial $|11\rangle$ state of the nuclei into $|00\rangle$, such that the subsequent Toffoli gate is not activated. Due to the more complicated dynamics induced by the multi-spin gates, this requirement is not satisfied by the encoding/decoding CR$_{xz}$ gates alone. We resolve this issue by introducing unconditional $R_y(-\pi)$ gates on the nuclei in between the two encoding/decoding CR$_{xz}$ gates; this ensures that the encoding/decoding and $R_y(-\pi)$ gates compose together so as to flip the nuclei, and deactivate the subsequent Toffoli gate [see Appendix~\ref{App:Toffoli} for a proof]. 

The correction circuit is composed of unconditional nuclear and electron rotations, as well as CR$_x(\pi/2)$ gates. For simplicity, we will treat the additional R$_{y}(-\pi)$ rotations that we require as part of the encoding and the gates of the correction circuit as ideal. We do not find the optimal parameters to perform the correction gates, since we would numerically optimize and implement them in the same way for both the sequential and the multi-spin schemes. The $R_y(-\pi)$ rotations can be implemented by direct driving of the nuclei or composed through unconditional $R_x$ and $R_z$ gates obtained via dynamical decoupling sequences~\cite{CramerThesis2016}, through appropriate tuning of the interpulse spacing of the sequence.

A bit-flip on the electron makes the rotation that each nucleus undergoes during the encoding differ from the one it undergoes during the decoding. The success of our protocol lies in the fact that now the CR$_{xz}$ and $R_y(-\pi)$ gates combine to rotate the nuclei approximately about the $z$-axis. This means that the nuclei return close to the $|11\rangle$ state, activating the subsequent Toffoli gate. The evolution of the nuclei up to the decoding involves also a non-vanishing $x$-axis rotation. Consequently, at the end of the decoding, the nuclei are not fully disentangled from the electron. However, the $x$-rotation is quadratically suppressed by the nuclear Larmor frequency [see Appendix~\ref{App:Toffoli}], meaning that the recovery operation brings the electron close to its initial state, but as we will quantify shortly, the electron's final state is slightly mixed.

To illustrate the performance of the multi-spin QEC scheme, we start with the recovery of the electron state $|y\rangle$  from a bit-flip error. We consider case $\#$22 and $k=4$ of the multi-spin gates of Fig.~\ref{fig:27Spins}, for which we entangle the electron with nuclei C10 and C12. The gate error due to residual entanglement with unwanted spins is $1-F\approx 0.04$, and the gate time is $T\approx 645.6~\mu$s.  In Fig.~\ref{fig:3QubitQEC}(a), we show the coefficients of the three-qubit state at each step of the circuit, prior to the encoding and up to the correction step. We find that the probability of recovering the electron's state is $99.63\%$. The electron's reduced density matrix [Fig.~\ref{fig:3QubitQEC}(b)] after tracing out the two nuclei verifies that it is close to the desired $|y\rangle\langle y|$ state; the purity is found to be $99.26\%$. In Fig.~\ref{fig:3QubitQEC}(c) we show the error probability, defined as $1-|\langle \psi_{\text{el},0}|\psi_\text{final}\rangle|^2$ ($|\psi_{\text{el,0}}\rangle$ is the electron's initial state and $|\psi_\text{final}\rangle$ the final three-qubit state) for arbitrary initial states $|\psi_{\text{el},0}\rangle=\cos(\gamma/2)|0\rangle+e^{i\delta}\sin(\gamma/2)|1\rangle$. We find that in all cases, we recover the electron's state with an error on the order of $\sim 10^{-3}$. 

We perform a similar analysis for the recovery of the $|\pi/3\rangle=\cos(\pi/6)|0\rangle+\sin(\pi/6)|1\rangle$ state, now for case $\#$13 and $k=4$ of Fig.~\ref{fig:27Spins}. For this realization, we again entangle the electron with nuclei C10 and C12; the gate duration is $T\approx 827~\mu$s, and the gate error due to residual entanglement is $1-F\approx 0.0152$. In Fig.~\ref{fig:3QubitQEC}(d), we show the coefficients of the three-qubit state, and in Fig.~\ref{fig:3QubitQEC}(e) the electron's reduced density matrix, whose purity is $99\%$. We find that the recovery probability is $99.5\%$. In Fig.~\ref{fig:3QubitQEC}(f), we show the error probability for arbitrary initial states of the electron. In Fig.~\ref{fig:3QubitQEC}(g) and (h), we show the evolution of each nuclear spin up to the decoding step. The blue arrows indicate the initial state of the nuclei, which is the $|1\rangle$ state. The green (red) curves show the path each nucleus traces on the Bloch sphere if the electron starts from the $|0\rangle$ ($|1\rangle$) state and undergoes a bit-flip. The final green/red arrows indicate that the nuclei return approximately to the $|11\rangle$ state, such that the Toffoli gate then corrects the electron's bit-flip. In the case when no bit-flip occurs, both nuclei traverse a great arc on the Bloch sphere and end up exactly in the $|0\rangle$ state at the end of the decoding [see Appendix~\ref{App:Toffoli}].

We now compare our direct multi-spin protocol with the sequential three-qubit QEC code. For a fair comparison, we impose constraints on the sequential entangling gates that are similar to those of the multi-spin operation. By searching over the first ten resonances of C12 or C10 we find a list of acceptable CR$_x(\pi/2)$ gates [see Table~\ref{tab:12} of Appendix~\ref{App:OptimalCRx}]. For C12, the CR$_x(\pi/2)$ gate can be implemented with error $1-F=0.0238$ due to unwanted residual entanglement and duration of 449.4277~$\mu$s. This gate is faster than the two cases of multi-spin operations mentioned previously [although faster multi-spin gates were found in Fig.~\ref{fig:27Spins}], with an error lower than case $\#22$ and $k=4$, but higher than case $\#13$ and $k=4$. Note that in Fig.~\ref{fig:27Spins}, the multi-spin gates were restricted to $k\leq 5$, but to implement the CR$_x(\pi/2)$ gate reliably, we expanded the search over $k\geq 5$, as higher-order resonances are needed for improved selectivity for the sequential scheme. Addressing the C10 nucleus is much more challenging than addressing C12. In the time constraint of $1.5$~ms, the lowest infidelity is $\sim  0.384$; imposing a new constraint of 5~ms, we find that the CR$_x(\pi/2)$ gate can be implemented for a duration of $\sim 3$~ms with an infidelity of $\sim 0.106$.

The sequential scheme can, in principle, succeed with a recovery probability of $100\%$, assuming all gates are error-free, since the disentanglement in the decoding step can be perfect [see Appendix~\ref{App:Toffoli}]. Nevertheless, errors due to unresolved residual entanglement reduce the probability of recovering the electron's initial state. That is, tracing out unwanted spins and the nuclei of the protocol yields in general a mixed density matrix for the electron. Thus, in cases when cross-talk errors cannot be resolved by the sequential scheme, the recovery probability is expected to be smaller for the sequential protocol compared to the multi-spin scheme, and the electron's reduced density matrix more mixed at the end of the correction.

For both protocols, it is necessary to implement the correction CR$_x(\pi/2)$ gates reliably. The advantage of the multi-spin QEC scheme lies in the fact that it can reduce the encoding and decoding durations by utilizing the CR$_{xz}$ operations, while to ensure reliable CR$_{x}(\pi/2)$ correction gates, we can allow more relaxed time constraints for the Toffoli implementation. In this way, we save time during the first two parts of the QEC scheme. On the other hand, the entire sequential QEC scheme relies on the successful performance of the CR$_x(\pi/2)$ gates, which are implemented using the same optimal sequence parameters for all parts of the circuit. Thus, in the sequential QEC scheme, one might have to trade off gate fidelity with speed of operations, and the total duration of the gates can quickly exceed the coherence times. 

Interestingly, both protocols can be combined to provide optimal performance of the QEC codes. For example, reliable and fast CR$_{x}(\pi/2)$ encoding/decoding gates could be combined with CR$_{xz}$ encoding/decoding gates to address subsets of nuclei that cannot be resolved individually within given time constraints. Considering that the number of spinful nuclei in experimental conditions could be hundreds, it is highly likely that particular CR$_x(\pi/2)$ gates will fail to provide both speed of operation and selectivity of a single spin. This was verified, for example, in Ref.~\cite{BradleyPRX19}, wherein certain electron-nuclear Bell-state fidelities were as low as $63\%$ due to unresolved cross-talk arising from nearby nuclei, combined with loss of coherence due to long two-qubit operations. Inability to address nuclei individually means that they would have to be excluded from any protocol (i.e., decoupled such that they don't induce errors) but could become a valuable resource using the multi-spin gates. The CR$_{xz}$ encoding/decoding gates would be accompanied by $R_y(-\pi)$ unconditional rotations on these nuclear spin subsets, which, as we mentioned previously, are required for the multi-spin QEC scheme.

Our analysis shows that the multi-spin entangling gates can drastically reduce the entanglement generation time and mitigate dephasing issues. In a measurement-free QEC scheme, the entanglement generation speed-up could be crucial for protecting the logical state; leaving it unprotected for a shorter duration reduces the probability of errors occurring during the decoding step.  Additionally, the synchronous controlled gates can outperform the sequential entanglement schemes, especially when we cannot resolve cross-talk issues. An interesting future direction would be to examine further the utility of CR$_{xz}$ gates for QEC protocols, and potentially adjust the correction circuit to account for the imperfect disentanglement at the end of the decoding.

\section{Conclusions}

Nuclear spins are an essential component of spin-based solid-state platforms for quantum networks. Harnessing their full potential to create large-scale quantum networks requires a detailed understanding of and precise control over the entanglement distribution in the system. We showed how to quantify the entanglement in a multi-nuclear spin register coupled to a single electron qubit and presented a faithful metric for nuclear spin selectivity. We studied the properties of CPMG, UDD$_3$, and UDD$_4$ sequences and extended their resonance conditions to arbitrary electron systems for applicability to any defect qubit in diamond or SiC. We further showed how to implement synchronous controlled gates on multiple nuclei by driving the electron appropriately. Such multipartite gates provide a speed-up over the conventional way of generating sequential entanglement links, especially for large nuclear spin registers, where the total sequence time can exceed the dephasing time. We quantify the performance of multipartite gates implemented by CPMG, UDD$_3$, or UDD$_4$ sequences in the presence of unwanted nuclear spins, revealing that the gate fidelity tends to decrease as the residual entanglement with the unwanted bath becomes significant. Using experimental parameters for 27 $^{13}$C atoms in close proximity to an NV center in  diamond, we have further verified that such multipartite gates can be performed reliably and with high fidelity, and can facilitate implementations of quantum error correction codes.

\begin{acknowledgments}
The authors would like to thank Vlad Shkolnikov for useful discussions. E.B. acknowledges support from NSF Grant No. 1847078. S.E.E. acknowledges support from NSF Grant No. 1838976.
\end{acknowledgments}

\appendix

\section{Mathematical description of multi-spin nuclear register }

\subsection{Evolution operator of multiple spins\label{App:EvolOper}}

We mentioned in the main text that $\pi$-pulse sequences generate an evolution operator which is a sum of terms, each of which includes an electron spin projector tensored with a product of single-qubit gates acting on the nuclei. Here, we show this explicitly. Let us consider for simplicity two nuclear spins, with HF parameters $A_l$ and $B_l$ [$l\in\{1,2\}$]. Neglecting inter-nuclear spin interactions, the secular Hamiltonian is given by:
\begin{equation}
\begin{split}
    H&=\frac{\omega_L}{2}(\mathds{1}\otimes\sigma_z\otimes\mathds{1}+\mathds{1}\otimes\mathds{1}\otimes\sigma_z)+\frac{A_1}{2}Z_e\otimes \sigma_z\otimes\mathds{1}
    \\&+\frac{B_1}{2}Z_e\otimes \sigma_x\otimes\mathds{1}
    +\frac{A_2}{2}Z_e\otimes \mathds{1}\otimes\sigma_z+\frac{B_2}{2}Z_e \otimes\mathds{1}\otimes\sigma_x
    \\&=\sum_{j\in\{0,1\}}\sigma_{jj}\otimes \Big[\frac{ \omega_L+s_jA_1}{2}\sigma_z\otimes\mathds{1}+\frac{s_j B_1}{2}\sigma_x\otimes\mathds{1}
    \\&~~~~~~~~~~~~~~~~~~~~+\frac{\omega_L+s_jA_2}{2}\mathds{1}\otimes\sigma_z +\frac{s_j B_2}{2}\mathds{1}\otimes\sigma_x\Big]
    \\&=\sum_{j\in\{0,1\}}\sigma_{jj}\otimes (H_{j}^{(1)}\otimes \mathds{1}+\mathds{1}\otimes H_j^{(2)}),
    \end{split}
\end{equation}
where we have defined $H_j^{(l)} $:
\begin{equation}
    H_j^{(l)}=\frac{\omega_L+s_j A_l}{2}\sigma_z^{(l)}+\frac{s_j B_l}{2}\sigma_x^{(l)},
\end{equation}
with $\sigma_x^{(l)}$ and $\sigma_z^{(l)}$ being the Pauli matrices which act on the $l$-th spin (and the identity acts on the other spin). As a concrete example, let us focus on the CPMG sequence ($t/4-\pi-t/2-\pi-t/4$). Its evolution operator over one unit of the sequence (which consists of two pulses) has the form:
\begin{equation}
    U= \sigma_{00} \otimes e^{-i\tilde{h}_0}e^{-2i\tilde{h}_1}e^{-i\tilde{h}_0}+\sigma_{11}\otimes e^{-i\tilde{h}_1}e^{-2i\tilde{h}_0}e^{-i\tilde{h}_1},
\end{equation}
where $\tilde{h}_j=t/4(H_j^{(1)}\otimes \mathds{1}+\mathds{1}\otimes H_j^{(2)})$. Notice that $[H_j^{(1)}\otimes \mathds{1},\mathds{1}\otimes H_j^{(2)}]=0$, and thus we can write down the total evolution operator as
\begin{equation}
    U=\sum_{j\in\{0,1\}} \sigma_{jj} \otimes R_{\textbf{n}_j}^{(1)}(\phi_j^{(1)}) \otimes R_{\textbf{n}_j}^{(2)}(\phi_j^{(2)}),
\end{equation}
where $R_{\textbf{n}_0}^{(1)}(\phi_0^{(1)})=e^{-i  H_0^{(1)}t/4}e^{-i  H_1^{(1)}t/2}e^{-i  H_0^{(1)}t/4}$ ($R_{\textbf{n}_1}^{(1)}(\phi_1^{(1)})=e^{-i  H_1^{(1)}t/4}e^{-i  H_0^{(1)}t/2}e^{-i  H_1^{(1)}t/4}$), and similarly for  $R_{\textbf{n}_j}^{(2)}(\phi_j^{(2)})$. Therefore, if more nuclear spins are considered, their Hamiltonians commute and thus, one obtains a tensor product of single-qubit rotations acting on the nuclei. 

\subsection{Kraus operators and gate fidelity \label{App:Kraus}}

In the main text, we mentioned that the unwanted nuclei affect the gate fidelity of target nuclei when the former have non-zero entanglement with the target subspace. Here we provide the steps to obtain the formula for the gate fidelity of the target subspace.

One way to describe the evolution of the target subspace in the presence of unwanted spins is by tracing out the latter. This procedure can be performed on the density matrix level, but this requires that we specify an initial state for the system. To avoid this limitation, we can instead describe the same partial-trace channel using the operator-sum representation~\cite{nielsen_chuang_2010}. The elements of the partial-trace channel are Kraus operators, defined via a chosen complete basis for the environment (i.e., the unwanted spins). Since one can choose any complete basis, the Kraus operators are not unique. Using the operator-sum representation then, one can naturally extend the fidelity of a general quantum operation into the form~\cite{Pedersen2007}:
\begin{equation}\label{Eq:Fid0}
    F=\frac{1}{m(m+1)}\sum_k \text{tr}[(U_0^\dagger E_k)^\dagger U_0^\dagger E_k]+|\text{tr}[U_0^\dagger E_k]|^2,
\end{equation}
where $m=2^{K+1}$ is the dimension of the target subspace (consisting of the electron and $K$ target spins), whereas $E_k$ are the Kraus operators of the quantum channel described by $\mathcal{E}(\rho)=\sum_{k}E_k\rho E_k^\dagger$, and they satisfy the completeness relation $\sum_k E_k^\dagger E_k=\mathds{1}$. 

We assume $L$ nuclear spins in total, with $K$ target ones and hence, $L-K$ unwanted. The environment is thus spanned by $2^{L-K}$ basis states. We further assume that we have permuted the total evolution operator $U$ such that the target spins appear first in the tensor product with the electron's projector and the unwanted spins appear in the last positions, i.e.:
\begin{equation}
    U=\sum_{j\in\{0,1\}}\sigma_{jj}\otimes_{k=1}^{K}R_{\textbf{n}_j^{(k)}}(\phi_{j}^{(k)})\otimes_{l=1}^{L-K}R_{\textbf{n}_j^{(K+l)}}(\phi_j^{(K+l)}).
\end{equation}
Without loss of generality, we consider the initial state of the environment to be $|e_0\rangle\equiv |0\rangle^{\otimes(L-K)}$, which when extended to the total space becomes $|e_0\rangle=\mathds{1}_{K+1\times K+1}\otimes|0\rangle^{\otimes(L-K)}$. Here $\mathds{1}_{K+1\times K+1}$ is the identity gate acting on the space of target spins and the electron. We further define the complete computational basis $\{|e_i\rangle\}_{i=0}^{2^{L-K}-1}$, where all $|e_i\rangle$ states correspond to all possible bit-strings of zeros and ones. The states $|e_i\rangle$ are again extended into the total space as $|\tilde{e}_i\rangle=\mathds{1}_{K+1\times K+1}\otimes|e_i\rangle$. With these definitions we are now ready to introduce the expression for the $i$-th Kraus operator of the partial-trace quantum channel:

\begin{equation}\label{Eq:Kraus0}
    \begin{split}
    E_i&=\langle \tilde{e}_i|U|e_0\rangle =
        \sum_{j\in\{0,1\}}\sigma_{jj}\otimes_{k=1}^K R_{\textbf{n}_j^{(k)}}(\phi_j^{(k)}) \Big\{
        \\&~~~\langle e_i|\Big[\otimes_{l=1}^{L-K}R_{\textbf{n}_j^{(K+l)}}(\phi_j^{(K+l)})\Big]|0\rangle^{\otimes(L-K)}\Big\}.        
    \end{split}
\end{equation}

If for the state $|e_i\rangle$ the $m$-th nuclear spin of the environment is in state $|0\rangle$ then we have:
\begin{equation}\label{Eq:Kraus1}
    \langle 0|R_{\textbf{n}_j^{(m)}}(\phi_j^{(m)})|0\rangle=\cos\frac{\phi_j^{(m)}}{2}-in_{z,j}^{(m)}\sin\frac{\phi_j^{(m)}}{2},
\end{equation}
and whenever the $m$-th ket is $|1\rangle$ we have:
\begin{equation}\label{Eq:Kraus2}
    \langle 1|R_{\textbf{n}_j^{(m)}}(\phi_j^{(m)})|0\rangle=-i(n_{x,j}^{(m)}+i n_{y,j}^{(m)})\sin\frac{\phi_j^{(m)}}{2}.
\end{equation}
Suppose that out of the $L-K$ spins in the environment $M$ of them are in $|0\rangle$ and the other $L-K-M$ are in state $|1\rangle$. Substituting Eq.~(\ref{Eq:Kraus1}) and Eq.~(\ref{Eq:Kraus2}) into Eq.~(\ref{Eq:Kraus0}) we obtain the final form of the $i$-th Kraus operator:
\begin{equation}\label{Eq:KrausOper}
    E_i=
    \sum_j c_{j}^{(i)}p_{j}^{(i)}\sigma_{jj}\otimes_{k=1}^K R_{\textbf{n}_j^{(k)}}(\phi_j^{(k)}),
\end{equation}
where we define $c_{j}^{(i)}\equiv\prod_{m=m_{1}}^{m_{M}}\Big[\cos\frac{\phi_j^{(m)}}{2}-in_{z,j}^{(m)}\sin\frac{\phi_j^{(m)}}{2}\Big]$ and $p_{j}^{(i)}\equiv\prod_{s=s_1}^{s_{L-K-M}}\Big[-i(n_{x,j}^{(s)}+in_{y,j}^{(s)})\sin\frac{\phi_j^{(s)}}{2})\Big]$, while $\{n_x,n_y,n_z\}$ correspond to the rotation axis components of each nuclear spin. In the case when $M=L-K$ (i.e., $|e_i\rangle=|0\rangle^{\otimes (L-K)}$) it holds that  $p_{j}^{(i)}=1$, and in the case when $M=0$ (i.e. $|e_i\rangle=|1\rangle^{\otimes (L-K)}$) it holds that $c_{j}^{(i)}=1$.

The last element we need to evaluate the expression of the gate fidelity for the target subspace is the target gate operation $U_0$. We take as our target gate the evolution operator of the $K$ target spins in the absence of the unwanted spins, i.e.,
\begin{equation}\label{Eq:TargetGateApp}
    U_0=\sum_{j\in\{0,1\}}\sigma_{jj}\otimes_{k=1}^{K}R_{\textbf{n}_j^{(k)}}(\phi_j^{(k)}).
\end{equation}
By substituting Eq.~(\ref{Eq:TargetGateApp}) and Eq.~(\ref{Eq:KrausOper}) into Eq.~(\ref{Eq:Fid0}), we find that the expression of the gate fidelity reads:
\begin{equation}\label{Eq:FidApp}
\begin{split}
    F&=\frac{1}{m(m+1)}\Big(\text{tr}[\sum_{k=1}^{2^{L-K}} E_k^\dagger E_k]+\sum_{k=1}^{2^{L-K}} |\text{tr}[U_0^\dagger E_k]|^2 \Big)
    \\&=\frac{1}{m(m+1)}\Big(m+\sum_{k}\Big|\text{tr}\Big[\sum_{j}c_{j}^{(k)}p_{j}^{(k)}\sigma_{jj}\otimes \mathds{1}_{2^{K}\times 2^{K}}\Big]\Big|^2\Big)
    \\&=\frac{1}{m(m+1)}\Big(m+\sum_k\Big|\text{tr}\Big[\sum_{j}c_{j}^{(k)}p_{j}^{(k)}\sigma_{jj}\Big]\text{tr}[\mathds{1}_{2^K\times 2^K}]\Big|^2\Big)
    \\&=\frac{1}{2^{K+1}(2^{K+1}+1)}\Big(2^{K+1}+2^{2K}\sum_{k}\Big|\sum_{j\in\{0,1\}}c_{j}^{(k)}p_{j}^{(k)}\Big|^2\Big)
    \\&=\frac{1}{2^{K+1}+1}\Big(1+2^{K-1}\sum_{k}\Big|\sum_{j\in\{0,1\}}c_{j}^{(k)}p_{j}^{(k)}\Big|^2\Big),
    \end{split}
\end{equation}
where we have used the fact that $U_0$ is a $2^{K+1}\times 2^{K+1}$ target gate, the Kraus operators, $E_k$, are projectors with dimension $2^{K+1}\times 2^{K+1}$, as well as the trace property of the Kronecker product $\text{tr}[A\otimes B]=\text{tr}[A]\text{tr}[B]$.

In Sec.~\ref{SubSec:CR2}, we mentioned that one can optimize the gate fidelity over the target gate. For a generic target gate, it is difficult to find a closed form expression of the gate fidelity. For this reason, we assume a target gate of the form:
\begin{equation}\label{Eq:NewTargetGate}
    U_0 = \sum_{\rho\in\{0,1\}} \sigma_{\rho\rho} \otimes_{k=1}^K R_{\textbf{n}'_\rho}(\phi_\rho'^{(k)}),
\end{equation}
where now one would have to optimize over the single qubit rotations that act on the target nuclear spins. Again, the first step is to calculate $U_0^\dagger E_i$ which gives:
\begin{widetext}
\begin{equation}
\begin{split}
    U_0^\dagger E_i&=\sum_\rho \sum_j p_{j}^{(i)}c_{j}^{(i)}\sigma_{\rho\rho }\sigma_{jj}\otimes_{k=1}^K R_{\textbf{n}_\rho'}^\dagger({\phi'}_\rho^{(k)})R_{\textbf{n}_j'}(\phi_j^{(k)})
    \\&=\sum_j p_{j}^{(i)}c_{j}^{(i)}\sigma_{jj}\otimes_{k=1}^K\Big\{ \Big[\cos\frac{{\phi'}_j^{(k)}}{2}+i\boldsymbol{\sigma}\cdot{\textbf{n}_j'}^{(k)}\sin\frac{{\phi'}_j^{(k)}}{2}\Big] 
    \Big[\cos\frac{{\phi}_j^{(k)}}{2}-i\boldsymbol{\sigma}\cdot\textbf{n}_j^{(k)}\sin\frac{{\phi}_j^{(k)}}{2}\Big] \Big\}
    \\&=\sum_{j}p_j^{(i)}c_{j}^{(i)}\sigma_{jj}\otimes_{k=1}^K \Big\{
    \cos\frac{{\phi'}_j^{(k)}}{2}\cos\frac{{\phi}_j^{(k)}}{2}
    +
    (\boldsymbol{\sigma}\cdot {\textbf{n}'_j}^{(k)})(\boldsymbol{\sigma}\cdot \textbf{n}_j^{(k)})\sin\frac{{\phi}_j^{(k)}}{2}\sin\frac{{\phi'}_j^{(k)}}{2}
    \\&-
    i\boldsymbol{\sigma}\cdot\Big[\textbf{n}_j^{(k)}\sin\frac{\phi_j^{(k)}}{2}\cos\frac{{\phi'}_j^{(k)}}{2}-{\textbf{n}'}_j^{(k)}\sin\frac{{\phi'}_j^{(k)}}{2}\cos\frac{\phi_j^{(k)}}{2}\Big]\Big\}
    \\&=\sum_{j}p_j^{(i)}c_{j}^{(i)}\sigma_{jj}\otimes_{k=1}^K C^{(k)}_j
    \Big\}.
    \end{split}
\end{equation}
Evaluating the trace gives:
\begin{equation}
\begin{split}
    \text{tr}[U_0^\dagger E_i]&=\sum_j \text{tr}[p_j^{(i)}c_j^{(i)}\sigma_{jj}]\prod_{k=1}^K \text{tr} [C_j^{(k)}]
   =\sum_j p_j^{(i)}c_j^{(i)}\prod_{k=1}^K 2\Big(\cos\frac{\phi_j^{(k)}}{2}\cos\frac{{\phi'}_j^{(k)}}{2}+\textbf{n}_j^{(k)}\cdot{\textbf{n}'}_j^{(k)}\sin\frac{\phi_j^{(k)}}{2}\sin\frac{{\phi'}_j^{(k)}}{2}\Big)
   \\&=2^K\sum_j p_j^{(i)}c_j^{(i)}f_j,
    \end{split}
\end{equation}
\end{widetext}
where we have have defined $f_j=\prod_{k=1}^K\Big(\cos\frac{\phi_j^{(k)}}{2}\cos\frac{{\phi'}_j^{(k)}}{2}+\textbf{n}_j^{(k)}\cdot{\textbf{n}'}_j^{(k)}\sin\frac{\phi_j^{(k)}}{2}\sin\frac{{\phi'}_j^{(k)}}{2}\Big)$. Finally, the fidelity expression reads:
\begin{equation}
\begin{split}
    F&=\frac{1}{2^{K+1}(2^{K+1}+1)}\Big(2^{K+1}+2^{2K}\sum_{i=1}^{2^{L-K}}\Big|\sum_j p_j^{(i)}c_j^{(i)}f_j\Big|^2\Big)
    \\&=\frac{1}{2^{K+1}}\Big(1+2^{K-1}\sum_{i=1}^{2^{L-K}}\Big|\sum_j p_j^{(i)}c_j^{(i)}f_j\Big|^2\Big).
    \end{split}
\end{equation}
Clearly, for ${\phi'}_j^{(k)}=\phi_j^{(k)}$ and ${\textbf{n}'}_j^{(k)}=\textbf{n}_j^{(k)}$, $f_j=1$ and we recover Eq.~(\ref{Eq:FidApp}). To find if there is a higher overlap with the target gate of Eq.~(\ref{Eq:NewTargetGate}), one would have to optimize over the set $\{{\phi'}_j^{(k)},{\textbf{n}'}_j^{(k)}\}$, which corresponds to the parameters of the single qubit rotations that act on the target subspace. Such a computation could be potentially performed via gradient-based optimization methods, supplemented by the Jacobian. If a target gate with better overlap is found, then the one-tangles can be re-evaluated using the optimized set $\{{\phi'}_j^{(k)},{\textbf{n}'}_{j}^{(k)}\}$ to obtain the entanglement distribution of the target subsystem.

\section{Resonance times \label{App:ResTimes}}

For completeness, we present the formula for the coherence function $P_x$; this function is used to derive the resonance times. The expressions we present below can also be found in Ref.~\cite{TaminiauPRL2012}. In DD protocols, the electron is initialized in the $|+\rangle$ state; assuming a single nuclear spin, the initial density matrix is given by:

\begin{equation}
    \rho_0=|+\rangle|\psi_n\rangle\langle +| \langle \psi_n|,
\end{equation}
where the tensor product is implied between kets and bras. The probability to find the electron in the $|+\rangle$ state after some time $t$ is $P_x=\langle +|\rho(t)|+\rangle$, where $\rho(t)=U\rho_0 U^\dagger$ is the time-evolved density matrix of the system. Further, $U=\sigma_{00}\otimes U_0+\sigma_{11}\otimes U_1$, with $U_j\equiv R_{\textbf{n}_j}(\phi_j)$. Calculating first $U\rho_0 U^\dagger$ we find:
\begin{equation}
\begin{split}
    U\rho_0 U^\dagger &=\frac{1}{2}\Big(
    U_0|0\rangle |\psi_n\rangle\langle \psi_n |\langle 0| U_0^\dagger
    + U_0|0\rangle |\psi_n\rangle\langle \psi_n|\langle 1| U_1^\dagger
    \\&+U_1|1\rangle |\psi_n\rangle \langle \psi_n| \langle 0| U_0^\dagger + U_1 |1\rangle |\psi_n  \rangle \langle \psi_n|\langle 1| U_1^\dagger
    \Big).
    \end{split}
\end{equation}
Evaluating $\langle + | U\rho_0 U^\dagger |+\rangle$ we obtain:
\begin{equation} \label{Eq:B3}
    \langle + |U \rho_0U^\dagger| +\rangle=\frac{1}{4}\sum_{i,j}U_i |\psi_n\rangle\langle \psi_n |U_j^\dagger.
\end{equation}
The probability to find the electron in the $|+\rangle$ state (irrespective of the nuclear spin state) is the trace of Eq.~(\ref{Eq:B3}) with respect to the nuclear spin state:

\begin{equation}
\begin{split}
    P_x&= \frac{1}{4}\sum_{i,j}\langle \psi_n |U_i U_j^\dagger |\psi_n\rangle
    =\frac{1}{2}(1+\Re \langle \psi_n |U_0^\dagger U_1 |\psi_n\rangle)\\&=
    \frac{1}{2}(1+M).
    \end{split}
\end{equation}
Since $U_0^\dagger U_1$ is unitary it can be written as $\begin{pmatrix}a & -b^* \\ b  & a^*\end{pmatrix}$. Hence for $\psi_n =[c_1~ c_2]^T$ we get:

\begin{equation}
\begin{split}
    &\Re(a|c_1|^2-b^*c_1^*c_2+bc_1c_2^*+a^*|c_2|^2)=\\&\Re(a)(|c_1|^2+|c_2|^2)+\Re(2i \Im(bc_1 c_2^*))=\Re(a),
    \end{split}
\end{equation}
and therefore, $M=\frac{1}{2}\Re\text{Tr}(U_0^\dagger U_1)=\frac{1}{2}\Re\text{Tr}(U_0 U_1^\dagger)$. Finally, by setting $U_j=R_{\textbf{n}_j}(\phi_j)$, $M$ becomes:
\begin{equation}\label{Eq:B6}
\begin{split}
    M&=\frac{1}{2}\Re\text{Tr}\Big[\prod_j\Big(\cos\frac{\phi_j}{2}+i(-1)^j\sin\frac{\phi_j}{2}\boldsymbol{\sigma}\cdot \textbf{n}_j\Big)\Big]
    \\&=\cos\frac{\phi_0}{2}\cos\frac{\phi_1}{2}+\textbf{n}_0\cdot \textbf{n}_1\sin\frac{\phi_0}{2}\sin\frac{\phi_1}{2}.
    \end{split}
\end{equation}
In the case of $\phi_0=\phi_1\equiv \phi$, $M$ can be re-written as:
\begin{equation}\label{Eq:B7}
    M=1-\sin^2\frac{\phi}{2}(1-\textbf{n}_0\cdot \textbf{n}_1).
\end{equation}
Using the explicit expression for $U_j$, one can derive the resonance condition by setting $\textbf{n}_0\cdot\textbf{n}_1=-1$ in Eq.~(\ref{Eq:B7}) for sequences that produce the same nuclear spin rotation angle. For sequences that produce different nuclear spin rotation angles (e.g. UDD$_4$) one would have to use Eq.~(\ref{Eq:B6}). 

\section{Nuclear spin rotation angles\label{App:RotationAngles}}

Here we provide the expressions for the nuclear spin rotation angles corresponding to 2-$\pi$, 3-$\pi$, and 4-$\pi$ sequences (meaning two, three, or four $\pi$ pulses in a single sequence unit). The analytical formulas are summarized below:

\begin{itemize}
    \item 2-$\pi$ sequence:
    \begin{equation}
        \phi_0=2\cos^{-1}[g(\omega_0,\omega_1)]
    \end{equation}
        \item 4-$\pi$ sequence:
        \begin{equation}
        \begin{split}
          \phi_0&=2\cos^{-1}\Big[g(\omega_0,\omega_1)\\&-2\sin^2\tilde{\theta}\sin\frac{q_2\omega_1}{2}\sin\frac{q_3\omega_0}{2}\sin\frac{q_4\omega_1}{2}\sin\frac{(q_1+q_5)\omega_0}{2}\Big]
          \end{split}
        \end{equation}
        \item 3$-\pi$ sequence (6-$\pi$ time-symmetric):
        \begin{equation}\label{Eq:UDD_3angle}
        \begin{split}
            \phi_0&=2\cos^{-1}\Big[g(\omega_0,\omega_1) \\&
            +4\cos\tilde{\theta}\sin^2\tilde{\theta}\sin(q_1\omega_0)\sin(q_1\omega_1)\sin^2\frac{q_2\omega_0}{2}\sin^2\frac{q_2\omega_1}{2} \\&
            -2\sin^2\tilde{\theta}\cos(q_1\omega_1)\sin(q_1\omega_0)\sin(q_2\omega_0)\sin^2\frac{q_2\omega_1}{2} \\&
            -2\sin^2\tilde{\theta}\sin(q_1\omega_1)\sin(q_2\omega_1)\sin\frac{q_2\omega_0}{2}\sin(q_1\omega_0+\frac{q_2\omega_0}{2})\Big]
            \end{split}
        \end{equation}
\end{itemize}
where we define $g(\omega_0,\omega_1)$ as:

\begin{equation}\label{Eq:gfunc}
\begin{split}
    g(\omega_0,\omega_1)&=\cos\frac{\sum_{j,\text{odd}} q_j\omega_0}{2}\cos\frac{\sum_{j,\text{even}}q_j\omega_1}{2}\\&-\cos\tilde{\theta}\sin\frac{\sum_{j,\text{odd}} q_j\omega_0}{2}\sin\frac{\sum_{j,\text{even}}q_j\omega_1}{2},
    \end{split}
\end{equation}
and $\omega_j =t \sqrt{(\omega_L+s_jA)^2+(s_jB)^2}$; note this is different from the definition in the main text where we defined $\omega_j$ as $\omega_j = \sqrt{(\omega_L+s_jA)^2+(s_jB)^2}$. The $\phi_1$ angles are found from $\phi_0$ with the replacements $\omega_1\mapsto \omega_0,~ \omega_0\mapsto\omega_1,~\tilde{\theta}\mapsto-\tilde{\theta}$. Here we defined $\tilde{\theta}=\theta_0-\theta_1$, where $\cos\theta_j=(\omega_L+s_jA)/\sqrt{(\omega_L+s_jA)^2+(s_jB)^2}$.

For example, for the CPMG sequence we have $q_1 t-q_2 t-q_3 t$ with $q_3=q_1=q_2/2$. This means that the odd summation in $g$ of Eq.~(\ref{Eq:gfunc}) is $q_1+q_3$ and the even is $q_2$. As we mentioned in the main text, for the CPMG sequence the rotation angles $\phi_0$ and $\phi_1$ are equal, but this is not the case for a 2-$\pi$ sequence with arbitrary $q_j$ that do not satisfy $q_1=q_3=q_2/2$. 

For the UDD$_n$ sequences the spacings are given by:
\begin{equation}\label{Eq:UDDspacings}
    q_s=\sin^2(\frac{\pi s}{2n+2})-\sin^2(\frac{\pi (s-1)}{2n+2}),
\end{equation}
where $s$ goes from 1 to $n+1$, since there are $n+1$ free evolution periods. The UDD$_4$ [$q_1 t-q_2t-q_3t-q_4t-q_5t$ with spacings given by Eq.~(\ref{Eq:UDDspacings})] sequence produces rotation angles $\phi_0$ and $\phi_1$ that are not equal. The UDD sequences (as the CPMG) are symmetric, i.e. in the UDD$_4$ case it holds that $q_5=q_1$, and $q_2=q_4$.

Regarding the UDD$_3$ sequence (or any odd-$\pi$ sequence), it needs to be repeated twice to form the basic unit. Specifically the initial block with spacings $q_1t-q_2t-q_3t-q_4t$ becomes a new unit with spacings $q_1t/2-q_2t/2-q_3t/2-(q_4+q_1)t/2-q_2t/2-q_3t/2-q_4t/2$, where we divide by a factor of two to make sure that the sum of all $q_j$ is equal to one and hence, the time of one unit is $t$. Again, for UDD$_3$ it holds that it is symmetric with $q_4=q_1$ and $q_3=q_2$.  Conversely to the UDD$_4$ sequence, UDD$_3$ produces rotation angles that are equal (i.e., $\phi_0=\phi_1$).

\section{UDD\texorpdfstring{$_4$}{4} jumps in the dot product\label{App:UDD4jumps}}

\begin{figure}[!htbp]
    \centering
    \includegraphics[scale=0.47]{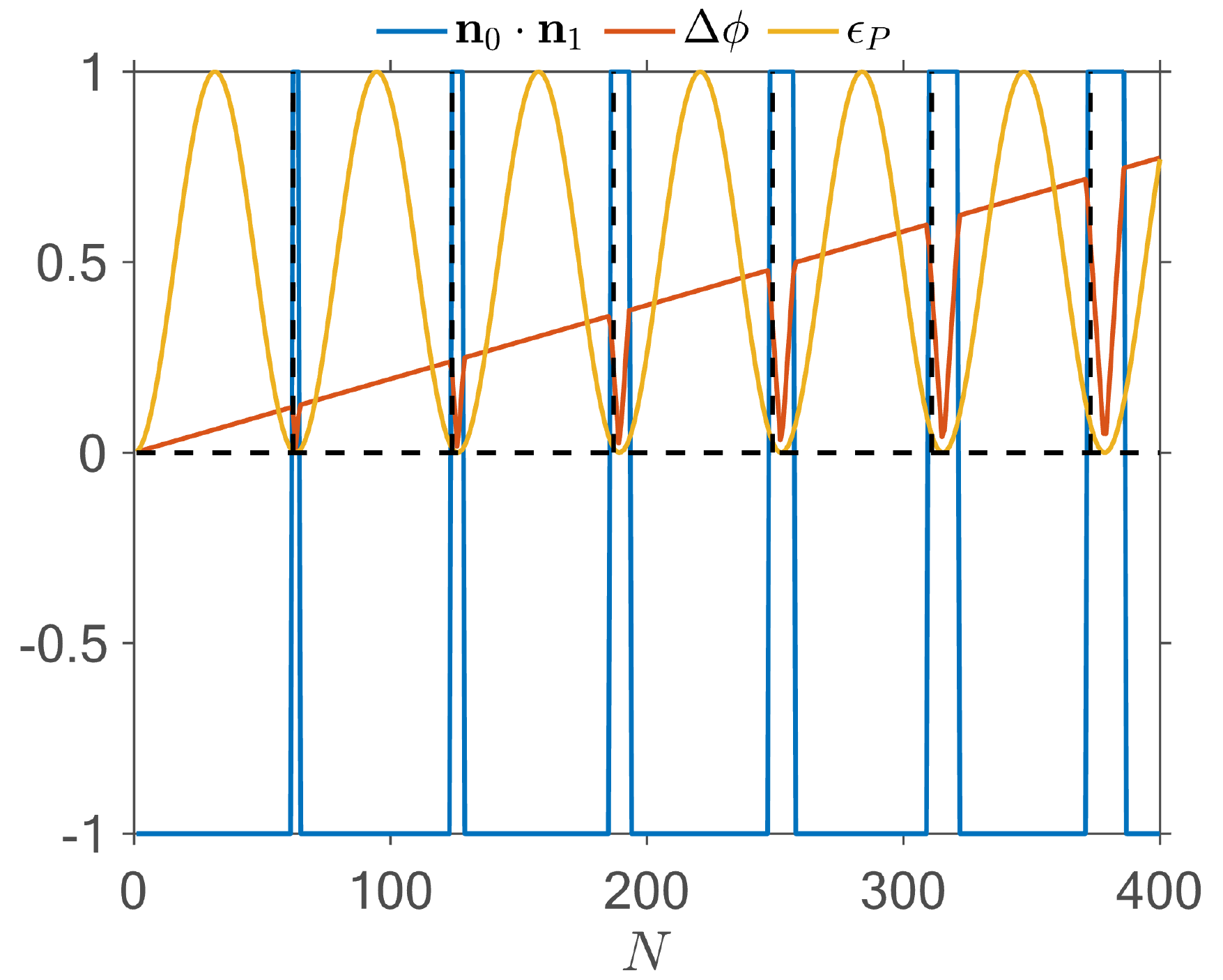}
    \caption{Jumps in the dot product of nuclear spin rotation axes for UDD$_4$ as a function of $N$ occur in the ranges when $\Delta\phi \rightarrow 0$; $\Delta \phi$ is defined as $|\phi_0-\phi_1|$. In these ranges, the entangling power  $\epsilon_P$ is zero. For this simulation, we set $(\omega_L,A,B)=2\pi\cdot(314,60,30)$~kHz, unit time of $t=3.1861~\mu$s and we considered a spin $S=1/2$ electron system. }
    \label{fig:UDD4Jumps}
\end{figure}

As we mentioned in Sec.~\ref{Sec:PropertiesSeqs}, the dot product of nuclear spin rotation axes in the case of UDD$_4$ shows a non-trivial behavior and depends on the number of iterations. We found that these jumps happen near values of $N$ for which $\Delta\phi=|\phi_0-\phi_1|=0$. Using the expression for $G_1$ and substituting $\Delta\phi=0$ and $\textbf{n}_0\cdot \textbf{n}_1=1$, we find that the jumps occur around $N=\text{round}[2\kappa \pi/(\phi_0+\phi_1)]$, where $\phi_j$ are the rotation angles in one iteration. We show this behavior in Fig.~\ref{fig:UDD4Jumps}, which captures all the jumps; in these ranges, the nuclear spin evolves trivially i.e. independent of the electron's state.

\section{Minimization of one-tangle for \texorpdfstring{$S=1$}{S=1} defect electron spin \label{App:Minimization_Tangle}}

\begin{figure}[!htbp]
    \centering
    \includegraphics[scale=0.5]{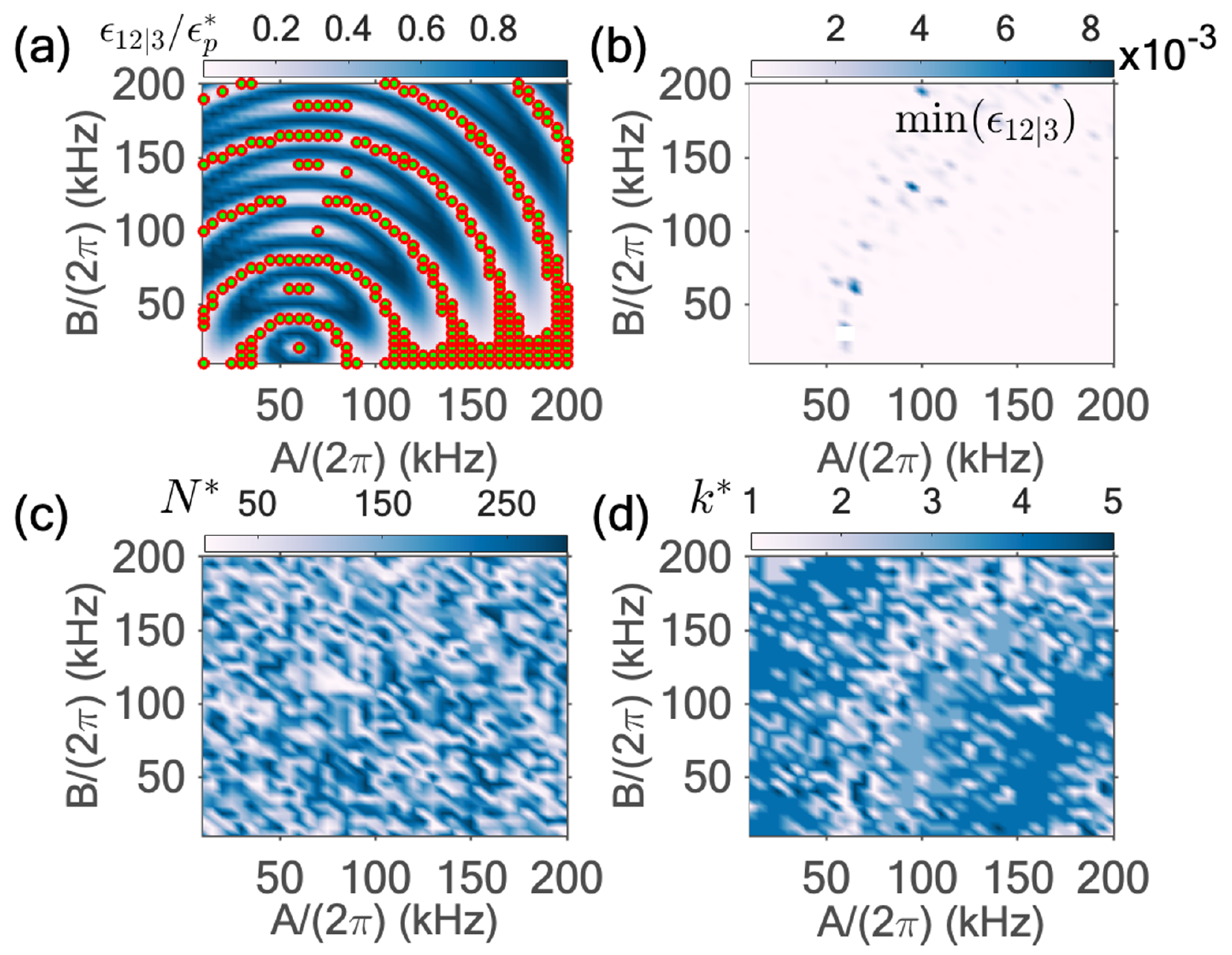}
    \caption{Controlling a target spin with  parameters $(A,B,\omega_L)=2\pi\cdot(60,30,314)$~kHz in the presence of an unwanted spin with varying HF parameters $\in 2\pi\cdot[10,200]$~kHz. (a) One-tangle of the unwanted spin, scaled by the maximum bound of $2/9$. The time of the unit is the first resonance time of the target spin and the number of iterations is $N=20$, which maximize its one-tangle. (b) Minimization of the one-tangle of the unwanted spin using the first five ($k=1,\dots,5$) resonances of the target spin, and up to 300 pulses on the electron. Optimal number of repetitions (c) and optimal resonance (d) to minimize the unwanted spin's one-tangle, while keeping the one-tangle of the target spin maximal. In all plots, we considered the CPMG sequence, and an electron spin $S=1$ ($s_0=0$ and $s_1=-1$). }
    \label{fig:App_One_Tangles}
\end{figure}

In this section, we consider a target nuclear spin with hyperfine (HF) parameters $(A,B)=2\pi\cdot (60,30)$~kHz and Larmor frequency $\omega_L=2\pi\cdot 314$~kHz as in Sec.~\ref{SubSec:One-tangles}. However, in this case, we assume a defect electron system $S=1$, and we define the qubit using the spin projections $s_0=0$ and $s_1=-1$. As in Sec.~\ref{SubSec:One-tangles}, we choose the CPMG sequence. First, we consider the $k=1$ resonance of the target spin i.e, we set the basic unit time to be $t=3.5102~\mu$s and set the number of iterations to be $N=20$, which gives rise to maximum one-tangle of the target spin. In Fig.~\ref{fig:App_One_Tangles}(a), we plot the one-tangle of an unwanted nuclear spin whose HF parameters could lie in the range $\in2\pi\cdot[10,200]$~kHz. We further indicate with circles the HF parameters of an unwanted spin, which satisfies approximately the condition for trivial evolution, presented in Sec.~\ref{SubSec:TrivialEvol}. To display these points, we set a bound for the unwanted one-tangle to be $\epsilon_{12|3}<0.02$ and a tolerance for satisfying the trivial evolution of $1.3\times 10^{-2}$. We see that indeed the minimal one-tangles correspond to nuclei that approximately evolve trivially.

In Fig.~\ref{fig:App_One_Tangles}(b), we minimize the unwanted one-tangle by searching over the first five resonances of the target spin and iterations of the basic unit that preserve maximum entanglement between the target register and the electron. The optimal repetitions of the basic CPMG unit, as well as the optimal resonances, are shown in Fig.~\ref{fig:App_One_Tangles}(c) and Fig.~\ref{fig:App_One_Tangles}(d) respectively.

\section{Minima of \texorpdfstring{$G_1$}{G1} for \texorpdfstring{$\textbf{n}_0 \cdot \textbf{n}_1\leq 0$}{n0n1 0} \label{App:Nestimate}}

In this section we provide the number of iterations that maximize the one-tangle of a nuclear spin, as long as $\textbf{n}_0\cdot \textbf{n}_1\leq 0$. Let us first consider the CPMG or UDD$_3$ sequences. For these sequences it holds that $\phi_0=\phi_1$ and $G_1$ of Eq.~(\ref{Eq:G1}) simplifies into:
\begin{equation}
    G_1(N) =[\cos^2(\phi_0(N)/2)+\textbf{n}_0\cdot \textbf{n}_1 \sin^2(\phi_0(N)/2) ]^2.
\end{equation}
Requiring that $G_1(N)=0$ we find:
\begin{equation}
\begin{split}
   &\cos^2\frac{N\phi_0}{2}+\textbf{n}_0\cdot \textbf{n}_1\sin^2\frac{N\phi_0}{2}=0 \Rightarrow
\\&\frac{-1}{\textbf{n}_0\cdot \textbf{n}_1}=\tan^2\frac{N\phi_0}{2}  \Rightarrow
\\&
N = \text{round}\Big[\frac{1}{\phi_0}\left(2\kappa \pi -2\tan^{-1}\sqrt{\frac{-1}{\textbf{n}_0\cdot \textbf{n}_1}}\right)\Big],\\&
N = \text{round}\Big[\frac{1}{\phi_0}\left((2\kappa-1) \pi +2\tan^{-1}\sqrt{\frac{-1}{\textbf{n}_0\cdot \textbf{n}_1}}\right)\Big],
    \end{split}
\end{equation}
where $\phi_0$ is the rotation angle in one unit. The two expressions hold as long as $\textbf{n}_0\cdot \textbf{n}_1< 0$ and $\phi_0(N)\neq (2\kappa+1)\pi$. For $\textbf{n}_0\cdot \textbf{n}_1\approx 0$, $G_1=0$ when $\phi_0(N)=(2\kappa+1)\pi$, and hence $N=\text{round}[\frac{(2\kappa+1)\pi}{\phi_0}]$. For $\textbf{n}_0\cdot \textbf{n}_1>0$, $G_1$ cannot go to zero. Regarding the UDD$_4$ sequence for which it holds that $\phi_0\neq\phi_1$, we cannot estimate analytically the repetitions $N$; some values are captured by the above expressions with the replacement $\phi_0\mapsto \phi_{0}+\phi_1$, but due to the complicated oscillations of $G_1$ these modified expressions for $N$ do not hold in all cases.

\section{Comparison of CPMG, UDD\texorpdfstring{$_3$}{3} and UDD\texorpdfstring{$_4$}{4} rotation angles \label{App:CPMGvsUDDRotAngle}}

To understand geometrically the rotation angle induced by each sequence we use the Rodrigues formula~\cite{Gray1980} for the composition of rotations. Two rotations of the form $R_\textbf{l}(\alpha)R_\textbf{m}(\beta)$ give rise to the total rotation $R_\textbf{n}(\gamma)$ for which the rotation angle is given by:
\begin{equation}
    \cos\frac{\gamma}{2}=\cos\frac{\alpha}{2}\cos\frac{\beta}{2}-\sin\frac{\alpha}{2}\sin\frac{\beta}{2}(\textbf{l}\cdot \textbf{m}),
\end{equation}
while the rotation axis is given by:
\begin{equation}
    \sin\frac{\gamma}{2}\textbf{n}=\sin\frac{\alpha}{2}\cos\frac{\beta}{2}\textbf{l}+\cos\frac{\alpha}{2}\sin\frac{\beta}{2}\textbf{m}+\sin\frac{\alpha}{2}\sin\frac{\beta}{2}(\textbf{l}\times \textbf{m}).
\end{equation}
We apply this composition law repeatedly to find the induced nuclear spin rotation after each free-evolution period of the sequence has passed. For the CPMG sequence, there are two compositions and three free-evolution periods, UDD$_4$ has four compositions and five free-evolution periods, and  UDD$_3$ has six compositions and seven free-evolution periods. UDD$_3$ has more free-evolution periods than UDD$_4$ because we repeat the basic unit twice to yield a new sequence unit. In this way, we ensure that the electron returns to its initial state since the new basic block of the sequence has an even number of $\pi$-pulses.

Without loss of generality, we consider an electron spin $S=1$, with $s_0=0$ and $s_1=-1$, and assume that the electron starts from the $|0\rangle$ state (similar analysis holds when the electron starts in $|1\rangle$) and is flipped repeatedly according to the number of $\pi$-pulses in the CPMG or UDD units. 

\begin{figure}[!htbp]
    \centering
    \includegraphics[scale=0.31]{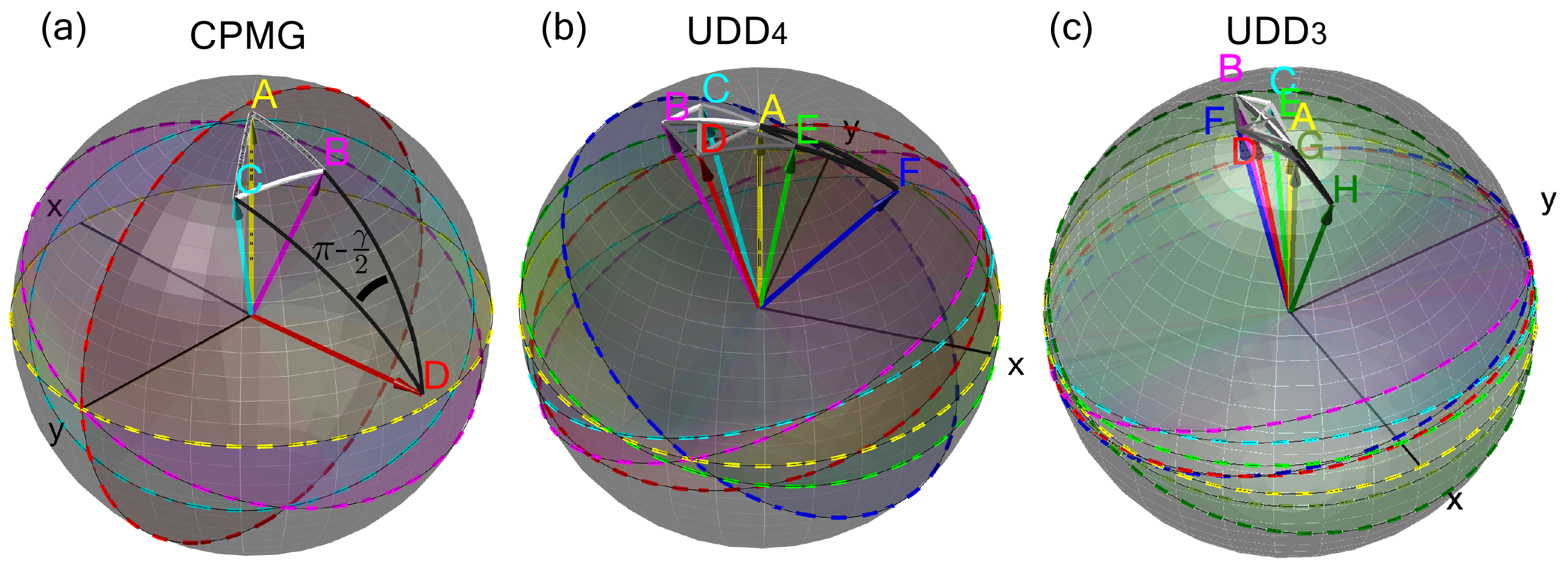}
    \caption{Comparison of CPMG (a) with UDD$_4$ (b) and UDD$_3$ (c) rotation angles. We compose the rotations of the free-evolution periods for one unit of the sequence using the Rodrigues formula. The last composition gives rise to the black spherical triangles.  In (a) we show  $\sphericalangle$ CDB which gives the total rotation angle $\gamma$. We find that geometrically the rotation angle of CPMG is larger.}
    \label{fig:RotAngComp}
\end{figure}

In Fig.~\ref{fig:RotAngComp} we show the rotation axes of a single nuclear spin $[(\omega_L,A,B)=2\pi\cdot(314,120,90)~\text{kHz}~\text{and}~t=3.7889~\mu\text{s}]$  after each composition. In each free-evolution period the nuclear spin rotates alternately about axis `A' and `B'. For example, in Fig.~\ref{fig:RotAngComp}(a) the nuclear spin first rotates about `A', then about `B' and again about `A'. The composition of `A' and `B' gives rise to the new axis `C', and the composition of `C' with `A' leads to the final axis `D'. The angle enclosed by the arcs CD and BD, gives the total rotation angle $\gamma$ in one CPMG unit. Similar analysis holds for Fig.~\ref{fig:RotAngComp}(b) and Fig.~\ref{fig:RotAngComp}(c)  where rotations follow the composition $[A][B][A][B][A]$ and $[A][B][A][B][A][B][A]$ respectively. We notice that it holds $\pi-\frac{\gamma}{2}^{\text{CPMG}}>\pi-\frac{\gamma}{2}^{\text{UDD}_4}>\pi-\frac{\gamma}{2}^{\text{UDD}_3}$ and so one would conclude that $\gamma^{\text{CPMG}}<\gamma^{\text{UDD}_4}<\gamma^{\text{UDD}_3}$. However, these rotation angles are close to $2\pi$ and hence we need to consider as actual rotation $\tilde{\gamma}=2\pi-\gamma$. Therefore, we find that it holds:
\begin{equation}
    \tilde{\gamma}^{\text{CPMG}}>\tilde{\gamma}^{\text{UDD}_4}>\tilde{\gamma}^{\text{UDD}_3}.
\end{equation}
Thus, UDD$_3$ produces the finest rotation angle of all three sequences, which can offer greater precision, but CPMG is the fastest of all.

\section{Derivation of one-tangles for the electron and a nuclear spin \label{App:DerOneTangle}}

In this section we prove that the nuclear spin one-tangle essentially reduces to the two-qubit entangling power. We start from the general expression for an arbitrary number $n$ of qubits (with $n-1$ nuclear spins):
\begin{widetext}
\begin{equation}
\begin{split}
    \epsilon_{p|q}^{\text{nuclear}}&=
    1-\prod_{i}\frac{d_i}{d_i+1}\Bigg\{1+\frac{1}{2^n}+\frac{n}{2^{n-1}}
    +
    \sum_{k=1}^{n-1}\frac{2^{k-1}}{2^{n-1}}\begin{pmatrix}n-2 \\k-1\end{pmatrix}(1+G_1)
    +
    \sum_{k=1,n>3}^{n-3}\frac{2^k}{2^{n-1}}\left[\begin{pmatrix}n-2 \\k\end{pmatrix}+\begin{pmatrix}n-1 \\k+1\end{pmatrix} \right]  \Bigg\}
    \\&=
    1- \frac{2^n}{3^n}\Bigg\{1+\frac{1}{2^n}+\frac{n}{2^{n-1}}
    +
    \frac{1}{2^{n-1}}\frac{1}{3^{-n+2}}(1+G_1)
    +
    \Theta(n-3)\frac{1}{2^n 3^2}[5\cdot 3^n -3^2(1+2n+2^n)]\Bigg\}
    \\&=1-\frac{2}{9}(1+G_1)-\frac{2^n}{3^n}\Bigg\{\frac{1+2n+2^n}{2^n} 
    +
    \Theta(n-3)\frac{1}{2^n}[5\cdot 3^{n-2}-(1+2n+2^n)]\Bigg\}
    \\&=1-\frac{2}{9}(1+G_1)-\frac{1}{3^n}\Bigg\{1+2n+2^n
    +
    \Theta(n-3)[5\cdot 3^{n-2}-(1+2n+2^n)]\Bigg\},
    \end{split}
\end{equation}
\end{widetext}
 where $\Theta(n-3)$ is the step function. Clearly for $n=3$ the last term vanishes and we recover $2/9(1-G_1)$. 
 For $n>3$ we have:
 \begin{equation}
     \epsilon_{p|q}^{\text{nuclear}}=\frac{7}{9}-\frac{2}{9}G_1-\frac{5 \cdot 3^{n-2}}{3^n}=\frac{7-5}{9}-\frac{2}{9}G_1=\frac{2}{9}(1-G_1),
 \end{equation}
which concludes our proof.

For the electron, we start from the expression:

\begin{widetext}
\begin{equation}
    \begin{split}
        \epsilon_{p|q}^{\text{electron}}&=1-\frac{2^n}{3^n}\Bigg\{
    \frac{1}{2}+\frac{1}{2^n}+\frac{n}{2^{n-1}}+
    \sum_{k=1}^{n-3}\frac{2^k}{2^{n-1}}\begin{pmatrix}n-1\\k+1\end{pmatrix}
    +
    \Big[\frac{1}{2^{n-1}}\sum_{j=1}^{n-1}(1+G_1^{(j)})
    +
    \frac{1}{2^{n-2}}\sum_{\substack{j_1,j_2\\j_2>j_1}}(1+G_1^{(j_1)}G_1^{(j_2)})
    \\&+\dots
    +
    \frac{1}{2^{n-(n-2)}}\sum_{j_1,\dots,j_{n-2}}(1+\prod_{i=1}^{n-2}G_1^{(j_i)})
    +
    \frac{1}{2^{n-(n-1)}}(1+\prod_{i=1}^{n-1}G_1^{(j_i)})\Big]\Bigg\},
    \end{split}
\end{equation}
which can be simplified into:

\begin{equation}\label{Eq:H4}
        \begin{split}
        \epsilon_{p|q}^{\text{electron}}&=1-\frac{1}{3^n}\Bigg\{
    2^{n-1}+1+2n
    +
    \Theta(n-3)(1-2n-2^{n-1}+3^{n-1})
    +\Big(\sum_{k=1}^{n-1}2^k \sum_{\substack{j_1,\dots,j_k\\j_{m+1}>j_m}}1\Big) + F(G_1)\Bigg\}
    \\&=1-\frac{1}{3^n}\Bigg\{
    2^{n-1}+1+2n
    +
    \Theta(n-3)(1-2n-2^{n-1}+3^{n-1})
    +\sum_{k=1}^{n-1}2^k\begin{pmatrix}n-1 \\ k\end{pmatrix} + F(G_1)\Bigg\}
    \\&=1-\frac{1}{3^n}\Bigg\{
    2^{n-1}+1+2n
    +
    \Theta(n-3)(1-2n-2^{n-1}+3^{n-1})
    +\frac{1}{3}(-3+3^n) + F(G_1)\Bigg\},
    \end{split}
\end{equation}
where we have defined $F(G_1)$ as:
\begin{equation}
    F(G_1) = \sum_{k=1}^{n-1}2^{k} \sum_{j_1,\dots j_k}\prod_{i=1}^k G_1^{(j_i)}.
\end{equation}
We note that $F(G_1)$ can be re-written as:
\begin{equation}
\begin{split}
    F(G_1) &= 2\Bigg\{\Big(G_1^{(1)}+\dots + G_1^{(n-1)}\Big) 
    +
    2\Big(G_1^{(1)}G_1^{(2)}+G_1^{(1)}G_1^{(3)}+\dots+G_1^{(1)}G_1^{(n-1)}
    +\dots + G_1^{(n-2)}G_1^{(n-1)}\Big)
    \\&+
    2^2\Big(G_1^{(1)}G_1^{(2)}G_1^{(3)}+G_1^{(1)}G_1^{(2)}G_1^{(4)}+\dots 
    + G_1^{(1)}G_1^{(2)}G_1^{(n-1)}+\dots \Big)+\dots
    \Bigg\}
    \end{split},
\end{equation}
which actually reduces to:
\begin{equation}
    -1+(1+2G_1^{(1)})\dots(1+2G_1^{(n-1)})=-1+\prod_{i=1}^{n-1}(1+2G_1^{(i)}).
\end{equation}
Finally, the electron's one-tangle reads:
\begin{equation}
\begin{split}
  \epsilon_{p|q}^{\text{electron}}&=  1-\frac{1}{3^n}\Bigg\{
    -1+2n+2^{n-1}+3^{n-1}
    +
    \Theta(n-3)(1-2n-2^{n-1}+3^{n-1})
    +\prod_{i=1}^{n-1}(1+2G_1^{(i)})\Bigg\}.
    \end{split}
\end{equation}
\end{widetext}
For $n>3$ it becomes:
\begin{equation}
\begin{split}
    \epsilon_{p|q}^{\text{electron}}&=1-\frac{1}{3^n}\Bigg\{2 \cdot 3^{n-1}+\prod_{i=1}^{n-1}(1+2G_1^{(i)})\Bigg\}\\&=\frac{1}{3}-\frac{1}{3^n}\prod_{i=1}^{n-1}(1+2G_1^{(i)}),
    \end{split}
\end{equation}
whereas for $n=3$ the term multiplying the $\Theta(n-3)$ function vanishes and we obtain:
\begin{equation}
\begin{split}
    \epsilon_{p|q}^{\text{electron}}&=1-\frac{18}{9\cdot 3}-\frac{1}{3^n}\prod_{i=1}^{n-1}(1+2G_1^{(i)})
    \\&=
    \frac{1}{3}-\frac{1}{3^n}\prod_{i=1}^{n-1}(1+2G_1^{(i)})
    \end{split}
\end{equation}

\section{Parameters of the multipartite gates  \label{App:HFparams} }

\subsection{Random generation of nuclei \label{App:HFparamsRandom}}
In Tables~\ref{tab:2},~\ref{tab:3},~\ref{tab:4} we list the HF parameters of the randomly generate nuclei (labeled by $\#$) we used in Sec.~\ref{SubSec:CR1}. We present their one-tangles, rotation angles, dot product of rotation axes, and positions compared to the vacancy site. To estimate the distances and polar angles of nuclei, we use the approach found in the supplemental of Ref.~\cite{BourassaNatMater2020,CarterPRB2015} and in Ref.~\cite{ZopesNatCommun2018}. Since we are studying weakly coupled spins far away from the electron, the interaction is well approximated by the dipole-dipole interaction, and the Fermi contact interaction is negligible. The hyperfine vector can be broken into parallel and perpendicular components which are related to the polar angle $\theta$ and the distance $R$ by:

\begin{equation}
    B=3A_0 \cos\theta\sin\theta,
\end{equation}
\begin{equation}
    A=A_0(3\cos^2\theta-1),
\end{equation}
with $A_0=\mu_0\gamma_n\gamma_e\hbar/(4\pi R^3)$. We solve these equations for $R$ and $\theta$, assuming $^{13}$C atoms and present the values in the tables.

\setlength{\arrayrulewidth}{0.001mm}
\renewcommand{\arraystretch}{1.2}

\begin{table}[!htbp]
\centering
\scalebox{0.92}{
\begin{tabular}{c|c|c|c|c|c|c|c}
\hline 
 \multicolumn{1}{c}{$k=1$} \\
\hline
    $\#$ & $\frac{A}{2\pi}$  & $\frac{B}{2\pi}$  & $\epsilon_{p|q}/\epsilon_p^*$  & $\phi_0/(\pi/2)$& $\textbf{n}_0\cdot \textbf{n}_1$  & $R^{^{13}C}$ & $\theta^{^{13}C}$  \\
     & (kHz) & (kHz) & & & & $(\angstrom)$ & (deg)\\

1 &  195.78 & 49.619 & 0.9986 & 0.97615 & -1 & 5.798 & 9.4595 \\
2 & 27.783 & 136.51 & 0.99906 & 1.0206 & -0.99667 & 6.0056 & 48.7804 \\
3 & 124.53 & 128.31 & 0.95585 & 0.86597 & -0.99715 & 5.8549 & 29.8479 \\
4 & 100.03 & 22.072 & 0.99978 & 0.99104 & -0.99812 & 7.2761 & 8.2809 \\
5 & 26.926 & 181.33 & 0.96884 & 1.1155 & -0.99328 & 5.4466 & 50.4027\\
6 & 65.726 & 128.15 & 0.99849 & 1.0257 & -0.99718 & 6.1206 & 40.0504 \\
7 & 63.767 & 74.919 & 0.99849 & 1.0257 & -0.99718 & 7.1056 & 32.1510\\
8 & 106.88 & 164.74 & 0.96352 & 1.1244 & -0.99462 & 5.5762 & 36.6319\\
9 & 193.4 & 122.07 & 0.97393 & 0.89743 & -0.99748 & 5.4859 & 21.2539\\
10 & 144.18 & 93.415 & 0.99612 & 0.96065 & -0.99893 & 6.0305 & 21.6910 \\                   \hline   
\multicolumn{1}{c}{$k=2$}\\
\hline
 
     $\#$ & $\frac{A}{2\pi}$  & $\frac{B}{2\pi}$  & $\epsilon_{p|q}/\epsilon_p^*$ & $\phi_0/(\pi/2)$ & $\textbf{n}_0\cdot \textbf{n}_1$ & $R^{^{13}C}$ & $\theta^{^{13}C}$   \\  
    & (kHz) & (kHz) & & & & $\angstrom$ &(deg) \\

1 & 188.85 & 131.4 & 0.99822 & 1.0268 & -1 & 5.4598 & 22.9024 \\
2 & 56.381 & 179.7 & 0.95301 & 1.4246 & -0.50348 & 5.4953 & 45.5628 \\
3 & 88.294 & 109.64 & 0.98979 & 0.94983 & -0.95158 & 6.2905 & 33.1012 \\
4 & 56.527 & 78.511 & 0.98742 & 1.1009 & -0.5337 & 7.0931 & 34.9707 \\
5 & 134.8 & 150.66 & 0.99807 & 1.1201 & -0.75814 & 5.6010 & 31.2819\\
6 & 82.906 & 187.71 & 0.99703 & 1.2709 & -0.33844 & 5.4062 & 41.9618 \\
7 & 121.1 & 73.468 & 0.99934 & 1.6293 & -0.11789 & 6.4427 & 20.6001 \\
8 & 10.288 & 157.82 & 0.99436 & 1.024 & -0.78257 & 5.6665 & 52.8493                          
 \\ 
\end{tabular}}
    \caption{Parameters for the $k=1$ CPMG resonances of Figs.~\ref{fig:MaximizationTangles}(a), (b) and the $k=2$ CPMG of Figs.~\ref{fig:MaximizationTangles}(c), (d).}
    \label{tab:2}
\end{table}

\setlength{\arrayrulewidth}{0.001mm}
\renewcommand{\arraystretch}{1.2}

\begin{table}[!htbp]
\centering
\scalebox{0.92}{
\begin{tabular}{c|c|c|c|c|c|c|c}
\hline 
 \multicolumn{1}{c}{$k=1$} \\
\hline
    $\#$ & $\frac{A}{2\pi}$  & $\frac{B}{2\pi}$  & $\epsilon_{p|q}/\epsilon_p^*$ & $\phi_0/(\pi/2)$&  $\textbf{n}_0\cdot \textbf{n}_1$ & $R^{^{13}C}$ & $\theta^{^{13}C}$   \\ 
    & (kHz)& (kHz) & & & & ($\angstrom$) & (deg)  \\

1 & 156.64 & 77.034 & 1 & 0.99888 & -1 & 6.0380 & 17.3233 \\
2 & 140.3 & 86.029 & 0.99978 & 1.0098 & -0.9988 &  6.1266 & 20.7752 \\
3 & 198.87 & 166.64 & 0.99937 & 1.0055 & -0.93307 & 5.2130 & 26.1614 \\
4 & 66.029 & 49.357 & 0.99992 & 1.0119 & -0.98071 & 7.6696 & 24.1452 \\
5 & 70.082 & 148.45 & 1 & 1.0151 & -0.9537 & 5.8393 & 41.1360\\
6 & 123.25 & 121.93 & 0.99991 & 1.013 & -0.97865 & 5.9252 & 29.1248 \\
7 & 26.135 & 112.96 & 0.9988 & 1.0269 & -0.98538 & 6.4041 & 47.9618 \\
8 & 41.644 & 103.96 & 0.99922 & 0.98495 & -0.99114 & 6.5907  & 43.0975 \\
9 & 159.78 & 104.28 & 0.99925 & 0.98541 & -0.99101 & 5.8221 & 21.8136  \\
10 & 61.719 & 76.878 & 0.99975 & 1.01 & -1 & 7.0826  & 33.1540 \\
11 & 45.081 & 191.59 & 0.99999 & 1.0355 & -0.90059 & 5.3707 & 47.8464
 \\   
    \hline   
\multicolumn{1}{c}{$k=3$}\\
\hline

    $\#$ & $\frac{A}{2\pi}$ & $\frac{B}{2\pi}$  & $\epsilon_{p|q}/\epsilon_p^*$ & $\phi_0/(\pi/2)$ & $\textbf{n}_0\cdot \textbf{n}_1$ &$R^{^{13}C}$ & $\theta^{^{13}C}$  \\  
    & (kHz) & (kHz) &  & & & ($\angstrom$) & (deg) \\

1 & 168.78 & 12.804 & 0.99937 & 1.016 & -1 & 6.1687 & 2.8916\\
2 & 82.989 & 158.3 & 0.99968 & 1.4125 & -0.22476 & 5.7008 & 39.7526\\
3 & 63.816 & 88.135 & 0.99638 & 1.4196 & -0.31499 & 6.8222 & 34.8782 \\
4 & 136.9 & 149.52 & 0.95987 & 1.5528 & -0.36149 & 5.6013 & 30.8775 \\
5 & 141.46 & 99.44 & 0.96969 & 1.0693 & -0.48989 & 6.0031 & 23.0786\\
6 & 142.11 & 76.191 & 0.99961 & 1.1399 & -0.60942 & 6.1886 & 18.6298 \\
7 & 186.1 & 56.749 & 0.99309 & 0.99304 & -0.85408 & 5.8622 & 11.2690\\
8 & 199.65 & 138.43 & 0.99634 & 1.0949 & -0.63609 & 5.3621 & 22.8424
 \\   

\end{tabular}}
    \caption{Parameters for the $k=1$ UDD$_3$ resonance of Figs.~\ref{fig:MaximizationTangles}(e), (f) and the $k=3$  UDD$_3$ resonance of Figs.~\ref{fig:MaximizationTangles}(g), (h).}
    \label{tab:3}
\end{table}

\setlength{\arrayrulewidth}{0.001mm}
\renewcommand{\arraystretch}{1.2}

\begin{table}[!htbp]
\centering
\scalebox{0.79}{
\begin{tabular}{c|c|c|c|c|c|c|c|c}
\hline 
 \multicolumn{1}{c}{$k=1$} \\
\hline
    $\#$ & $\frac{A}{2\pi}$ & $\frac{B}{2\pi}$ & $\epsilon_{p|q}/\epsilon_p^*$ & $\phi_0/(\pi/2)$ & $\phi_1/(\pi/2)$  & $\textbf{n}_0\cdot \textbf{n}_1$ &$R^{^{13}C}$ & $\theta^{^{13}C}$ \\  
    & (kHz) & (kHz) & & & & & ($\angstrom$) & (deg) \\

1 & 185.97 & 180.32 & 0.99726 & 0.0082 & 1.9251 & -1 & 5.1871 & 28.7667\\
2 & 66.715 & 101.62 & 0.99231 & 0.44477 & 1.5634 & -0.70796 & 6.5463 & 36.4471 \\
3 & 74.691 & 53.908 & 0.99997 & 1.6433 & 1.8615 & -0.037267 & 7.3993 & 23.5345 \\
4 & 142.18 & 92.353 & 0.99995 & 1.96 & 0.051616 & -0.60599 & 6.0567 & 21.7336\\
5 & 129.3 & 56.393 & 0.99998 & 1.0714 & 1.9898 & -0.013046 & 6.4964 & 15.6109\\
6 & 176.75 & 56.919 & 0.99306 & 0.84782 & 1.8841 & 0.019231 & 5.9511 & 11.8573 \\
7 & 53.599 & 136.86 & 0.95539 & 0.50412 & 1.8016 & -0.93027 & 6.0146 & 43.3476\\
8 & 22.803 & 92.34 & 0.99526 & 1.2626 & 0.92551 & -0.61159 & 6.8526 & 47.5050\\
9 & 36.541 & 194.7 & 0.99741 & 1.6852 & 0.38214 & -0.99322 & 5.3311 & 49.2472
 \\   
    \hline   
\multicolumn{1}{c}{$k=2$}\\
\hline
    $\#$ & $\frac{A}{2\pi}$  & $\frac{B}{2\pi}$  & $\epsilon_{p|q}/\epsilon_p^*$ & $\phi_0/(\pi/2)$ & $\phi_1/(\pi/2)$  & $\textbf{n}_0\cdot \textbf{n}_1$ &$R^{^{13}C}$ & $\theta^{^{13}C}$    \\ 
    & (kHz) & (kHz) & & & & & ($\angstrom$) & (deg) \\

1 & 57.301 & 157.25 & 0.99983 & 1.6338 & 0.34939 & -1 & 5.7448 & 44.115 \\
2 & 83.42 & 41.407 & 0.98732 & 0.33954 & 1.5977 & -0.74718 & 7.4432 & 17.4606\\
3 & 91.972 & 183.32 & 0.99433 & 1.4705 & 0.64575 & -0.96429 & 5.4352 & 40.3454 \\
4 & 167.87 & 70.649 & 0.99108 & 1.1752 & 1.0431 & -0.54509 & 5.9694 & 15.1266 \\
5 & 150.76 & 190.51 & 0.99943 & 1.2509 & 0.78326 & -0.89464 & 5.2403 & 33.3981\\
6 & 81.29 & 135.96 & 0.99609 & 0.74827 & 1.1903 & -0.96808 & 5.9687 & 37.8747 \\
7 & 165.25 & 99.13 & 0.99998 & 1.9905 & 0.092461 & -0.035242 & 5.8161 & 20.4161 \\
8 & 179.08 & 30.338 & 0.96742 & 0.09924 & 1.8562 & 0.88088  & 6.0179 & 6.40345
\\

\end{tabular}}
    \caption{Parameters for the $k=1$ UDD$_4$ resonance of Figs.~\ref{fig:MaximizationTangles}(i), (j) and the $k=2$ UDD$_4$ resonance of Figs.~\ref{fig:MaximizationTangles}(k), (l).}
    \label{tab:4}
\end{table}

\subsection{Gate error comparison for the three sequences\label{App:GateErrorComp}}

Here we provide a more thorough calculation of the gate error of multipartite gates under the CPMG, UDD$_3$, or UDD$_4$ evolution. Following a similar approach as in Sec.~\ref{SubSec:CR2}, we generate ensembles of $5\times 10^{5}$ unwanted nuclear spins with randomly distributed HF parameters and identify those with one-tangles in the range [0.1, 0.7]. We keep the same number of repetitions and gate time we considered for each sequence in Sec.~\ref{SubSec:CR2}. However, in this case, we repeat the random generation 8 times to produce 8 different ensembles of $5\times 10^{5}$ unwanted nuclei. For each of these 8 different ensembles we
repeat the same procedure as in Sec.~\ref{SubSec:CR2}; we gradually increase the size of the unwanted spin bath (which contains up to 6 spins) and calculate the gate error it induces on the target subspace.
\begin{figure}[!htbp]
    \centering
    \includegraphics[scale=0.7]{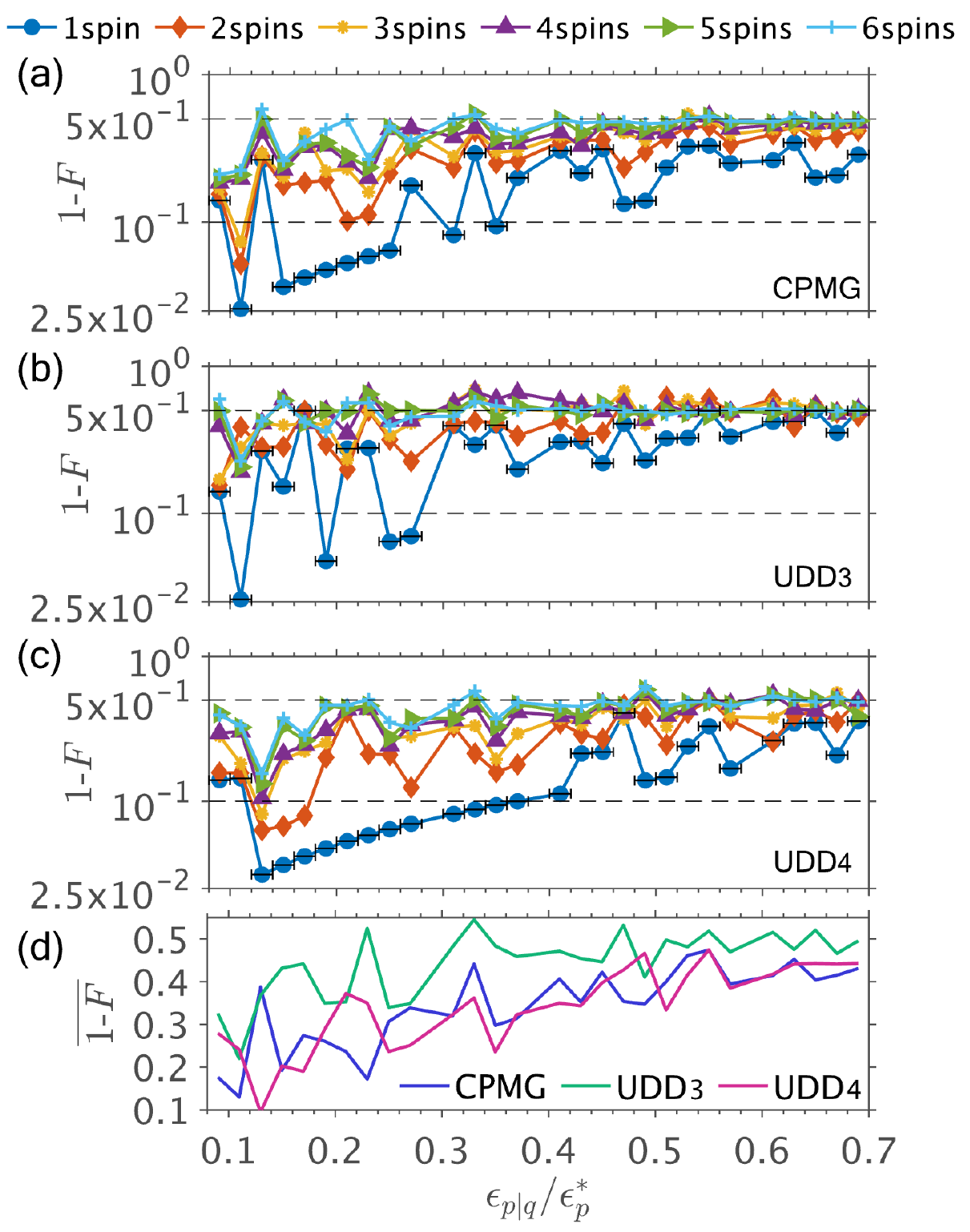}
    \caption{Gate error $1-F$ as a function of one-tangles of unwanted nuclei for (a) CPMG, (b) UDD$_3$ and (c) UDD$_4$. The labels in all graphs show up to how many spins were  ``traced-out'' from the total system. The unwanted spins have one-tangles in the range [0.1, 0.7]. The error-bars of the blue points show the intervals where we assign unwanted spins, and are the same for all differently colored lines. The gate error shown in (a), (b), (c) is the average over eight different ensembles of unwanted nuclei for each bath size. In (d) we take the gate error from plots (a), (b) and (c) respectively and further average over the six different unwanted spin baths for each one-tangle range [see text]. }
    \label{fig:GateErComp}
\end{figure}
At the end, we take the average of the error (over the 8 ensembles) for each case of unwanted spin bath size (from 1 up to 6 spins). Notice that the 8 ensembles are distinct for each sequence.

In Fig.~\ref{fig:GateErComp} we show the gate error averaged over the 8 different realizations of unwanted spin ensembles for CPMG (a), UDD$_3$ (b) and UDD$_4$ (c). We notice that CPMG performs in general on par with UDD$_4$, while UDD$_3$ fails to protect the target subspace as effectively as the other two sequences. Although in this scenario UDD$_3$ has the longest total gate time of all three sequences (as we mentioned in the main text), we see that choosing a longer sequence does not always ensure improved performance. To provide a comparison of the gate error we further evaluate the average gate error in each range of one-tangles. That is, we average the gate error for each fixed interval of one-tangles as $\overline{1-F}=1/6(1-F^{1\text{spin}}+1-F^{2\text{spins}}+\dots +1-F^{6\text{spins}})$. The results are shown in Fig.~\ref{fig:GateErComp}(d) where we see that UDD$_3$ underperforms the other two sequences.

\subsection{Parameters for 27 nuclear spins \label{App:27Spins}}

The HF parameters for the 27 nuclear spins~\cite{BradleyThesis2021} we considered in Sec.~\ref{SubSec:27Spins} are listed in Table~\ref{tab:5}. In addition, in Tables~\ref{tab:6}, \ref{tab:7}, \ref{tab:8}, \ref{tab:9} we list the target spins of Fig.~\ref{fig:27Spins} for each of the 27 different realizations of each resonance $k$, their one-tangles and the gate error.

\begin{table}[!htbp]
\centering
\begin{tabular}{c|c|c||c|c|c}
     C atoms & $\frac{A}{2\pi}$ (kHz) & $\frac{B}{2\pi}$ (kHz) & C atoms & $\frac{A}{2\pi}$ (kHz) & $\frac{B}{2\pi}$ (kHz) \\  

C1 & -20.72 & 12 & C14 & -19.815 & 5.3\\
C2 & -23.22 & 13 & C15 & -13.971 & 9\\
C3 & -31.25 & 8 & C16 & -4.66 & 7\\
C4 & -14.07 & 13 & C17 & -5.62 & 5\\
C5 & -11.346 & 59.21 & C18 & -36.308 & 26.62\\
C6 & -48.58 & 9 & C19 & 24.399 & 24.81\\
C7 & -8.32 & 3 & C20 & 2.690 & 11\\
C8 & -9.79 & 5 & C21 & 1.212 & 13\\
C9 & 213.154 & 3 & C22 & 7.683 & 4\\
C10 & 17.643 & 8.6 & C23 & -3.177 & 2\\
C11 & 14.548 & 10 & C24 & -4.225 & 0\\
C12 & 20.569 & 41.51 & C25 & -3.873 & 0\\
C13 & 8.029 & 21.0 & C26 & -3.618 & 0\\
C27 & -4.039 & 0
\\
\end{tabular}
    \caption{Hyperfine parameters of the $^{13}$C atoms we considered in Sec.~\ref{SubSec:27Spins}.}
    \label{tab:5}
\end{table}


\begin{table*}[!htbp]
\centering
\scalebox{0.87}{
\begin{tabular}{c|c|c|c}
     Case, resonance & Target spins &   $\epsilon_{p|q}^T/\epsilon_p^*$  & Gate error \\  
(1,1) & \{C5,C10,C11,C12,C19\} & \{0.99992, 0.98416, 0.94779, 0.99407, 0.99794\}  & 0.13441\\
(2,1) & \{C5,C10,C11,C12,C19\} & \{0.99971, 0.95059, 0.86247, 0.98284, 0.99901\} & 0.1162\\
(3,1) & \{C5,C11,C12,C13,C19,C21,C22\} & \{0.8453, 0.99463, 0.96933, 0.98594, 0.98129, 0.97588, 0.99683\} & 0.12447 \\
(4,1) & \{C5,C10,C11,C12,C19\} & \{0.99723, 0.99945, 0.99104, 0.94326, 0.95514\} & 0.15166\\
(5,1) & \{C5,C11,C12,C13,C19,C21,C22\} & \{0.84556, 0.99468, 0.96925, 0.98591, 0.98134, 0.97584, 0.99677\} & 0.1245\\
(6,1) & \{C5,C10,C11,C12,C13,C19\} & \{0.92365, 0.96192, 0.84689, 0.97734, 0.96071, 0.89176\} & 0.15378\\
(7,1) & \{C5,C10,C11,C12,C13,C19\} & \{0.91651, 0.92315, 0.89504, 0.92934, 0.97713, 0.95719\} & 0.16815\\
(8,1) & \{C5,C10,C11,C12,C19\} & \{0.99845, 0.99321, 0.97571, 0.99713, 0.99518\} & 0.15631\\
(9,1) & - & - & -\\
(10,1) & \{C5,C10,C11,C12,C19\} & \{1, 0.97895, 0.93309, 0.99234, 0.9985\} & 0.12788\\
(11,1) & \{C5,C10,C11,C12,C13,C19\} & \{0.83532, 0.95652, 0.86071, 0.90097, 0.9606, 0.95554\} & 0.14142\\
(12,1) & \{C5,C10,C11,C12,C19\} & \{0.99973, 0.95397, 0.87263, 0.98432, 0.99905\} & 0.11661\\
(13,1) & \{C5,C10,C11,C12,C19\} & \{0.99971, 0.95199, 0.8657, 0.98332, 0.99902\} & 0.1163\\
(14,1) & \{C5,C10,C11,C12,C13,C19\} & \{0.88769, 0.97141, 0.81981, 0.96851, 0.9474, 0.89137\} & 0.14173\\
(15,1) & \{C5,C10,C11,C12,C19\} & \{0.99678, 0.99878, 0.98662, 0.94073, 0.9591\} & 0.14517\\
(16,1) & \{C5,C10,C11,C12,C19\} & \{0.99997, 0.97499, 0.92239, 0.99102, 0.99847\} & 0.12438\\
(17,1) & \{C5,C10,C11,C12,C19\} & \{0.99934, 0.9901, 0.96567, 0.99606, 0.99661\} & 0.14638\\
(18,1) & \{C5,C10,C11,C12,C19\} & \{0.99995, 0.98334, 0.94544, 0.9938, 0.99806\} & 0.13321\\
(19,1) & \{C5,C10,C11,C12,C19\} & \{0.99654, 0.99819, 0.98328, 0.93901, 0.96139\} & 0.14137\\
(20,1) & \{C5,C10,C11,C12,C19\} & \{0.99998, 0.98185, 0.94118, 0.9933, 0.99823\} & 0.13119\\
(21,1) & \{C5,C10,C11,C12,C19\} & \{0.98923, 0.99974, 0.99523, 0.82026, 0.88772\} & 0.13951\\
(22,1) & \{C5,C11,C12,C13,C19,C21,C22\} & \{0.86639, 0.99771, 0.96268, 0.98338, 0.98524, 0.97219, 0.99093\} & 0.1276\\
(23,1) & \{C5,C10,C11,C12,C19\} & \{0.99765, 0.92023, 0.82014, 0.99831, 0.99323\} & 0.1157\\
(24,1) & \{C5,C10,C11,C12,C13,C19\} & \{0.89227, 0.97047, 0.8232, 0.9696, 0.94914, 0.8914\} & 0.14306\\
(25,1) & \{C5,C10,C11,C12,C19\} & \{0.99972, 0.9529, 0.86779, 0.98362, 0.99903\} & 0.11638\\
(26,1) & \{C5,C11,C12,C13,C19,C21,C22\} & \{0.83905, 0.99353, 0.97109, 0.98664, 0.98005, 0.97684, 0.99795\} & 0.12382\\
(27,1) & \{C5,C10,C11,C12,C19\} & \{0.99975, 0.95795, 0.87966, 0.98532, 0.99906\} & 0.11708\\
\end{tabular}}
    \caption{Target spins, one-tangles and gate error for each case $\#$ for the $k=1$ resonance of Fig.~\ref{fig:27Spins}.}
    \label{tab:6}
\end{table*}


\begin{table*}[!htbp]
\centering
\scalebox{0.87}{
\begin{tabular}{c|c|c|c}
     Case, resonance & Target spins &   $\epsilon_{p|q}^T/\epsilon_p^*$  & Gate error \\  

(1,2) & \{C12,C13,C20,C21\} & \{0.82937, 0.99907, 0.91408, 0.96915\}  & 0.13008\\
(2,2) & \{C12,C13,C20,C21\} & \{0.84017, 0.99543, 0.91192, 0.96435\} & 0.12729\\
(3,2) & \{C1,C5,C14\} & \{0.97821, 0.99698, 0.97014\} & 0.13208 \\
(4,2) & \{C10,C11,C12,C19\} & \{0.98693, 0.98302, 0.98753, 0.86079\} & 0.065161\\
(5,2) & \{C12,C13,C20,C21\} & \{0.80845, 0.99156, 0.93499, 0.97261\} & 0.10642\\
(6,2) & \{C3,C5, C18\} & \{0.99834, 0.92935, 0.96076\} & 0.087809\\
(7,2) & \{C12,C13,C20,C21\} & \{0.82692, 0.9933, 0.98608, 0.84837\} & 0.055868\\
(8,2) & \{C3,C15,C15\} & \{0.9983, 0.9294, 0.9608\} & 0.074\\
(9,2) & - & - & -\\
(10,2) & \{C12,C13,C20,C21\} & \{0.80833, 0.99161, 0.93501, 0.97265\} & 0.10642\\
(11,2) & \{C12,C13,C20,C21\} & \{0.90511, 0.92114, 0.98317, 0.81608\} & 0.070816\\
(12,2) & \{C10,C11,C12,C19\} & \{0.98109, 0.98771, 0.9929, 0.86759\} & 0.057813\\
(13,2) & \{C12,C13,C20,C21\} & \{0.83675, 0.99692, 0.91267, 0.96598\} & 0.12814\\
(14,2) & \{C12,C13,C20,C21\} & \{0.80291, 0.99763, 0.91743, 0.97761\} & 0.13794\\
(15,2) & \{C12,C13,C20,C21\} & \{0.84717, 0.99125, 0.91016, 0.96064\} & 0.12567\\
(16,2) & \{C12,C13,C20,C21\} & \{0.8637, 0.91792, 0.80065, 0.97057\} & 0.079901\\
(17,2) & \{C12,C13,C20,C21\} & \{0.8985, 0.93371, 0.98448, 0.82279\} & 0.066377\\
(18,2) & \{C3,C5, C18\} & \{0.86836, 0.94177, 0.97611\} & 0.095639\\
(19,2) & \{C12,C13,C20,C21\} & \{0.85612, 0.98321, 0.90738, 0.95496\} & 0.12389\\
(20,2) & \{C12,C13,C20,C21\} & \{0.84067, 0.99518, 0.9118, 0.96411\} & 0.12717\\
(21,2) & \{C10,C11,C12,C19\} & \{0.93295, 0.99959, 0.99859, 0.87879\} & 0.03653\\
(22,2) & \{C12,C13,C20,C21\} & \{0.84877, 0.99004, 0.90971, 0.95971\} & 0.12532\\
(23,2) & \{C12,C13,C20,C21\} & \{0.83847, 0.98904, 0.98655, 0.84717\} & 0.055117\\
(24,2) & \{C12,C13,C20,C21\} & \{0.80217, 0.99742, 0.91749, 0.9778\} & 0.13817\\
(25,2) & \{C10,C11,C12,C19\} & \{0.9923, 0.97629, 0.97869, 0.84976\} & 0.075422\\
(26,2) & \{C12,C13,C20,C21\} & \{0.81029, 0.99069, 0.93467, 0.97199\} & 0.10638\\
(27,2) & \{C10,C11,C12,C19\} & \{0.94817, 0.99822, 0.99993, 0.8788\} & 0.039691\\
\end{tabular}}
    \caption{Target spins, one-tangles and gate error for each case $\#$ for the $k=2$ resonance of Fig.~\ref{fig:27Spins}.}
    \label{tab:7}
\end{table*}

\begin{table*}[!htbp]
\centering
\scalebox{0.87}{
\begin{tabular}{c|c|c|c}
     Case, resonance & Target spins &   $\epsilon_{p|q}^T/\epsilon_p^*$  & Gate error \\  

(1,3) & \{C1,C2,C5\} & \{0.99802, 0.95519, 0.98551\}  & 0.076654\\
(2,3) & \{C4,C5,C15\} & \{0.98289, 0.98811, 0.99895\} & 0.090689\\
(3,3) & \{C1,C2,C5\} & \{0.99973, 0.92645, 0.98235\} & 0.085654 \\
(4,3) & \{C4,C5,C15\} & \{0.96005, 0.99248, 0.97739\} & 0.091492\\
(5,3) & \{C5,C16,C17\} & \{0.97177, 0.9997, 0.95842\} & 0.045708\\
(6,3) & \{C3, C18\} & \{0.93876, 0.88274\} & 0.046505\\
(7,3) & \{C1,C2,C5\} & \{0.99357, 0.96903, 0.98701\} & 0.071505\\
(8,3) & \{C5,C16,C17\} & \{0.97121, 0.99998, 0.95796\} & 0.045857\\
(9,3) & \{C6, C9\} & \{0.85118, 0.89631\} & 0.065242 \\
(10,3) & \{C12,C19\} & \{0.99978, 0.99996\} & 0.037397\\
(11,3) & \{C12,C19\} & \{0.99999, 0.99925\} & 0.03834\\
(12,3) & \{C10,C11,C12\} & \{0.825, 0.97102, 0.99984\} & 0.063887\\
(13,3) & \{C5,C16,C17\} & \{0.93999, 0.96379, 0.99759\} & 0.13709\\
(14,3) & \{C4,C5,C15\} & \{0.98331, 0.99733, 0.95522\} & 0.096136\\
(15,3) & \{C4,C5,C15\} & \{0.99941, 0.99571, 0.99854\} & 0.067369\\
(16,3) & \{C4,C5,C15\} & \{0.99994, 0.9966, 0.99758\} & 0.067963\\
(17,3) & \{C4,C5,C15\} & \{0.97391, 0.98469, 0.99369\} & 0.080809\\
(18,3) & \{C1,C2,C5\} & \{0.99948, 0.92296, 0.98196\} & 0.086644\\
(19,3) & \{C10,C11,C12\} & \{0.86242, 0.95745, 0.98541\} & 0.068323\\
(20,3) & \{C5,C16,C17\} & \{0.97418, 0.97614, 0.96638\} & 0.047334\\
(21,3) & \{C4,C5,C15\} & \{0.98699, 0.98929, 0.94684\} & 0.072855\\
(22,3) & \{C5,C16,C17\} & \{0.97887, 0.99543, 0.94663\} & 0.065835\\
(23,3) & \{C4,C5,C15\} & \{0.99994, 0.99662, 0.99756\} & 0.067977\\
(24,3) & \{C4,C5,C15\} & \{0.98578, 0.99781, 0.95156\} & 0.096777\\
(25,3) & \{C4,C5,C15\} & \{0.99899, 0.98888, 0.9961\} & 0.082447\\
(26,3) & \{C5,C16,C17\} & \{0.97262, 0.99841, 0.95937\} & 0.045567\\
(27,3) & \{C4,C5,C15\} & \{0.9882, 0.98958, 0.99968\} & 0.064595\\
\end{tabular}}
    \caption{Target spins, one-tangles and gate error for each case $\#$ for the $k=3$ resonance of Fig.~\ref{fig:27Spins}.}
    \label{tab:8}
\end{table*}
\begin{table*}[!htbp]
\centering
\scalebox{0.87}{
\begin{tabular}{c|c|c|c|c|c|c|c}
     Case, resonance & Target spins &   $\epsilon_{p|q}^T/\epsilon_p^*$  & Gate error & Case, resonance & Target spins &   $\epsilon_{p|q}^T/\epsilon_p^*$  & Gate error \\  

(1,4) & \{C4,C5,C15\} & \{0.99781, 0.99636, 0.99831\}  & 0.053818 & (1,5) & \{C4,C5,C15\} & \{0.9144, 1, 0.98698\}  & 0.025322 \\
(2,4) & \{C1,C5,C14\} & \{0.97919, 0.99746, 0.99552\} & 0.074751 & (2,5) & \{C4,C5,C15\} & \{0.99903, 0.98728, 0.99631\} & 0.022729\\
(3,4) & \{C1,C5,C14\} & \{0.9942, 0.99517, 0.98206\} & 0.055318 & (3,5) & \{C1,C5,C14\} & \{0.99818, 0.87003, 0.85005\} & 0.056428  \\
(4,4) & \{C4,C5,C15\} & \{0.97017, 0.99492, 0.98328\} & 0.071316 & (4,5) & \{C4,C5,C15\} & \{0.92756, 0.99942, 0.98465\} & 0.025334\\
(5,4) & \{C4,C5,C15\} & \{0.99734, 0.98992, 0.99943\} & 0.053567 & (5,5) & \{C4,C5,C15\} & \{0.99889, 0.99463, 0.91721\} & 0.022455\\
(6,4) & \{C6, C9\} & \{0.84938, 0.89701\} & 0.065279 & (6,5) & - & - & -\\
(7,4) & \{C1,C5,C14\} & \{0.99783, 0.998, 0.96511\} & 0.034986 & (7,5) & \{C4,C5,C15\} & \{0.99773, 0.99232, 0.9129\} & 0.022384\\
(8,4) & \{C4,C5,C15\} & \{0.95584, 0.99667, 0.97678\} & 0.071062 & (8,5) & \{C4,C5,C15\} & \{0.99709, 0.98972, 0.99443\} & 0.022413\\
(9,4) & \{C12, C19\} & \{0.94943, 0.9978\} & 0.055554 & (9,5) & - & - & -  \\
(10,4) & \{C12,C19\} & \{0.99965, 0.99888\} & 0.019047 & (10,5) & \{C10,C12\} & \{0.99749, 0.99963\} & 0.087708\\
(11,4) & \{C12,C19\} & \{0.99983, 0.99806\} & 0.018949 & (11,5) & \{C10,C12\} & \{0.97477, 0.9974\} & 0.047869\\
(12,4) & \{C12,C19\} & \{0.99561, 0.99718\} & 0.039132 & (12,5) & \{C10,C12\} & \{0.98442, 0.99569\} & 0.047928\\
(13,4) & \{C10,C12\} & \{ 0.99916, 0.99639\} & 0.015197 & (13,5) & \{C20,C21\} & \{ 0.99996, 0.98958\} & 0.075203\\
(14,4) & \{C4,C5,C15\} & \{0.99333, 0.98922, 0.99489\} & 0.071773 & (14,5) & \{C4,C5,C15\} & \{0.99984, 0.98508, 0.99759\} & 0.022995\\
(15,4) & \{C4,C5,C15\} & \{0.98711, 0.99892, 0.99249\} & 0.05407 & (15,5) & \{C4,C5,C15\} & \{0.98548, 0.99883, 0.99586\} & 0.019793\\
(16,4) & \{C4,C5,C15\} & \{0.99907, 0.99156, 0.9999\} & 0.053608 & (16,5) & \{C4,C5,C15\} & \{0.97624, 0.99743, 0.98204\} & 0.021121\\
(17,4) & \{C4,C5,C15\} & \{0.95775, 0.99875, 0.99976\} & 0.077026 & (17,5) & \{C4,C5,C15\} & \{0.91076, 0.99995, 0.98755\} & 0.025321\\
(18,4) & \{C2,C5\} & \{1, 0.90604\} & 0.094235 & (18,5) & - & - & -\\
(19,4) & \{C12,C19\} & \{0.99572, 0.99716\} & 0.039168 & (19,5) & \{C10,C12\} & \{0.99986, 0.99962\} & 0.037721\\
(20,4) & \{C16,C17\} & \{0.99994, 0.99937\} & 0.076236 & (20,5) & \{C20,C21\} & \{0.99803, 0.99765\} & 0.069339\\
(21,4) & \{C16,C17\} & \{0.9741, 0.99881\} & 0.046931 & (21,5) & \{C20,C21\} & \{1, 0.98865\} & 0.041343\\
(22,4) & \{C10,C12\} & \{0.99998, 0.99975\} & 0.040316 & (22,5) & \{C10,C12\} & \{0.99592, 0.99997\} & 0.083296\\
(23,4) & \{C4,C5,C15\} & \{0.9695, 0.99502, 0.98297\} & 0.071304 & (23,5) & \{C4,C5,C15\} & \{0.99612, 0.98956, 0.90837\} & 0.022317\\
(24,4) & \{C4,C5,C15\} & \{0.99134, 0.99004, 0.99379\} & 0.071728 & (24,5) & \{C4,C5,C15\} & \{0.99928, 0.98678, 0.99663\} & 0.022791\\
(25,4) & \{C4,C5,C15\} & \{0.9693, 0.99901, 0.99898\} & 0.041347 & (25,5) & \{C4,C5,C15\} & \{0.96988, 0.99834, 0.97894\} & 0.020891\\
(26,4) & \{C4,C5,C15\} & \{0.99922, 0.99176, 0.99993\} & 0.053613 & (26,5) & \{C4,C5,C15\} & \{0.99746, 0.99185, 0.91208\} & 0.022372\\
(27,4) & \{C4,C5,C15\} & \{0.98215, 0.99949, 0.99279\} & 0.075682 & (27,5) & \{C4,C5,C15\} & \{0.99687, 0.99081, 0.91036\} & 0.022346\\
\end{tabular}}
    \caption{Target spins, one-tangles and gate error for each case $\#$ for the $k=4$ and $k=5$ resonance of Fig.~\ref{fig:27Spins}.}
    \label{tab:9}
\end{table*}

\clearpage

\subsection{Parameters for comparison of multi-spin operations with sequential entanglement generation\label{App:SeqVsMulti}}

\begin{table}[!htbp]
\centering
\begin{tabular}{c|c||c|c||c|c}
\hline 
 \multicolumn{1}{c}{C4} \\
\hline
     C atoms & $\epsilon_{p|q}/\epsilon_p^*$  & C atoms & $\epsilon_{p|q}/\epsilon_p^*$ & C atoms & $\epsilon_{p|q}/\epsilon_p^*$ \\  
     
     C1  & 0.0498 & C2  & 0.0968 & C3 & 0.0001 \\
     C4  & 0.9993 & C5  & 0.0645 & C6 & 0.0002 \\
     C7  & 0.0062 & C8  & 0.0062 & C9 & 0.0001 \\
     C10 & 0.0005 & C11 & 0.0044 & C12 & 0.0565 \\
     C13 & 0.0198 & C14 & 0.0490 & C15 & 0.1767\\
     C16 & 0.0289 & C17 & 0.0002 & C18 & 0.0546\\
     C19 & 0.0065 & C20 & 0.0031 & C21 & 0.0098\\
     C22 & 0.0007 & C23 & 0.0005 & C24 & 0\\
     C25 & 0      & C26 & 0      & C27 & 0 \\
     \hline
      \multicolumn{1}{c}{C5} \\
\hline 
     C atoms & $\epsilon_{p|q}/\epsilon_p^*$  & C atoms & $\epsilon_{p|q}/\epsilon_p^*$ & C atoms & $\epsilon_{p|q}/\epsilon_p^*$ \\  
     
     C1  & 0.0925 & C2  & 0.0967 & C3  & 0.0173 \\
     C4  & 0.1130 & C5  & 1      & C6  & 0.0011 \\
     C7  & 0.0048 & C8  & 0.0147 & C9  & 0.0001 \\
     C10 & 0.0003 & C11 & 0      & C12 & 0.0132 \\
     C13 & 0.0148 & C14 & 0.0191 & C15 & 0.0551\\
     C16 & 0.0193 & C17 & 0.0108 & C18 & 0.0541\\
     C19 & 0.0096 & C20 & 0.0162 & C21 & 0.0300\\
     C22 & 0.0005 & C23 & 0.0013 & C24 & 0\\
     C25 & 0      & C26 & 0      & C27 & 0 \\
           \hline
      \multicolumn{1}{c}{C15} \\
      \hline
           C atoms & $\epsilon_{p|q}/\epsilon_p^*$  & C atoms & $\epsilon_{p|q}/\epsilon_p^*$ & C atoms & $\epsilon_{p|q}/\epsilon_p^*$ \\  
     C1  & 0.1296 & C2  & 0.1054 & C3 & 0.0135 \\
     C4  & 0.3086 & C5  & 0.0030 & C6 & 0.0030 \\
     C7  & 0.0031 & C8  & 0.0115 & C9 & 0.0003\\
     C10 & 0.0014 & C11 & 0.0011 & C12 & 0.0143 \\
     C13 & 0.0378 & C14 & 0.0036 & C15 & 0.9999\\
     C16 & 0.0201 & C17 & 0.0149 & C18 & 0.0901\\
     C19 & 0.0132 & C20 & 0.0072 & C21 & 0.0115\\
     C22 & 0.0014 & C23 & 0.0012 & C24 & 0\\
     C25 & 0      & C26 & 0      & C27 & 0 \\
\end{tabular}
    \caption{Nuclear one-tangles when we aim to entangle only C4 or C5 or C15 atom with the electron. The optimal parameters for C4 are: $(N^*,k^*)=(82,3)$, $1-F=0.1133$ and $T^*\approx 0.9337$~ms. There is no other optimal case for C4 within the time constraint of 1.5~ms and unwanted one-tangles below 0.2. The optimal parameters for C5 are: $(N^*,k^*)=(6,3)$, $T^*\approx 68.24~\mu$s, and $1-F=0.1045$; there are other cases that satisfy the time constraint and tolerance of unwanted one-tangles of 0.14, but we selected the fastest option. The optimal parameters for C15 are $(N^*,k^*)=(118,3)$, $T^*\approx 1.3439~$ms, and $1-F=0.1421$; the tolerance for unwanted one-tangles for C15 was 0.31. For the time constraint of 1.5~ms, we found no other optimal case to address only C15.}
    \label{tab:10}
\end{table}

Here we provide the parameters we considered in Sec.~\ref{SubSec:SeqVsMulti}.
To obtain the optimal sequential CR$_x(\pi/2)$ gates with C4, C5 and C15, we set the time constraint of 1.5~ms for each sequential gate. We further require that the unwanted one-tangles of the remaining 26 unwanted spins are below 0.14-0.4
(we cannot satisfy the unwanted one-tangle bound of 0.14 for all cases of addressing each Cj nucleus). The unwanted one-tangles for the optimal choices we found are listed in Table~\ref{tab:10}. The rotation angles and axes for the sequential and multi-spin gates of Fig.~\ref{fig:SeqVsMulti} are listed in Table~\ref{tab:11}.

\begin{table}[!htbp]
\centering
\begin{tabular}{c|c|c|c}
\hline
      \multicolumn{1}{c}{Multi-spin gate} \\
\hline 
       & C4  & C5 & C15 \\  
     $n_{x,0}$ & -0.7532 & 0.9468 & 0.6832 \\
     $n_{y,0}$ & 0 & 0 & 0 \\
     $n_{z,0}$ & -0.6566 & 0.3219 & 0.7302 \\
     $n_{x,1}$ & 0.7754  & -0.9844 & -0.6996 \\
     $n_{y,1}$ & 0 & 0 & 0 \\
     $n_{z,1}$ & -0.6314 & 0.1758 & 0.7145 \\     
     $\phi/(\pi/2)$ & 1.5136 & 1.0028 & 1.9021 \\
     \hline
      \multicolumn{1}{c}{Sequential gates} \\
\hline 
& C4  & C5 & C15 \\  
     $n_{x,0}$ & -1 & 0.99996 & -1 \\
     $n_{y,0}$ & 0 & 0 & 0 \\
     $n_{z,0}$ & 0 & 0.00905 & 0 \\
     \midrule
     $n_{x,1}$ & 0.99952  & -0.98996 & 0.99977 \\
     $n_{y,1}$ & 0 & 0 & 0 \\
     $n_{z,1}$ & 0.031 & -0.14314 & 0.021 \\  
     $\phi/(\pi/2)$ & 0.98282 & 1.0048 & 0.099227 \\
\end{tabular}
    \caption{Nuclear rotation axes and rotation angles for the multi-spin operation (case $\#$ 23, $k=3$), and the sequential gates we discussed in Sec.~\ref{SubSec:SeqVsMulti}.}
    \label{tab:11}
\end{table}

\subsection{QEC with CR\texorpdfstring{$_{xz}$}{xz} multi-spin gates \label{App:Toffoli}}

\begin{figure*}[!htbp]
    \centering
    \includegraphics[scale=0.6]{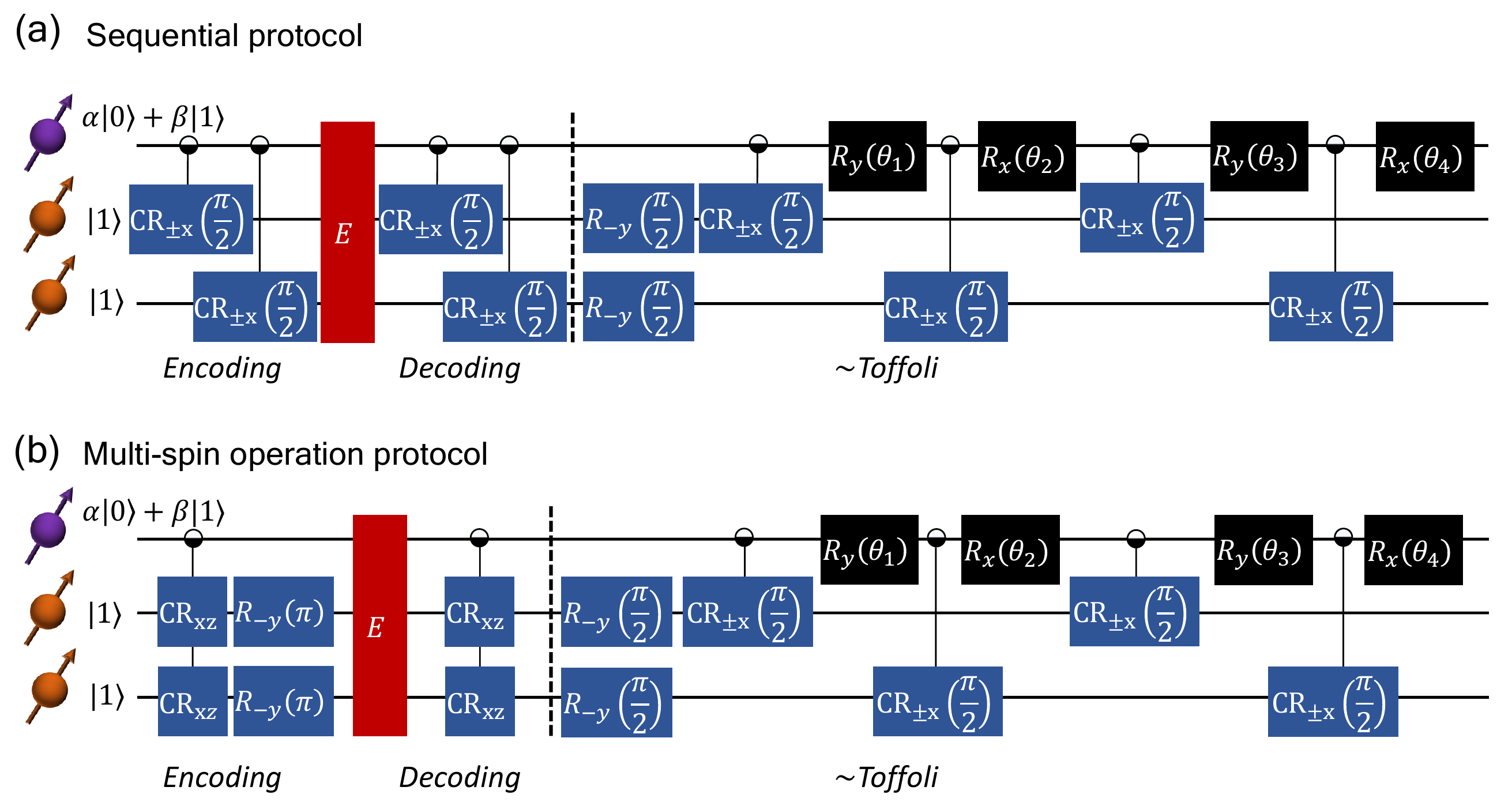}
    \caption{Circuit diagrams to correct a bit-flip on the electron for (a) the sequential approach that utilizes the CR$_{x}(\pi/2)$ gates and (b) the multi-spin operation protocol. $E$ denotes the bit-flip error, which can happen either on the electron or on one of the nuclei. In (b), we further require the $R_{-y}(\pi)$ unconditional rotations on the nuclear spins for the encoding step. These rotations can be either performed at the encoding step after the CR$_{xz}$ gate or the decoding step before the CR$_{xz}$ gate. The half-white, half-black circles of the control of the CR$_{\pm x}(\pi/2)$ gates indicate that depending on the electron's state, the nuclear spin rotates by $R_{x}(\pi/2)$ or $R_{x}(-\pi/2)$. For the CR$_{xz}$ gates, the half-white, half-black notation means that depending on the elctron's state, the nuclear spin rotates by $R_{\textbf{n}_0}$ or $R_{\textbf{n}_1}$.}
    \label{fig:QEC_Circuits}
\end{figure*}

In this section, we provide the details of how we implement the three-qubit bit-flip code using the CR$_{xz}$ gates. First, we will explain the three-qubit bit-flip code that utilizes the sequential CR$_{\pm x}(\pi/2)$ gates. The circuit to implement the QEC code using the latter scheme is shown in Fig.~\ref{fig:QEC_Circuits}(a); the CNOT gates used in the usual QEC code have been expressed in terms of the CR$_{\pm x}(\pi/2)$ gates as well as initialization of the nuclei into the $|1\rangle$ state, while the Toffoli gate has been decomposed into single-qubit gates and CR$_{\pm x}(\pi/2)$ gates. The CR$_{\pm x}(\pi/2)$ shown in Fig.~\ref{fig:QEC_Circuits} are given by:
\begin{equation}
    \text{CR}_{\pm x}(\pi/2)=\sigma_{00}\otimes R_{x}(\pi/2)+\sigma_{11}\otimes R_{-x}(\pi/2),
\end{equation}
where we define $R_{x}(\phi)=e^{-i(\phi/2) \sigma_x}$. The half-white, half-black control implies that when the electron is in the $|0\rangle$ state, the nuclear spin still undergoes a rotation, but it differs from the one it undergoes when the electron is in the $|1\rangle$ state. The electron rotation angles $\theta_j$ have to satisfy particular relations to ensure recovery of the electron's state.

Let us start with the sequential protocol. If a bit-flip happens on the electron, then the final (after the correction) three-qubit state has the form [in the basis $\{|000\rangle,|001\rangle,|010\rangle,|011\rangle,|100\rangle,|101\rangle,|110\rangle,|111\rangle\}$]:
\begin{equation}
    |\psi_f\rangle=\frac{1}{2}\begin{pmatrix}
    -\beta \cos(\frac{\Theta}{2})+i\alpha \sin(\frac{\Theta}{2}) \\
    -\beta \cos(\frac{\Theta}{2})+i\alpha\sin(\frac{\Theta}{2})\\
    -\beta \cos(\frac{\Theta}{2})+i\alpha \sin(\frac{\Theta}{2}) \\
    -\beta \cos(\frac{\Theta}{2})+i\alpha \sin(\frac{\Theta}{2}) \\
    -\alpha \cos(\frac{\Theta}{2})+i\beta \sin(\frac{\Theta}{2}) \\
    -\alpha \cos(\frac{\Theta}{2})+i\beta \sin(\frac{\Theta}{2})\\
    -\alpha \cos(\frac{\Theta}{2})+i\beta \sin(\frac{\Theta}{2})\\
    -\alpha \cos(\frac{\Theta}{2})+i\beta \sin(\frac{\Theta}{2}),
    \end{pmatrix},
\end{equation}
where we have defined $\Theta=\theta_1-\theta_2-\theta_3+\theta_4$. Clearly, in order to recover the initial state of the electron it has to hold that $\Theta=(2k+1)\pi$. Under this condition one can verify that the final state is $(\alpha|0\rangle+\beta|1\rangle)|x\rangle|x\rangle$, with $|x\rangle=(|0\rangle+|1\rangle)/\sqrt{2}$.

If no error occurs, then the final state is:
\begin{equation}
    |\psi_f\rangle=\frac{1}{2}\begin{pmatrix}
    \alpha \cos(\frac{\tilde{\Theta}}{2})+i\beta\sin(\frac{\tilde{\Theta}}{2}) \\
    -\alpha \cos(\frac{\tilde{\Theta}}{2})-i\beta\sin(\frac{\tilde{\Theta}}{2})\\
    -\alpha \cos(\frac{\tilde{\Theta}}{2})-i\beta\sin(\frac{\tilde{\Theta}}{2})\\
    \alpha \cos(\frac{\tilde{\Theta}}{2})+i\beta\sin(\frac{\tilde{\Theta}}{2})\\
    \beta \cos(\frac{\tilde{\Theta}}{2})+i\alpha\sin(\frac{\tilde{\Theta}}{2})\\
    -\beta \cos(\frac{\tilde{\Theta}}{2})-i\alpha\sin(\frac{\tilde{\Theta}}{2})\\
    -\beta \cos(\frac{\tilde{\Theta}}{2})-i\alpha\sin(\frac{\tilde{\Theta}}{2})\\
    \beta \cos(\frac{\tilde{\Theta}}{2})+i\alpha\sin(\frac{\tilde{\Theta}}{2}),
    \end{pmatrix},
\end{equation}
where we have defined $\tilde{\Theta}=\theta_1+\theta_2-\theta_3-\theta_4$. Clearly in order to preserve the initial state of the electron it has to hold that $\tilde{\Theta}=2\kappa\pi$. In this case, the final state is $(\alpha|0\rangle+\beta|1\rangle)|\bar{x}\rangle|\bar{x}\rangle$, where $|\bar{x}\rangle=(|0\rangle-|1\rangle)/\sqrt{2}$.

Based on the above observations we find that we can satisfy both conditions for $\Theta$ and $\tilde{\Theta}$ if we chose the $\theta_j$ to satisfy:
\begin{equation}
    \theta_2=\theta_1,~~~\theta_3=\theta_1-3\pi/2,~~~\theta_4=\theta_3+\pi.
\end{equation}
It is clear that in the sequential protocol if no error occurs on the electron's state, the CR$_{\pm x}(\pi/2)$ gates of the encoding add up with those of the decoding step to produce CR$_{\pm x}(\pi)$ gates which flip both nuclear spins into the $|00\rangle$ state and hence the subsequent Toffoli gate is not activated. We should further mention, that in the case where a single bit-flip happens on either the first or second nuclear spin, and we are interested in preserving the initial state of the electron, then the $\theta_j$ need to be constrained further. That is, we have two more conditions, namely if the bit-flip happens on the first nucleus, then the angles need to satisfy $\theta_{1}+\theta_2+\theta_3+\theta_4=2\pi$, whereas if the bit-flip happens on the second nucleus, the angles need to satisfy $\theta_1-\theta_2+\theta_3-\theta_4=2\pi$. Combining these constraints with the $\theta_j$ constraints when a bit-flip or no bit-flip happens on the electron, we find that $\theta_j$ need to satisfy
\begin{equation}
\theta_1=\theta_4=-\pi/4=-\theta_2=-\theta_3.  
\end{equation}

\begin{figure}[!htbp]
    \centering
    \includegraphics[scale=0.6]{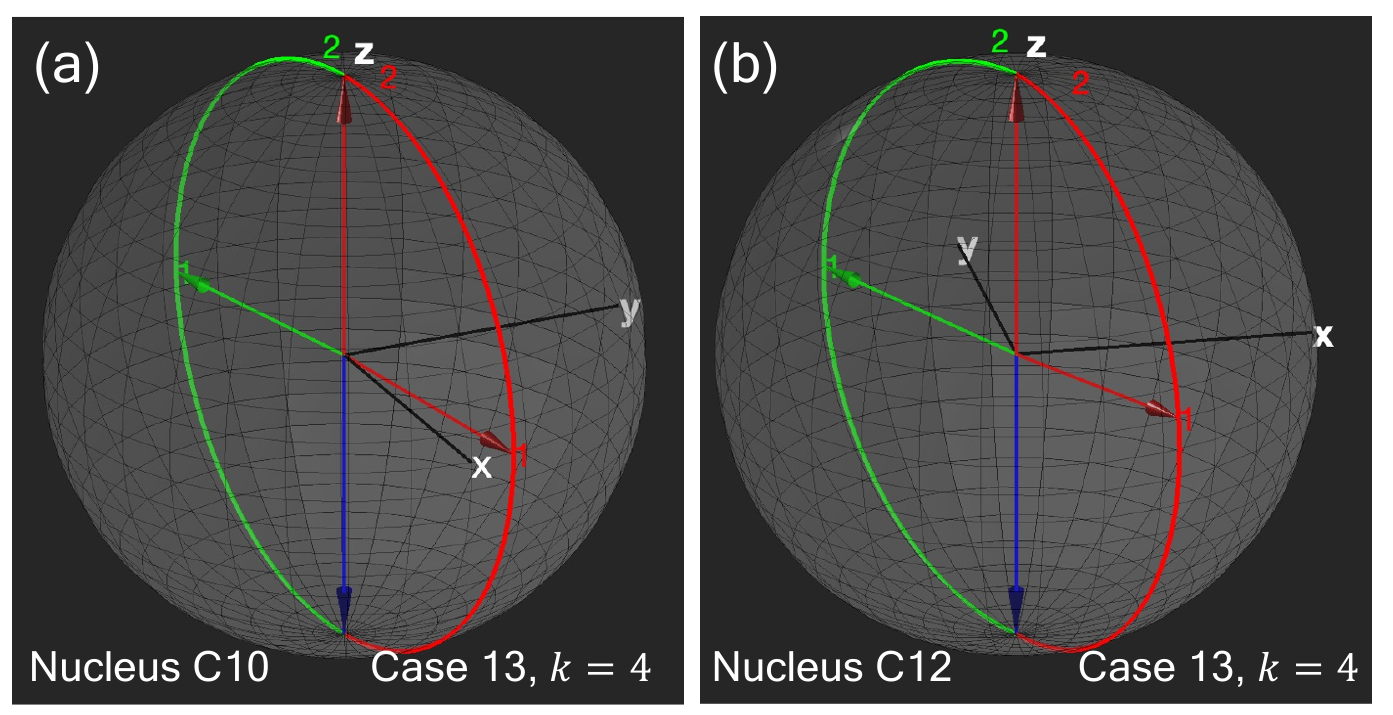}
    \caption{Evolution of nuclear spins C10 (a) and C12 (b) up to the decoding, if no error occurs on the electron during the CR$_{xz}$ QEC protocol. Initially, both nuclei are in the $|1\rangle$ state (blue arrow). If the electron starts in the $|0\rangle$ ($|1\rangle$) state, the nuclear spins follow the green (red) trajectory. The final state of each spin is indicated with the red arrow pointing to the north pole of the Bloch sphere (the final green and red arrows coincide).}
    \label{fig:App_BlochSphere}
\end{figure}

Let us return to the CR$_{xz}$ QEC protocol and consider first the case where no error happens on the electron. Now, the CR$_{xz}$ gates of the encoding and decoding would again add up, but in general, the total gate would not be equivalent to a bit-flip operation that brings the state $|11\rangle$ of the nuclei into the $|00\rangle$ state. However, if we put $R_y(-\pi)$ gates on the nuclei at the encoding step and right after the CR$_{xz}$ gate, we would then have the total gate CR$_{xz}[\mathds{1}_{2\times 2}\otimes R_y(-\pi)\otimes R_y(-\pi)]$CR$_{xz}$. If we consider the case when the electron is in the $|0\rangle$ state and consider the part of the gate acting on the first nuclear spin, we find that the total gate is:

\begin{widetext}
\begin{equation}
\begin{split}
    R_{\textbf{n}_0}^{(0)}R_y(-\pi)R_{\textbf{n}_0}^{(0)}&=[\cos\frac{\phi}{2}-i\sin\frac{\phi}{2}(n_{x,0}\sigma_x+n_{z,0}\sigma_z)][i\sigma_y][\cos\frac{\phi}{2}-i\sin\frac{\phi}{2}(n_{x,0}\sigma_x+n_{z,0}\sigma_z)]\\&=
    [\cos\frac{\phi}{2}i\sigma_y+i\sin\frac{\phi}{2}(n_{x,0}\sigma_z-n_{z,0}\sigma_x)][\cos\frac{\phi}{2}-i\sin\frac{\phi}{2}(n_{x,0}\sigma_x+n_{z,0}\sigma_z)]
    \\&=\cos^2\frac{\phi}{2}i\sigma_y+\sin^2\frac{\phi}{2}(n_{x,0}\sigma_z-n_{z,0}\sigma_x)(n_{x,0}\sigma_x+n_{z,0}\sigma_z)
    \\&=
    \cos^2\frac{\phi}{2}i\sigma_y+\sin^2\frac{\phi}{2}(n^2_{x,0}+n_{z,0}^2)i\sigma_y=i\sigma_y,
    \end{split}
\end{equation}
\end{widetext}
where in the last line we have used the fact that time-symmetric $\pi$-sequences do not produce an $n_y$ rotation component and thus, $n_x^2+n_z^2=1$. (Note that we consider here a $\pi$-pulse sequence that produces the same rotation angles irrespective of the electron's state i.e., $\phi_0=\phi_1\equiv \phi$.) The same analysis follows for the second nuclear spin, and for the case when the electron is in state $|1\rangle$ and thus, CR$_{xz}[\mathds{1}_{2\times 2}\otimes R_y(-\pi)\otimes R_y(-\pi)]$CR$_{xz}$ is equivalent to $i^2\mathds{1}_{2\times 2}\otimes \sigma_y\otimes \sigma_y$ if no error occurs on the electron. Thus, we see that the CR$_{xz}[\mathds{1}_{2\times 2}\otimes R_y(-\pi)\otimes R_y(-\pi)]$CR$_{xz}$ gate leads to the desired bit-flip operation of the nuclei, deactivating the subsequent Toffoli-gate. This is verified schematically in Figs.~\ref{fig:App_BlochSphere}(a) and (b), where we show the Bloch sphere evolution of nuclear spins C10 and C12 [that we considered in Sec.~\ref{SubSec:SeqVsMulti}] respectively, up to the decoding, assuming no error has occurred on the electron.

Now, let us assume that a bit-flip error happens on the electron. At the encoding step, which includes the $R_y(-\pi)$ rotations, the encoded state becomes:
\begin{equation}
\begin{split}
    |\psi_{\text{enc}}\rangle&=\alpha|0\rangle\otimes R_y(-\pi)R_{\textbf{n}_0}^{(1)}|1\rangle\otimes R_y(-\pi)R_{\textbf{n}_0}^{(2)}|1\rangle\\&+
    \beta|1\rangle\otimes R_y(-\pi)R_{\textbf{n}_1}^{(1)}|1\rangle\otimes R_y(-\pi)R_{\textbf{n}_1}^{(2)}|1\rangle.
    \end{split}
\end{equation}
After the bit-flip and the decoding step the state becomes:
\begin{equation}
\begin{split}
    &|\psi_{\text{dec}}\rangle=
    \\&
    ~~~\alpha|1\rangle\otimes R_{\textbf{n}_1}^{(1)}R_y(-\pi)R_{\textbf{n}_0}^{(1)}|1\rangle\otimes R_{\textbf{n}_1}^{(2)}R_y(-\pi)R_{\textbf{n}_0}^{(2)}|1\rangle
    \\&+
    \beta|0\rangle\otimes R_{\textbf{n}_0}^{(1)}R_y(-\pi)R_{\textbf{n}_1}^{(1)}|1\rangle\otimes R_{\textbf{n}_0}^{(2)}R_y(-\pi)R_{\textbf{n}_1}^{(2)}|1\rangle.
    \end{split}
\end{equation}
These four sets of gates approximately leave the nuclei at the state $|11\rangle$ such that we activate the Toffoli-gate, recovering the electron's state with high probability. To see this, let us consider $R_{\textbf{n}_0}^{(j)}R_y(-\pi)R_{\textbf{n}_1}^{(j)}$, which reads:
\begin{equation}\label{Eq.I9}
    \begin{split}
&R_{\textbf{n}_0}^{(j)}R_y(-\pi)R_{\textbf{n}_1}^{(j)}=       \hat{\textbf{y}}\cdot (\textbf{n}_1\times \textbf{n}_0)^{(j)}\sin^2\frac{\phi}{2}\mathds{1}
\\&+
i\sin\phi[(n_{x,0}^{(j)}-n_{x,1}^{(j)})\sigma_z-(n_{z,0}^{(j)}-n_{z,1}^{(j)})\sigma_x]
\\&+
i\sqrt{G_1^{(j)}}\sigma_y,
    \end{split}
\end{equation}
where $G_1^{(j)}=\cos^2\frac{\phi^{(j)}}{2}+\sin^2\frac{\phi^{(j)}}{2}(\textbf{n}_0\cdot\textbf{n}_1)^{(j)}$. Since for the multi-spin gates we choose the number of iterations $N$ such that $G_1^{(j)}$ is minimized for all $j$-nuclear spins (i.e., $G_1^{(j)}\approx 0$, $\forall j$), then the $y$-component of the composite rotation vanishes. Further, for the CR$_{xz}$ gates, and considering the CPMG sequence, it holds that $n_{x,0}\cdot n_{x,1}<0$, and that $n_{z,0}\approx n_{z,1}+\delta$, where $\delta$ is small, as we will show shortly [for brevity, we drop superscripts $j$ which refer to the $j$-th spin]. Let us further consider the action of the CR$_{xz}$ gate on a single nuclear spin (similar analysis holds for more nuclei). As we mentioned in Appendix~\ref{App:EvolOper}, the evolution of a nuclear spin over one unit of the CPMG sequence is defined by the Hamiltonians $H_j = \frac{1}{2}[(\omega_L+s_jA)Z+(s_jB)X]$, where the nuclear rotations over one unit of the sequence are: $R_{\textbf{n}_0}=e^{-iH_0 t/4}e^{-iH_1 t/2}e^{-i H_0 t/4}$ and $R_{\textbf{n}_1}=e^{-i H_1 t/4}e^{-i H_0 t/2}e^{-i H_1 t/4}$. Letting $\cos\theta_j=(\omega_L+s_jA)/\omega_j$ [where $\omega_j=\sqrt{(\omega_L+s_jA)^2+(s_jB)^2}$] and $\sin \theta_j = s_jB/\omega_j$, we find the SU(2) decomposition of Eq.~(\ref{Eq.I9}), and focusing on the $z$-components we find that it holds:

\begin{equation}
\begin{split}
    \sin\frac{\phi}{2}(n_{z,1}-n_{z,0})&=
    2\sin(\theta_0-\theta_1)\Big[\sin\theta_1\sin\frac{t \omega_0}{4}\sin^2\frac{t\omega_1}{8}
    \\&+
    \sin\theta_0 \sin\frac{t\omega_1}{4}\sin^2\frac{t\omega_0}{8}\Big]
    \end{split}.
\end{equation}
Note that $\sin(\theta_0-\theta_1)=\omega_L B (s_1-s_0)/(\omega_0\omega_1)$, and that $\sin\theta_j = s_jB/\omega_j$, meaning that for $\omega_L\gg A,B$, we have:

\begin{equation}
\begin{split}
    \sin\frac{\phi}{2}(n_{z,1}-n_{z,0})&\approx 
    2\frac{(s_1-s_0)B}{\omega_L}\Big[\frac{s_1 B}{\omega_L}\sin\frac{t \omega_0}{4}\sin^2\frac{t\omega_1}{8}
    \\&+
    \frac{s_0 B}{\omega_L} \sin\frac{t\omega_1}{4}\sin^2\frac{t\omega_0}{8}\Big] 
    \end{split},
\end{equation}
or:
\begin{equation}
    \sin\frac{\phi}{2}(n_{z,1}-n_{z,0})\propto  \left(\frac{B}{\omega_L}\right)^2.
\end{equation}
Thus, in Eq.~(\ref{Eq.I9}), we will have suppressed $x$-component of rotation, meaning that each nucleus will rotate approximately around the $z$-axis irrespective of the electron's state. Since each nucleus is initialized in the $|1\rangle$ state, an $R_z$ rotation will only approximately lead to a global phase. The non-zero difference of the $z$-axis components is what makes our CR$_{xz}$ QEC protocol probabilistic, since the disentanglement at the decoding step is imperfect, but it succeeds with high probability because the difference in the $z$-components is, in general, small (the external $B$-field is typically chosen such that $\omega_L\gg A,B$).


\subsection{Parameters for three-qubit QEC with CR\texorpdfstring{$_x(\pi/2)$}{x pi/2} \label{App:OptimalCRx}}

In Table~\ref{tab:12}, we provide a list of the optimal CR$_x(\pi/2)$ gates for nuclear spins C10 and C12 using the sequential entanglement scheme, we considered in Sec.~\ref{SubSec:SeqVsMulti}.

\begin{table}[!htbp]
\centering
\scalebox{0.85}{
\begin{tabular}{c|c|c|c|c|c}
\hline 
 \multicolumn{1}{c}{C12} \\
\hline
       & Gate time ($\mu$s)  & Gate error & $N^*$ & $k^*$ & $\phi/(\pi/2)$ \\  
     $\#1$ &  170.3095 & 0.1080 & 8 & 5 & 0.97567 \\
     $\#2$ &  208.156  & 0.0867 & 8 & 6 & 0.95198 \\
     $\#3$ &  276.7529 & 0.0732 & 9 & 7 & 1.0393 \\
     $\#4$ &  319.33   & 0.0319 & 9 & 8 & 1.0027  \\
     $\#5$ &  361.9076  & 0.0273 & 9 & 9 & 0.96137  \\
     $\#6$ &  449.4277  & 0.0238 & 10 & 10 & 1.0172   \\
     $\#7$ &  1054.9725  & 0.2428 & 446 & 1 & 1.066   \\
     $\#8$ &  1125.9347  & 0.0800 & 28 & 9 & 1.0091  \\
     $\#9$ &  1303.3404  & 0.0549 & 29 & 10 &  1.05 \\
     $\#10$ &  1457.092  & 0.0604 & 88 & 4 &  1.0578 \\
     \hline 
 \multicolumn{1}{c}{C10} \\
\hline 
       & Gate time ($\mu$s)  & Gate error & $N^*$ & $k^*$ & $\phi/(\pi/2)$ \\  
      $\#1$ &  645.0498 & 0.3839 & 39 & 4 & 1.0038 \\
      $\#2$ &  850.6152 & 0.4150 & 40 & 5 & 1.0118 \\
      $\#3$ &  985.2959 & 0.4208 & 417 & 1 & 0.98772  \\
      $\#4$ &  1039.6407 & 0.4292 & 40 & 6 & 0.98976  \\
      $\#5$ &  1290.0997 & 0.4821 & 42 & 7 & 1.0117  \\
      $\#6$ &  1358.6214 & 0.4732 & 115 & 3 & 1.0016  \\
      $\#7$ &  1524.0188 & 0.5455 & 43 & 8 & 1.0032  \\
      $\#8$ &  1767.3893 & 0.4753 & 44 & 9 & 0.98882  \\
      $\#9$ &  2065.1046 & 0.5144 & 46 & 10 & 0.98996  \\
      $\#10$ &  2431.3417 & 0.3214 & 49 & 11 & 1.0034  \\
      $\#11$ &  2530.5801 & 0.2248 & 119 & 5 & 0.98991   \\
      $\#12$ &  2825.9326 & 0.1953 & 52 & 12 & 1.006    \\
      $\#13$ &  3144.9133 & 0.1058 & 121 & 6 & 1.006    \\
      $\#14$ &  4075.8643 & 0.1124 & 345 & 3 & 0.99506    \\
      $\#15$ &  4453.9155 & 0.1326 & 65 & 15 & 1.0064     \\
      $\#16$ &  4572.0565 & 0.1679 & 129 & 8 & 0.9904   \\

\end{tabular}}
    \caption{Optimal iterations ($N^*$) and resonances ($k^*$) to perform CR$_x(\pi/2)$ between the electron and C12 or C10. We provide a list of optimal cases for $T\leq 1.5$~ms for C12. For C10 we could not satisfy the bound of unwanted one-tangles for this time constraint, so we further list cases for $T$ up to 5~ms.}
    \label{tab:12}
\end{table}

\clearpage

\end{document}